\documentclass[fleqn,10pt]{wlscirep}
\usepackage{amsthm}
\usepackage{silence}
\usepackage[utf8]{inputenc}
\usepackage[T1]{fontenc}

\usepackage{graphicx}
\usepackage{multirow}
\usepackage{amsmath,amssymb,amsfonts}
\usepackage{mathrsfs}
\usepackage[title]{appendix}
\usepackage{xcolor}
\usepackage{textcomp}
\usepackage{manyfoot}
\usepackage{booktabs}
\usepackage{algorithm}
\usepackage{algorithmicx}
\usepackage{algpseudocode}
\usepackage{listings}
\usepackage{subcaption}
\usepackage{float}
\usepackage[left]{lineno}

\title{AI-Driven Lumped-Element Modeling of Human Respiratory System for Studying Voice Mechanics}

\author[1,2]{Maruf Md Ikram}
\author[2,*]{Maryam Naghibolhosseini}
\author[1]{Mohsen Zayernouri}

\affil[1]{Department of Mechanical Engineering, Michigan State University, East Lansing, MI 48824, USA}

\affil[2]{Department of Communicative Sciences and Disorders, Michigan State University, East Lansing, MI 48824, USA}

\affil[*]{Corresponding author: naghib@msu.edu}

\begin{abstract}
A predictive physics-based model of human respiratory, phonatory, and articulatory subsystems is developed to simulate voice production. Representing lungs, compressible airways, and vocal folds as spring-damper-mass controlled piston-cylinder systems, our mathematical model robustly captures the intricate dynamics of airways during phonation. The nonlinear viscoelastic properties of lung tissues and compressible airways were investigated, yielding a responsive and expressive baseline respiratory model with the capability to further extend into a patient-specific model for both respiration and phonation. The resulting framework was subsequently integrated with a mechanical representation of the vocal tract, governed by the glottal area waveform (GAW) capturing the motion of vocal folds during sustained phonation. The GAW is extracted from laryngeal high-speed videoendoscopy data of a normophonic participant using deep learning. Our novel paradigm transcends beyond modeling the respiratory system, enabling AI-driven modeling of vocalization, including vocal fold dynamics, interactions with flow aerodynamics, and flow resistances, induced by the oscillatory behavior of vocal folds. Our investigation leads to the first-ever simulation of respiratory dynamics for vocalization, directly advancing the prediction of subglottal pressure distributions, impossible to measure directly and noninvasively in humans, dynamic resistances, and energy transfer mechanisms during phonation in voice mechanics.

\end{abstract}

\begin{document}

\flushbottom
\maketitle
%
%

\thispagestyle{empty}


\section*{Introduction}

Human voice production is a complex process emerging from the closely coupled interactions of the respiratory, phonatory, and articulatory subsystems. The respiratory subsystem generates the driving force for phonation by regulating airflow and subglottal pressure through the biomechanics of lung expansion/contraction, and airways flow resistances \cite{zhang2016mechanics}. The phonatory subsystem, within the larynx, interacts with the airflow, leading to self-sustained oscillations of vocal folds and producing a quasi-periodic acoustic signal \cite{luizard2023flow}. Finally, the articulatory subsystem, consisting of the supraglottal vocal tract influences the voice by creating resonances and anti-resonances via changing its shape \cite{weerathunge2022auditory}. Each subsystem contributes distinctively to phonation by their interdependent but collaborative dynamics. Hence, to create an accurate model of phonation, these subsystems should collectively be included within the modeling framework. Despite this interdependence, computational models of phonation have historically only emphasized the phonatory and articulatory subsystems. As a result, these models often considered a uniform subglottal inlet condition using an averaged or sinusoidal flow to drive the pressure profiles \cite{naseri2023towards}. Although such simplifications capture gross features of glottal flow, they overlook subject-specific respiratory mechanics and fail to reproduce the irregularities, asymmetries, and nonlinear variations inherent in true phonation dynamics. Physiological and experimental studies have shown that the glottal and supraglottal flow fields are highly sensitive to inlet conditions, and even small variations in subglottal pressure/velocity can alter jet asymmetry, turbulence intensity, and acoustic loading \cite{bodaghi2021effect, schickhofer2019compressible, zhang2016mechanics}. To address these challenges, the present study focuses on developing a holistic lumped-element model, including all voice production subsystems, to generate subject-specific inlet conditions, essential for developing accurate and clinically meaningful computational models of phonation.
\par
As the airflow during exhalation enters the glottis, it develops into complex flow phenomena including jet formation, vortex shedding, turbulence, and the Coandă effect, all of which shape the resulting sound field and phonation stability \cite{vsvec2025application, inoue2024nonlinear}. These behaviors can be visualized most clearly in computational modeling studies, which provide access to detailed velocity fields, pressure distributions, and temporal flow evolution that cannot be captured in vivo in humans due to experimental limitations \cite{sadeghi2019computational, dollinger2023computational, cveticanin2012review}. Such computational models have revealed that glottal jet asymmetry, supraglottal recirculation, and acoustic loading are strongly dependent on how the inlet condition, especially it's temporal evolution, is prescribed \cite{tao2007asymmetric, sidlof2010finite, de2015computational, xue2010computational}. Therefore, including a realistic inlet condition is critical toward having a robust and stable model of phonation. This underscores the need for considering the respiratory subsystem modeling to generate subject-specific time-dependent boundary conditions for the larynx. While researchers have recently incorporated time-dependent inlet conditions in phonation modeling, the implementation was done using a uniform subglottal pressure/velocity \cite{lehoux2021subglottal, alipour2004flow, hofmans2003unsteady, alipour1995experimental}. These studies showed the glottal jet could still develop asymmetries, producing lateral skewing, wall attachment consistent with the Coandă effect, and vortex structures altering acoustic loading and phonatory stability \cite{zheng2011computational, tao2007asymmetric, xue2014subject}. Later investigations demonstrated that even slight time-dependent non-uniformities at the inlet could significantly change the jet trajectory, turbulence onset, and the strength of supraglottal vortical structures \cite{zheng2011computational, sidlof2010finite, scherer2001intraglottal}. More recently, the (semi-)sinusoidal inlet conditions presenting a time-varying pressure/velocity at the subglottal boundary were introduced to approximate the impact of rhythmic breathing in voice production \cite{jiang2017computational, jiang2022computational}. While this represented a progress over the use of a steady inflow, such sinusoidal inputs imposed artificial regularity and could not reproduce the nonlinear and irregular characteristics of true subglottal dynamics \cite{zhang2016cause}. Moreover, sinusoidal conditions are idealistic and therefore disregard subject-specific variations linked to tbe respiratory effort, vocal fold vibratory dynamics, and supraglottal flow resistance. \cite{ibarra2021estimation, howe2013voicing, abur2022impact}. These limitations highlight the need for more advanced approaches that can generate inlet conditions based on the subject's respiratory function rather than relying on idealized/averaged assumptions.
\par
Lumped-element models (LEMs) have  been used to study respiratory mechanics during breathing \cite{marconi2020silico, bersani2017interaction, athanasiades2000energy, sharp1969total}. Early work by Sharp et al. established methods to quantify respiratory compliance and resistance from pressure-volume relationships \cite{sharp1969total}. Later, Athanasiades et al. developed a nonlinear LEM of the respiratory system \cite{athanasiades2000energy}. Later, Bersani et al. proposed a Windkessel-type multi-compartment 0-D models of lung and airways mechanics dividing the respiratory system into the rigid upper airways, the compliant intrapleural space, and the alveolar gas-exchange region \cite{bersani2017interaction}. Their framework explicitly coupled airways flow resistance, tissue compliance, and alveolar gas dynamics, highlighting the suitability of lumped models for patient-specific applications. More recently, Marconi and Lazzari expanded the model of respiration by Athanasiades et al. and applied it to healthy and pathological conditions, capturing essential features of breathing cycles, such as alveolar volume changes, intrapleural pressure dynamics, and flow-volume loops \cite{marconi2020silico}. These studies established LEMs as a justifiable approach to capture pressure-flow-volume behaviors of the respiratory system. 
\par
While LEMs have been used to model lung/airways mechanics, their applications have been limited to breathing. In this study, we present the first LEM that extends beyond modeling respiration and explicitly captures phonation mechanics of human's sustained phonation production. Our approach models the function of the lungs and respiratory system, with active dynamic vocal folds, bridging the gap between pulmonary airflow models and phonatory dynamics. This approach leads to an accurate and robust reconstruction of time-dependent glottal inlet conditions, generating a subject-specific computational model of vocalization. This work establishes a benchmark study with subject-specific data from a normophonic participant, which creates a foundational model that can be adjusted in the future to generate the subglottal conditions for various pathologies affecting voice production.
\par
\section*{Methods}

\subsection*{Data Collection and Processing}

\begin{figure}[htbp]
\centering
\includegraphics[width=1\textwidth]{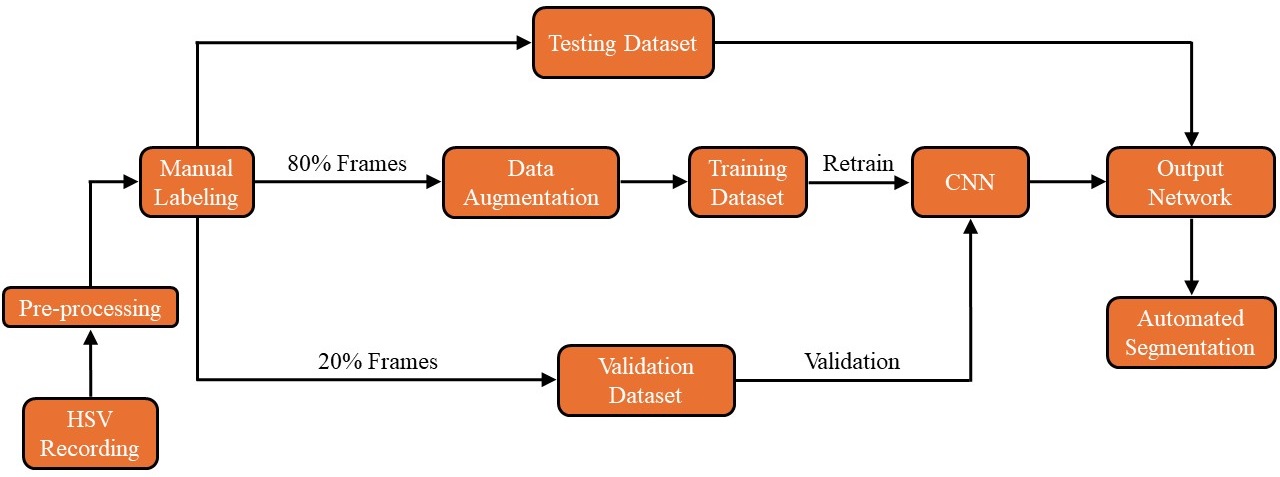}
\caption{Workflow of the convolutional neural network (CNN)–based framework used for automated glottal area segmentation from laryngeal high-speed videoendoscopy (HSV) recordings. HSV recordings are first pre-processed and manually labeled to generate ground-truth masks. The labeled frames are then divided into training, validation, and testing subsets. Eighty percent of the labeled frames are used for data augmentation and training, while the remaining twenty percent are reserved for the validation. The trained network produces an output segmentation model, which is subsequently applied to the independent testing dataset to perform a fully automated glottal segmentation.}
\label{fig:hsv_segmentation_pipeline}
\end{figure}

The development of the present unified lump-element modeling framework begins with the recording of laryngeal high-speed videoendoscopy (HSV) data. HSV data provide the raw imaging data from which the time-resolved glottal area waveform is extracted. This process involves several steps, starting from the image acquisition, preprocessing and image enhancement, deep-learning–based segmentation, and quality control. An overview of the HSV segmentation workflow is shown in Fig.~\ref{fig:hsv_segmentation_pipeline}. The pipeline illustrates the complete processing sequence from raw HSV recordings through the preprocessing, manual annotation, dataset partitioning, data augmentation, and convolutional neural network (CNN) training, to automated glottal segmentation. These interconnected steps establish the experimental foundation that provides the physiological input to the lumped-element model. 

\subsection*{Participant Recruitment and Ethics}
The present study analyzed HSV data from a 49-year-old male participant with normophonic voice characteristics. The participant was recruited as part of a larger investigation to study vocal fold dynamics during connected speech production. The data collection was conducted at Mayo Clinic- Scottsdale, AZ, under an Insitutional Review Board (IRB) protocol approved by Mayo Clinic. The data usage and analysis was approved by the IRB at Michigan State University. 

\subsection*{Laryngeal High-Speed Videoendoscopy (HSV) Acquisition} 
HSV recordings were acquired using a Photron FASTCAM mini AX200 monochrome high-speed camera (Photron Inc., San Diego, CA, USA) coupled with a flexible nasolaryngoscope (Kay Pentax, Lincoln Park, NJ, USA). The camera was configured to capture images at 4{,}000 frames per second (fps) with a spatial resolution of $256 \times 224$ pixels and 12-bit grayscale depth. A 300-W xenon light source provided continuous illumination throughout the recording session to ensure consistent image intensity and minimize motion artifacts. The flexible nasolaryngoscope was inserted transnasally and positioned approximately 1-2 cm above the glottal plane to optimize visualization of the vocal fold edges. Real-time monitoring during insertion ensured an adequate capturing of both vocal folds and surrounding laryngeal structures. The participant was instructed to read standardized speech stimuli from the Consensus Auditory-Perceptual Evaluation of Voice (CAPE-V) protocol~\cite{kempster2009consensus} and the Rainbow Passage~\cite{fairbanks1940voice}. Recording sessions yielded approximately 200{,}000-400{,}000 frames per participant, corresponding to 50-100 seconds of continuous laryngeal imaging during natural speech production.

\subsection*{Image Preprocessing and Enhancement}
Raw HSV image sequences were exported in uncompressed 12-bit grayscale format for subsequent analysis. To enhance glottal edge visibility and standardize image quality across frames, a multi-step preprocessing pipeline was applied. First, Gaussian spatial filtering with a kernel size of $3 \times 3$ pixels and standard deviation $\sigma = 0.5$ was used to reduce high-frequency noise while preserving edge information. Pixel intensity values were then linearly rescaled to span the full dynamic range $[0,255]$ using 
\begin{equation} 
I_{\mathrm{norm}}(x,y) = \frac{I(x,y) - I_{\mathrm{min}}}{I_{\mathrm{max}} - I_{\mathrm{min}}} \times 255
\label{eq:intensity_normalization} 
\end{equation} 
where $I(x,y)$ represents the original pixel intensity at coordinates $(x,y)$, and $I_{\mathrm{min}}$ and $I_{\mathrm{max}}$ denote the minimum and maximum intensities within each frame, respectively. Finally, gamma correction with $\gamma = 1.2$ was applied to enhance mid-tone contrast: 
\begin{equation} 
I_{\mathrm{gamma}}(x,y) = 255 \times \left( \frac{I_{\mathrm{norm}}(x,y)}{255} \right)^{1/\gamma}
\label{eq:gamma_correction} 
\end{equation}

\subsection*{Deep Neural Network Architecture for Glottal Segmentation}
We employed a U-Net convolutional neural network architecture\cite{ronneberger2015u} for automated pixel-wise segmentation of the glottal area in HSV frames. The U-Net consists of a contracting encoder path that captures hierarchical image features and an expansive decoder path that enables precise localization. The network architecture comprised five encoding blocks and five corresponding decoding blocks, with skip connections concatenating feature maps between corresponding encoder and decoder layers to preserve spatial information. Each encoding block contained two convolutional layers with 3 $\times$ 3 kernels, batch normalization, rectified linear unit (ReLU) activation functions, and a 2 $\times$ 2 max pooling operation for downsampling. The number of feature channels doubled after each pooling operation, following the sequence: 64, 128, 256, 512, 1024. Each decoding block performed 2 $\times$ 2 transposed convolution for upsampling, followed by concatenation with corresponding encoder features, and two convolutional layers with ReLU activations. The final layer employed a 1 $\times$ 1 convolution with softmax activation to produce binary segmentation masks distinguishing glottal area pixels (class 1) from background pixels (class 0).

\subsection*{Network Training and Validation}
The network was trained on a manually annotated dataset comprising 738 HSV frames selected from six participants (three normophonic controls and three patients with adductor laryngeal dystonia) to capture diverse glottal configurations including abducted, adducted, and obstructed states. Two experienced raters independently delineated glottal boundaries using custom MATLAB software (MathWorks Inc., Natick, MA, USA). Inter-rater reliability was assessed using the Dice similarity coefficient, yielding a mean score of 0.94 $\pm$ 0.03, indicating excellent agreement. The annotated dataset was randomly partitioned into training (80\%, $n = 590$ frames) and validation (20\%, $n = 148$ frames) subsets. Network training employed the Adam optimizer with an initial learning rate of $1 \times 10^{-4}$, batch size of 16, and 100 epochs. The loss function combined binary cross-entropy and Dice loss to address class imbalance:
\begin{equation}
\mathcal{L} = -\frac{1}{N}\sum_{i=1}^{N} \left[ y_i \log(\hat{y}_i)
+ (1 - y_i)\log(1 - \hat{y}_i) \right]
+ \left( 1 - \frac{2\sum_{i=1}^{N} y_i \hat{y}_i}{\sum_{i=1}^{N} y_i
+ \sum_{i=1}^{N} \hat{y}_i} \right)
\label{eq:loss_machine_learning}
\end{equation}
where $y_i$ and $\hat{y}_i$ represent ground truth and predicted labels for pixel $i$, and $N$ is the total number of pixels. Data augmentation techniques including random rotation ($\pm 15^\circ$), horizontal flipping, and elastic deformations were applied during training to improve generalization. The model achieving the lowest validation loss was selected for inference on test data. 
\par
Following completion and validation of the convolutional neural network, the trained model was applied to the HSV recordings from a single 49-year-old male participant with normophonic voice characteristics. The glottal area waveform extracted from this participant was used as the physiological input for the subsequent analyses. Glottal area waveforms derived from the remaining participants were used only for network development and validation and were not included in the downstream modeling or analysis.

\subsection*{Glottal Area Extraction}
The trained U-Net model was applied to the entire HSV sequence from the target participant to generate binary segmentation masks for all frames. For each mask, the glottal area $A_g(t)$ at time $t$ was computed as:

\begin{equation}
A_g(t) = \sum_{x=1}^{W} \sum_{y=1}^{H} M(x,y,t)\, S^{2}
\label{eq:glottal_area}
\end{equation}
where $M(x,y,t)$ is the binary mask value at pixel coordinates $(x,y)$ and time $t$, $W$ and $H$ are the image width and height, and $S$ is the spatial calibration factor (mm/pixel) determined from known anatomical dimensions. The resulting glottal area waveform (GAW) $A_g(t)$ provides a quantitative time-series representation of vocal fold kinematics throughout the speech sample. 

\subsection*{Quality Control and Validation}
To ensure segmentation accuracy, two independent raters visually inspected a random sample of 20,000 frames (approximately 10\% of the total dataset) representing diverse phonetic contexts and glottal configurations. Raters assessed whether automatically generated masks accurately captured glottal boundaries using a binary accept/reject criterion. Segmentation quality was deemed acceptable when at least 80\% of reviewed frames received positive ratings from both raters, consistent with validation metrics reported in prior work\cite{yousef2023spatial, yousef2022deep, yousef2021hybrid}. Frames with poor segmentation quality (e.g., due to excessive mucus accumulation or specular reflections) were flagged for manual correction or exclusion from downstream analyses.

\subsection*{Lumped-Element System Overview}

\begin{figure}[htbp]
    \centering
    \includegraphics[width=\textwidth]{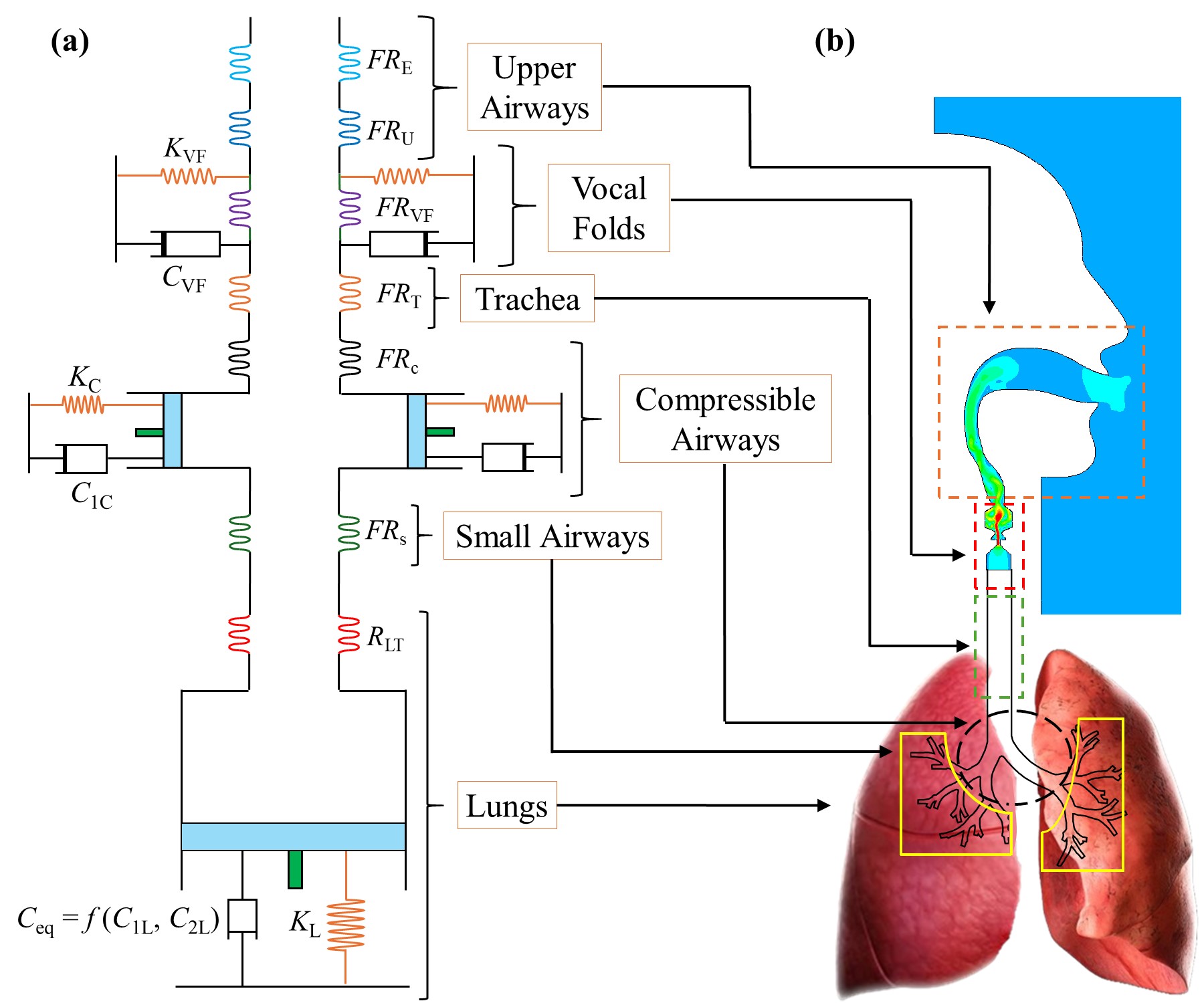}
    \caption{Unified mechanical-physiological representation of the human phonation system depicting: \textbf{(a)} Lumped-element architecture comprising three coupled subsystems of the respiratory subsystem (lungs, small airways, compressible airways, and trachea), the phonatory subsystem (vocal folds), and the articulatory subsystem (upper airways). Airflow is generated by lung contraction and propagates sequentially through the small airways, compressible airways, vocal folds, and upper airways before exiting at the mouth. Along this pathway, the flow encounters lung tissue resistance \(R_{\mathrm{LT}}\), small airways flow resistance \(FR_{\mathrm{s}}\), compressible airways resistance \(FR_{\mathrm{c}}\), tracheal flow resistance \(FR_{\mathrm{T}}\), vocal folds flow resistance \(FR_{\mathrm{VF}}\), upper airways flow resistance \(FR_{\mathrm{U}}\), and expiratory flow resistance \(FR_{\mathrm{E}}\); \textbf{(b)} Anatomical correspondence of the respiratory, phonatory, and articulatory subsystems, illustrating the mapping between the lumped-element representation and the underlying human airways anatomy.}
    \label{fig:unified_phonation_model}
\end{figure}

Figure~\ref{fig:unified_phonation_model} illustrates the unified lumped-element architecture developed in this study to describe human phonation as a coupled mechanical–aerodynamic process. The model comprises three dynamically interacting subsystems: the respiratory subsystem, the phonatory subsystem, and the articulatory subsystem. Together, these subsystems form a continuous airflow pathway extending from the lungs to the ambient environment. Each anatomical segment is represented using lumped mechanical elements and flow resistances that capture its dominant physiological function.

\par
The modeling begins with the respiratory subsystem, which consists of the lungs, small airways, compressible airways, and the trachea. In the present formulation, the two lungs are combined into a single equivalent cylindrical chamber whose volumetric expansion and compression are governed by a piston. This unified representation allows the total airflow generated by both lungs to be treated as a single volumetric source, thereby neglecting the phase lag between the left and right lungs. This assumption is justified because inter-lung phase differences, occur on time scales associated with slow respiratory mechanics. In contrast, phonation is governed by much faster aerodynamic and biomechanical processes. Therefore, the influence of the phase lag between the lungs on the overall phonation flow dynamics is assumed to be insignificant. Although the two lungs are combined into a single equivalent chamber in the present formulation, the airflow in the physiological system is inherently distributed between the left and right lungs and each lung responds independently to the applied intrapleural pressure. To preserve this physiological behavior, the present model treats the combined lung chamber as an energetically equivalent system, where the total airflow and pressure response generated by the unified lung representation matches the aggregate behavior of the two lungs acting simultaneously. In this way, the model retains the correct global pressure–flow characteristics of the respiratory subsystem without explicitly resolving left–right lung asymmetries. The motion of this unified lung system is governed by a spring–mass–damper system. The spring stiffness \(K_{\mathrm{L}}\) represents the elastic recoil of the lung parenchyma arising from the tissue elasticity. The damping coefficients \(C_{1\mathrm{L}}\) and \(C_{2\mathrm{L}}\) represent the viscous dissipation and nonlinear energy losses associated with the tissue viscosity and lung deformation, respectively. The lumped mass \(M_{\mathrm{L}}\) accounts for the total mass of the lungs. The external mechanical force imposed upon the lung piston is the intrapleural force, $F_{pl, \, L}$, which is caused due to the intrapleural pressure, $(P_{pl})$, signifying the pressure within the intrapleural cavity situated between the lungs and the thoracic wall. As a consequence of the intrapleural force \(F_{\mathrm{pl,L}}\), the lung piston undergoes axial motion, resulting in volumetric compression of the lung chamber and the expulsion of air from the lungs. As the airflow exits the lung chamber, it first encounters the resistance associated with the deformation of the lung tissue itself. This resistance is represented in the model by the lung tissue resistance \(R_{\mathrm{LT}}\), which accounts for pressure losses arising from the viscoelastic deformation of the parenchymal tissue. Following this initial resistance, the airflow passes into the small airways region, which physiologically corresponds to the bronchioles. These airways are characterized by their relatively small diameters, which cause significant viscous energy dissipation in this section. As the airflow passes this region, velocity gradients near the airway walls give rise to shear stresses that result in additional pressure losses. In the current framework, these pressure losses are incorporated through the small airways flow resistance \(FR_{\mathrm{s}}\).
\par
Above the lung chamber lies the compressible airways, which represent the peripheral airways together with the two bronchi. This segment undergoes changes in luminal cross-sectional area in response to transmural pressure. In the present framework, this compartment is modeled using two identical piston–cylinder assemblies arranged in parallel, reflecting the bilateral structure of the bronchial tree. Each piston is governed by a spring–mass–damper formulation, in which the spring stiffness \(K_{\mathrm{C}}\) represents the elastic behavior of the airway walls, the damping coefficient \(C_{1\mathrm{C}}\) accounts for viscous dissipation within the airway tissue, and the lumped mass \(M_{\mathrm{C}}\) represents the inertial contribution of the airway walls. Dynamic variations in airways volume are captured through the axial motion of the pistons. Since these airways are embedded within the intrapleural space, their lateral surfaces are subjected to the same intrapleural pressure \(P_{\mathrm{pl}}\) that drives the lung motion. The pressure loss associated with the airflow through this region is incorporated into the lumped-element framework through the compressible airways flow resistance \(FR_{\mathrm{c}}\), which accounts for the energy dissipation arising from the airway deformation.
Downstream of the compressible airways, airflow passes through the trachea, which is modeled as a rigid conduit. Pressure losses in this segment are incorporated through the tracheal flow resistance \(FR_{\mathrm{T}}\). 
\par
Past the trachea, the airflow enters the glottal region, constituting the phonatory subsystem. In the lumped-element representation, the vocal folds are modeled as a single effective mass–spring–damper system. The spring stiffness \(K_{\mathrm{VF}}\) represents the elastic properties of the vocal fold tissues, while the damping coefficient \(C_{\mathrm{VF}}\) accounts for the energy dissipation arising from the tissue viscosity, internal friction, and collision losses during the vocal folds contact. The aerodynamic interaction between the airflow and the oscillating vocal folds generates a time-varying transglottal pressure drop. This pressure loss is incorporated into the model through the vocal fold flow resistance \(FR_{\mathrm{VF}}\), which represents the instantaneous aerodynamic impedance of the glottis. 
\par
Downstream of the phonatory subsystem, the airflow enters the articulatory subsystem, which corresponds to the upper airways extending from the supraglottal region to the oral cavity. This segment includes the pharyngeal and oral passages that shape the airflow prior to the sound radiation. Aerodynamic losses within the upper airways are represented by the upper airways flow resistance \(FR_{\mathrm{U}}\), which accounts for pressure losses arising from viscous effects, the flow separation, and the geometric complexity of the supraglottal tract. Finally, the airflow is expelled from the oral cavity into the ambient environment. During sustained phonation, this outflow does not occur as a steady, freely diffusing stream. Instead, it must be actively driven through the lips. This behavior becomes evident when placing a hand in front of the mouth and perceiving discrete puffs of air during phonation. Such tactile sensations indicate that the pressure at the lips transiently exceeds atmospheric pressure, resulting in a jet-like expulsion of the airflow during each phonatory cycle. The additional aerodynamic loading associated with the driving airflow through the oral outlet against the atmospheric pressure is represented in the model by the expiratory flow resistance \(FR_{\mathrm{E}}\). Together, the resistive elements form the network that governs the pressure–flow relationship throughout the respiratory–phonatory–articulatory subsystems. The mathematical expressions for these resistances and their implementation are provided in the supplementary materials under the section compressible airways spring model and compliance.

\subsection*{Lung Geometry and Kinematics}

\begin{figure}[htbp]
\centering
\includegraphics[width=0.5\textwidth]{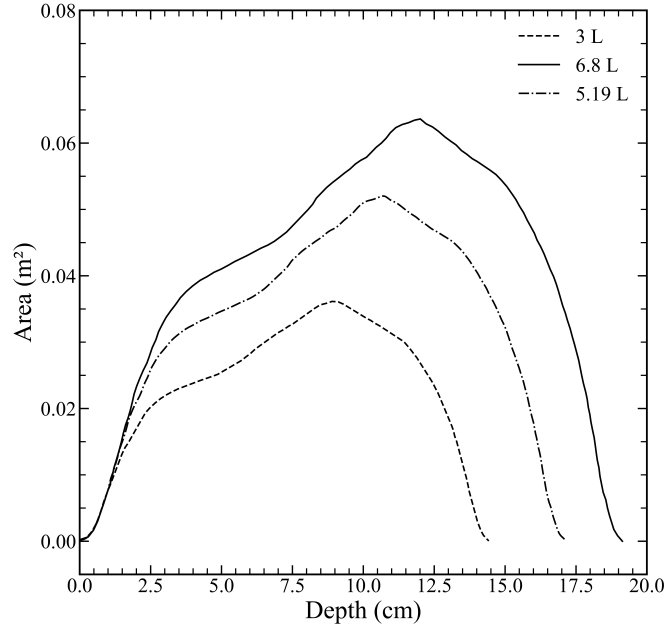}
\caption{Lung cross-sectional area $A$ as a function of the physical depth. Experimentally measured area–depth profiles for a total lung capacity of $3.0~\mathrm{L}$ and $6.8~\mathrm{L}$, reported by Mogensen \textit{et al.}~\cite{mogensen2011physiological}, are shown together with the interpolated profile at $5.19~\mathrm{L}$, produced in the present study.}
\label{fig:lung_cross_section_interp}
\end{figure}

In the present modeling framework, the two lungs are combined into a single equivalent piston–cylinder representation, characterized by an effective cross-sectional area $\alpha_L$ and diameter $D_L$. The total moving lung mass is taken as $M_L = 1.2~\mathrm{kg}$~\cite{mubbunu2018correlation}, and the total lung capacity (TLC), defined as the maximum volume of air contained in the lungs at full inspiration, is taken to be $5.19~\mathrm{L}$ in the present study. The average lung cross-sectional area $\alpha_L$ was obtained by interpolating between experimentally measured lung area–depth profiles at total lung capacities of $3.0~\mathrm{L}$ and $6.8~\mathrm{L}$, reported by Steimle \textit{et al.}~\cite{mogensen2011physiological}. Since direct area–depth measurement for a total lung capacity of $5.19~\mathrm{L}$ was not available, an interpolation procedure was adopted to construct an area–depth profile for this lung volume (see Fig.~\ref{fig:lung_cross_section_interp}).

\begin{figure}[htbp]
\centering
\includegraphics[width=0.7\textwidth]{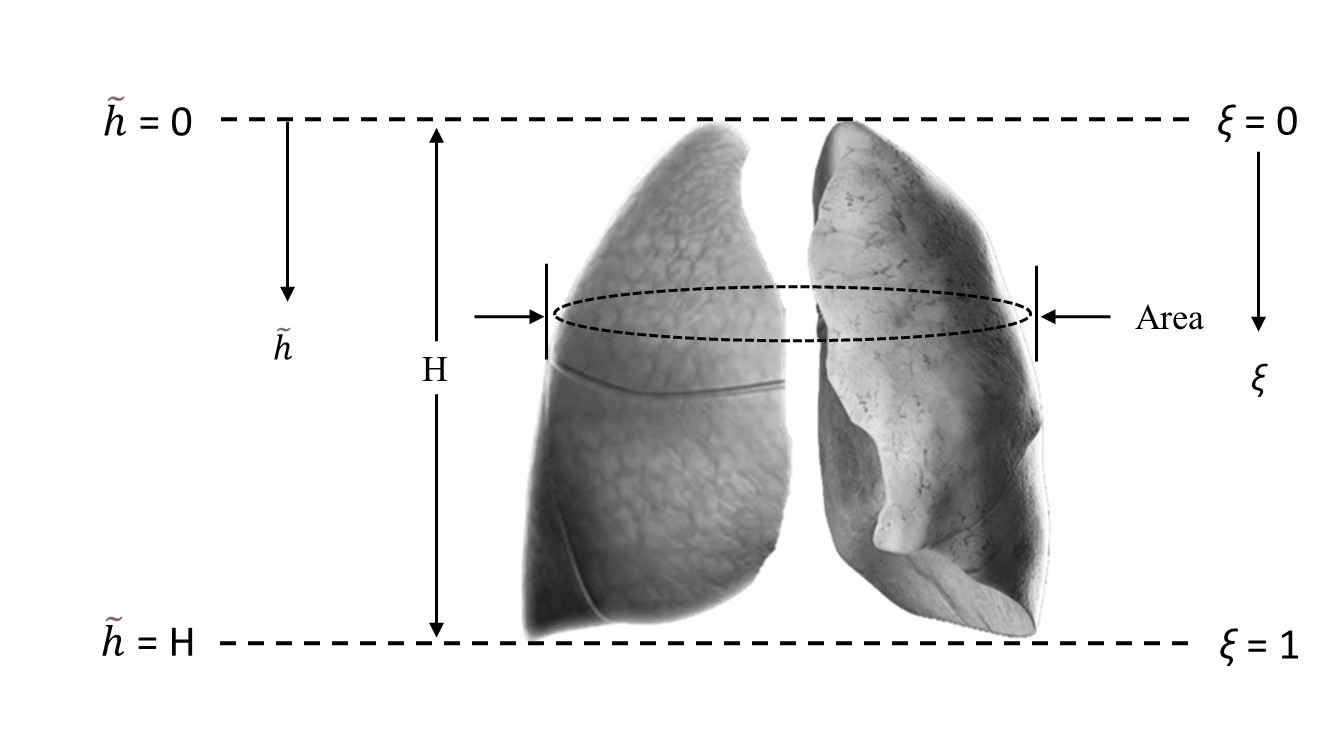}
\caption{Schematic illustration of the axial depth coordinate and cross-sectional area definition used for the lung geometry interpolation. The physical axial coordinate $\tilde{h}$ is measured from the lung apex ($\tilde{h}=0$) to the lung base ($\tilde{h}=H$), where $H$ denotes the total lung height at a given lung volume. The axial coordinate is normalized using the dimensionless variable $\xi=\tilde{h}/H$, such that $\xi=0$ corresponds to the apex and $\xi=1$ to the base. At each normalized depth $\xi$, a transverse lung cross-sectional area is defined, shown by the dashed \textit{Area}.}
\label{fig:lung_depth_normalization}
\end{figure}

To ensure a geometric consistency between the two datasets, the axial coordinate was expressed in terms of a normalized depth variable $\xi$, defined as
\begin{equation}
\xi = \frac{\widetilde{h}}{H}, \qquad \xi \in [0,1],
\end{equation}
where $\widetilde{h}$ denotes the physical depth measured from the lung apex and $H$ is the total lung height at the corresponding total lung capacity, as illustrated in Fig.~\ref{fig:lung_depth_normalization}. Under this normalization, $\xi = 0$ corresponds to the apex and $\xi = 1$ to the lung base. This transformation maps area–depth profiles for both total lung capacities onto a common scale.
At each normalized depth $\xi$, the lung cross-sectional area $A$ corresponding to the intermediate lung volume was obtained by a linear interpolation in the lung volume,
\begin{equation}
A_{5.19}(\xi) = A_{3.0}(\xi) + \frac{5.19 - 3.0}{6.8 - 3.0}\left[A_{6.8}(\xi) - A_{3.0}(\xi)\right],
\end{equation}
where $A_{3.0}(\xi)$ and $A_{6.8}(\xi)$ denote the experimentally measured cross-sectional area distributions at total lung capacities of $3.0~\mathrm{L}$ and $6.8~\mathrm{L}$, respectively, and $A_{5.19}(\xi)$ is the interpolated distribution at a total lung capacity of $5.19~\mathrm{L}$. This procedure produces a smooth area–depth profile for $5.19~\mathrm{L}$ that lies between the two experimental configurations, as shown in Fig.~\ref{fig:lung_cross_section_interp}.
\par

From the interpolated area distribution, an equivalent constant cross-sectional area $\alpha_L$ was computed by averaging over the lung height according to
\begin{equation}
\alpha_L = \frac{1}{H}\int_0^{H}A(\widetilde{h})\,\mathrm{d}\widetilde{h}=\int_0^{1}.
A(\xi)\,\mathrm{d}\xi
\end{equation}

Based on this formulation, the average cross-sectional area was obtained as $\alpha_L = 3.3873 \times 10^{-2}~\mathrm{m}^2$. The corresponding effective lung diameter is therefore
\begin{equation}
D_L = 2\sqrt{\frac{\alpha_L}{\pi}} = 0.2076~\mathrm{m} \;(20.76~\mathrm{cm}).
\end{equation}

\begin{figure}[htbp]
    \centering
    \includegraphics[width=0.6\textwidth]{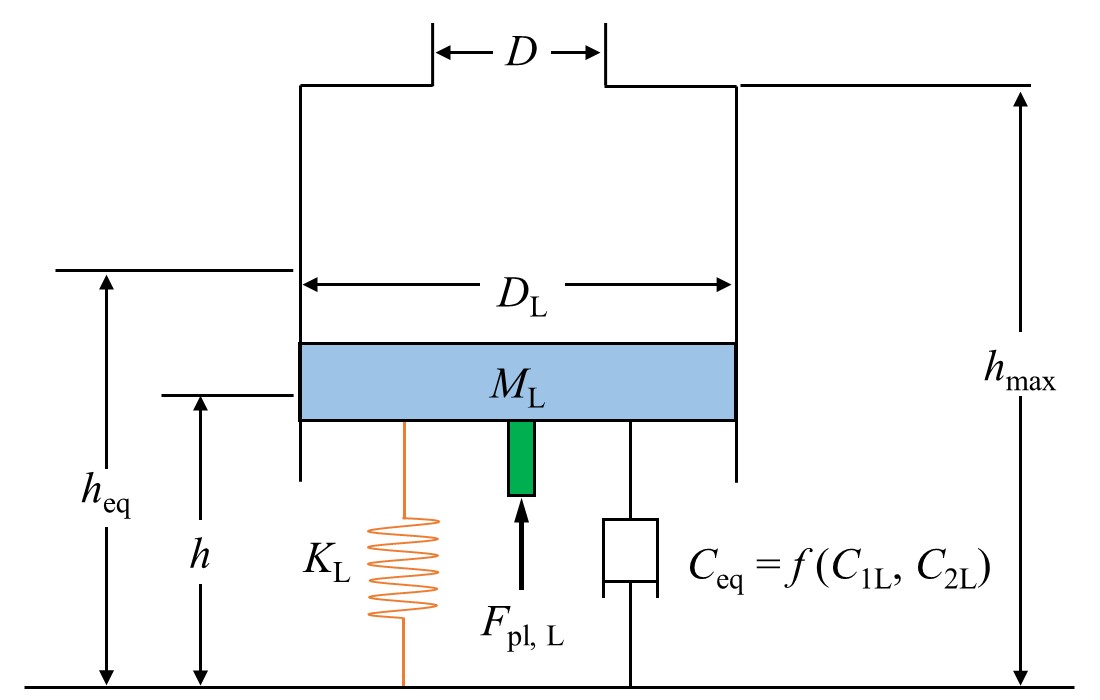}
    \caption{%
Lumped-element representation of the human lungs. The left and right lungs are combined into a single equivalent piston-cylinder system with lumped mass \(M_{\mathrm{L}}\). The axial piston position \(h\) governs the instantaneous lung volume. The spring stiffness \(K_{\mathrm{L}}\) represents the elastic recoil of the lung parenchyma, while the damping elements \((C_{1\mathrm{L}},\,C_{2\mathrm{L}})\) capture linear and nonlinear dissipative effects. The piston is driven by the intrapleural force \(F_{\mathrm{pl,L}}\); and the geometric parameters \(D_{\mathrm{L}}\), \(h_{\mathrm{eq}}\), and \(h_{\max}\) denote the effective lung diameter, equilibrium piston position, and maximum allowable piston excursion, respectively.}
    \label{fig:lung_model}
\end{figure}

Using this value of $\alpha_L$, the instantaneous lung volume $V_L$ is expressed as
\begin{equation}
V_L = \frac{\pi D_L^{2}}{4}\, \bigl(h_{\max}-h\bigr) =\alpha_L\,\bigl(h_{\max}-h\bigr),
\end{equation}
where $h$ denotes the instantaneous axial position of the lung piston measured from the reference, as shown in Fig.~\ref{fig:lung_model} and $h_{\max}$ represents the maximum allowable piston travel. For a total lung capacity of $5.19~\mathrm{L}$ and a cross-sectional area $\alpha_L = 3.3873 \times 10^{-2}~\mathrm{m}^2$, the maximum allowable piston travel can be calculated as
\begin{equation}
h_{\max} = \frac{{\mathrm{TLC}}}{\alpha_L} = 0.153~\mathrm{m} \;(15.3~\mathrm{cm}).
\label{eq:h_max}
\end{equation}

\subsection*{Lung Spring Model and Compliance}
The lung elastic recoil pressure $P_{el,\,L}$ quantifies the inward-directed mechanical pressure generated by the lung parenchyma at a given lung volume. This pressure arises from the intrinsic elasticity of the lung tissues. Physiologically, lung elastic recoil is governed by the balance between the inward recoil of the parenchymal tissue, driven primarily by elastin and collagen fibers, and the outward recoil imposed by the thoracic cage. Variations in $P_{el,\,L}$ therefore directly regulate lung volume by modifying the net mechanical load acting on the lung surface. An increase in the elastic recoil pressure enhances the inward-directed force on the lungs, leading to a reduction in the lung volume, whereas a decrease in elastic recoil pressure allows the lungs to expand. The sensitivity of lung volume to changes in the elastic recoil pressure is characterized by the lung compliance, which is defined as
\begin{equation} 
C_L = \left| \frac{\Delta V_L}{\Delta P_{el,\,L}} \right|,
\label{eq:lung_compliance_def} 
\end{equation}
where $\Delta V_L$ denotes the change in the lung volume produced by a corresponding change $\Delta P_{el,\,L}$ in the elastic recoil pressure. A larger value of $C_L$ indicates more distensible lungs that undergo substantial volume changes for small pressure variations, whereas a smaller value corresponds to stiffer lungs requiring higher pressures to achieve the same volume change. In the present lumped-element framework, the lung compliance is mapped onto a nonlinear spring that generates an elastic recoil force $F_{el,\,L}$, as shown in Fig.~\ref{fig:lung_model}. This force acts on the lung piston and serves as the mechanical analog of the elastic recoil pressure, with the relationship $P_{el,\,L} = F_{el,\,L} / \alpha_L$, where $\alpha_L$ is the effective lung cross-sectional area. In direct analogy with the physiological response of the lung volume to the elastic recoil pressure, the axial displacement of the piston responds to the applied elastic recoil force. For such a spring element, an incremental change in the applied force produces a corresponding incremental displacement. Accordingly, the effective stiffness $K_L$ of the lung spring is defined through
\begin{equation} 
\frac{1}{K_L} = \left| \frac{\Delta h}{\Delta F_{el,\,L}} \right|,
\label{eq:spring_compliance} 
\end{equation} 
which represents the differential form of Hooke’s law. Here, $\Delta F_{el,\,L}$ denotes the change in the elastic recoil force acting on the piston, and $\Delta h$ is the resulting incremental piston displacement. Since the lung volume in the piston–cylinder model is given by $V_L = \alpha_L (h_{\max}-h)$, the differentiation yields $\Delta V_L = -\alpha_L\,\Delta h$. Combining this relationship with equation~\eqref{eq:spring_compliance} and the definition of lung compliance in equation~\eqref{eq:lung_compliance_def} leads to
\begin{equation} 
C_L = \frac{\alpha_L^{2}}{K_L}
\label{eq:compliance_stiffness_relation},
\end{equation} 
which establishes a direct connection between the physiological lung compliance and the stiffness of the mechanical spring used in the model. To describe the nonlinear elastic behavior of lung tissues, the static elastic recoil pressure is modeled following the framework introduced by Le~Rolle \textit{et~al.}~\cite{le2013mathematical}. In this formulation, the elastic recoil pressure is expressed as an exponential function of the lung volume,
\begin{equation} 
P_{el,\,L}(V_L) = A_L\,e^{\bar{k}_L V_L} - B_L,
\label{eq:lung_recoil_pressure}
\end{equation} 
where $A_L$, $B_L$, and $\bar{k}_L$ are physiological parameters characterizing the intrinsic elastic response of the lung tissues. The values $A_L = 19.61~\mathrm{Pa}$, $B_L = 49.03~\mathrm{Pa}$, and $\bar{k}_L = 1000.0~\mathrm{m^{-3}}$ are adopted from~\cite{marconi2020silico}. This exponential form reflects a progressive tissue stiffening at higher lung volumes, where collagen fibers increasingly straighten and restrict further expansion. Differentiating equation~\eqref{eq:lung_recoil_pressure} with respect to the lung volume and taking the inverse provides the nonlinear lung compliance, 
\begin{equation} 
C_L(P_{el,\,L}) = \frac{1}{\bar{k}_L\,(P_{el,\,L} + B_L)},
\label{eq:nonlinear_lung_compliance} 
\end{equation}
which explicitly captures the volume-dependent reduction in the compliance at elevated lung pressures. Substituting equation~\eqref{eq:nonlinear_lung_compliance} into equation~\eqref{eq:compliance_stiffness_relation} yields the corresponding nonlinear stiffness of the lung spring, 
\begin{equation} 
K_L = \alpha_L^{2}\,\bar{k}_L\,(P_{el,\,L} + B_L) = \alpha_L^{2}\,\bar{k}_L\,A_L\,e^{\bar{k}_L V_L},
\label{eq:lung_stiffness} 
\end{equation} 
which governs the elastic response of the lung piston within the proposed unified mechanical framework.

\subsection*{Lung Damping Model and Tissue Dissipation}
The dissipative behavior of the lung in the present piston-cylinder representation is modeled through a nonlinear damping law that accounts for the tissue viscosity and the progressive increase in the viscous resistance associated with lung tissues stiffening during deformation. As the lungs expand and recoil, the internal friction within the parenchymal tissues and supporting collagen network converts mechanical energy into heat, giving rise to a velocity-dependent energy dissipation. In its most general form, the damping force acting on the lung piston may be expressed as an infinite series expansion in terms of the piston velocity,
\begin{equation}
F_{\mathrm{damp,\,L}}(\dot{h}) = a_{1L}\,\dot{h} + a_{2L}\,\dot{h}^{2} + a_{3L}\,\dot{h}^{3} + \cdots,
\label{eq:lung_damping_general}
\end{equation}
where $\dot{h}$ denotes the piston velocity and the coefficients $a_{1L}$, $a_{2L}$, $a_{3L}$, \ldots\ quantify successive orders of damping. The linear term $a_{1L}\dot{h}$ represents classical viscous losses associated with tissue viscosity, while the quadratic term $a_{2L}\dot{h}^{2}$ captures an enhanced viscoelastic resistance arising from the progressive collagen fiber recruitment and strain-dependent stiffening as lung deformation increases. Higher-order terms correspond to additional nonlinear dissipative mechanisms linked to geometric nonlinearity, tissue rearrangement, and the microstructural strain stiffening. Identification of all coefficients in the full expansion of equation~\eqref{eq:lung_damping_general} would require detailed in vivo characterization of lung tissues dissipation across a wide range of deformation rates, which is not available in the literature. Thus, we truncate the series and retain only the dominant contributions governing the lung energy dissipation during sustained phonation. This leads to the reduced damping representation of
\begin{equation}
F_{\mathrm{damp,\,L}}(\dot{h}) \approx a_{1L}\,\dot{h} + a_{2L}\,\dot{h}^{2},
\label{eq:lung_damping_reduced}
\end{equation}
which preserves the essential linear and nonlinear viscous effects. To ensure that the nonlinear damping contribution always opposes the direction of motion and that the effective damping remains positive for any nonzero velocity, the quadratic term is reformulated in a sign-preserving manner. The final damping expression implemented in the model is therefore given as
\begin{equation}
F_{\mathrm{damp,\,L}}(\dot{h}) = - C_{1L}\,\dot{h} - C_{2L}\,\frac{\dot{h}^{3}}{|\dot{h}|},
\label{eq:lung_damping_final}
\end{equation}
where $C_{1L}$ and $C_{2L}$ denote the linear and nonlinear viscous damping coefficients, respectively, as shown in Fig.~\ref{fig:lung_model}. The velocity ratio $\dot{h}^{3}/|\dot{h}|$ is equivalent to $\dot{h}^{2}\,\mathrm{sign}(\dot{h})$, which ensures that the nonlinear term consistently resists the piston motion and remains positive-definite for all nonzero velocities. The combined effect of the linear and nonlinear contributions may be expressed through a single velocity-dependent effective damping coefficient,
\begin{equation}
C_{\mathrm{eq}} = C_{1L} + C_{2L}\,|\dot{h}|,
\label{eq:lung_damping_effective}
\end{equation}
which provides a compact representation of the total dissipative behavior of the lungs.

\subsection*{Governing Equation of the Lung Piston-Cylinder Model}

\begin{figure}[htbp]
\centering
\includegraphics[width=0.3\textwidth]{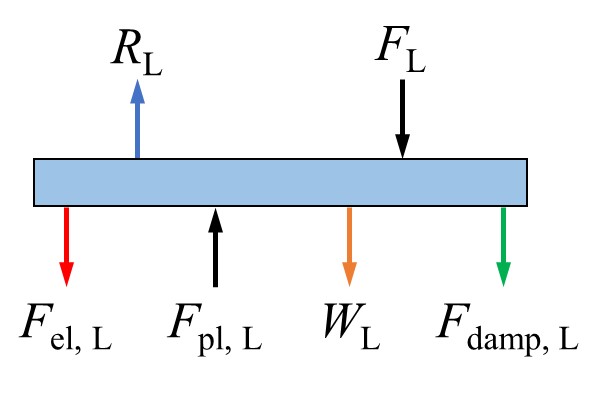}
\caption{Free-body diagram of the lung piston in the lumped piston-cylinder representation. The diagram illustrates all forces acting on the equivalent lung piston along the axial direction. Downward forces include the elastic recoil force of the lung tissues $F_{\mathrm{el,L}}$, which resists deformation from the equilibrium position, the gravitational force $W_{L} = M_{L} g$ associated with the lumped lung mass, the internal lung pressure force $F_{L} = \alpha_{L} P_{L}$, and the viscous damping force $F_{\mathrm{damp,L}}$, which accounts for linear and nonlinear tissue dissipation during the lung motion. Upward forces include the intrapleural force $F_{\mathrm{pl,L}} = \alpha_{L} P_{\mathrm{pl}}$, generated by the intrapleural pressure acting on the lung surface, and the supportive force $R_{L}$, representing structural support from the rib cage, diaphragm, and surrounding anatomical constraints.}
\label{fig:lung_fbd}
\end{figure}

As illustrated in Fig.~\ref{fig:lung_fbd}, multiple forces act on the lung piston in the piston-cylinder model as it moves within the lung piston-cylinder system. The downward gravitational force is given by the piston weight,
\(W_{L} = M_{L} g\), where \(M_{L}\) is the lumped lung mass and \(g\) is the gravitational acceleration.
The elastic recoil force generated by the lung tissue is denoted by \(F_{\mathrm{el,\,L}}\) and is modeled as
\begin{equation}
F_{\mathrm{el,\,L}}(t) = -K_{L}(t)\,\big(h(t) - h_{\mathrm{eq}}\big),
\label{eq:lung_elastic_force}
\end{equation}
where \(h_{\mathrm{eq}}\) denotes the equilibrium piston position. Mathematically, this equilibrium position corresponds to the piston location at which the elastic recoil force is absent. Physiologically, \(h_{\mathrm{eq}}\) represents the functional residual capacity (FRC) of the lungs, at which the inward elastic recoil of the lung tissues is balanced by the outward elastic recoil of the chest wall. At this resting configuration, no muscular effort is required to maintain the lung volume. The dissipative force arises from the tissue viscosity and is captured by the damping force \((F_{\mathrm{damp,\,L}})\). In addition, the piston experiences an upward intrapleural force \(F_{\mathrm{pl,\,L}} = \alpha_{L}\,P_{\mathrm{pl}}\), where \(P_{\mathrm{pl}}\) is the intrapleural pressure acting on the lung surface and \(\alpha_{L}\) is the effective lung cross-sectional area. This pressure is generated primarily by diaphragmatic and thoracic motion and serves as the primary driving mechanism for the lung deformation during breathing. Conversely, the internal lung pressure \(P_{L}\) exerts a downward force \(F_{L} = \alpha_{L}\,P_{L}\), which represents the resistance imposed by the compressed air within the lungs. Finally, the supportive force \(R_{L}\) accounts for the mechanical support provided by the rib cage, diaphragm, and surrounding thoracic structures that guide the axial motion of the piston.

Applying Newton’s second law of motion and summing all forces acting on the piston in Fig.~\ref{fig:lung_fbd} yields the governing equation of motion
\begin{equation}
M_{L}\,\ddot{h}(t) = F_{\mathrm{el,\,L}}(t) + F_{\mathrm{pl,\,L}}(t) + F_{\mathrm{damp,\,L}}(t) + W_{L} + R_{L}(t) + F_{L}(t).
\label{eq:lung_eom_general}
\end{equation}

Substituting the explicit expressions for each force gives
\begin{equation}
\begin{aligned}
M_{L}\,\ddot{h}(t) &= - K_{L}(t)\,\big(h(t)-h_{\mathrm{eq}}\big) + \big(P_{\mathrm{pl}}(t) - P_{L}(t)\big)\,\frac{\pi D_{L}^{2}}{4} \\ &\quad - C_{1L}\,\dot{h}(t) - C_{2L}(t)\,\frac{\dot{h}(t)^{3}}{|\dot{h}(t)|} - M_{L} g + R_{L}(t).
\end{aligned}
\label{eq:lung_eom_explicit}
\end{equation}

Expressing the dynamics in terms of the lung volume using the kinematic relation \(V_{L}(t) = \alpha_{L}\,\big(h_{\max} - h(t)\big)\), the governing equation can be recast directly in the volume form as
\begin{equation}
\begin{aligned}
-\frac{M_{L}}{\alpha_{L}}\,\ddot{V}_{L}(t) =&\; K_{L}(t)\!\left( \frac{V_{L}(t)}{\alpha_{L}} + h_{\mathrm{eq}} - h_{\max} \right) + \alpha_{L}\,\big(P_{\mathrm{pl}}(t) - P_{L}(t)\big) \\ &\; - M_{L} g + \frac{C_{1L}(t)}{\alpha_{L}}\,\dot{V}_{L}(t) + \frac{C_{2L}(t)}{\alpha_{L}^{2}}\,\frac{\dot{V}_{L}(t)^{3}}{|\dot{V}_{L}(t)|} + R_{L},
\end{aligned}
\label{eq:lung_volume_eom}
\end{equation}
which constitutes the general governing equation of the lung piston-cylinder system in the present lumped-element framework.
\par
The governing equation derived above is general and can accommodate a range of respiratory and phonatory conditions. In the present study, the model is evaluated under physiological respiratory patterns for which time-resolved reference data are available in the literature. Two representative breathing conditions are considered for our study. They are quiet breathing, corresponding to a relaxed baseline state with low tidal volume and minimal muscular effort, and regular breathing, representing a moderate respiratory demand with larger tidal volume. Marconi and De Lazzari~\cite{marconi2020silico} reported the time-resolved pressure and flow waveforms for both cases. Substituting the corresponding waveforms into our general model gives
\begin{equation}
\label{eq:lung_governing_regular}
\begin{aligned}
-\frac{M_{L}}{\alpha_{L}}\,\ddot{V}_{L}^{r}(t) =\; & K_{L}^{r}(t)\Bigg(\frac{V_{L}^{r}(t)}{\alpha_{L}} + h_{\mathrm{eq}} - h_{\mathrm{max}}\Bigg) + \alpha_{L}\Big(P_{\mathrm{pl}}^{r}(t) - P_{L}^{r}(t)\Big) \\ &\quad - M_{L}g + \frac{C_{1L}(t)}{\alpha_{L}}\,\dot{V}_{L}^{r}(t) + \frac{C_{2L}(t)}{\alpha_{L}^{2}}\,\frac{\dot{V}_{L}^{r}(t)^{3}}{\left|\dot{V}_{L}^{r}(t)\right|} + R_{L}^{r}
\end{aligned}
\end{equation}
and
\begin{equation}
\label{eq:lung_governing_quiet}
\begin{aligned}
-\frac{M_{L}}{\alpha_{L}}\,\ddot{V}_{L}^{q}(t) =\; & K_{L}^{q}(t)\Bigg(\frac{V_{L}^{q}(t)}{\alpha_{L}} + h_{\mathrm{eq}} - h_{\mathrm{max}}\Bigg) + \alpha_{L}\Big(P_{\mathrm{pl}}^{q}(t) - P_{L}^{q}(t)\Big) \\ &\quad - M_{L}g + \frac{C_{1L}(t)}{\alpha_{L}}\,\dot{V}_{L}^{q}(t) + \frac{C_{2L}(t)}{\alpha_{L}^{2}}\,\frac{\dot{V}_{L}^{q}(t)^{3}}{\left|\dot{V}_{L}^{q}(t)\right|} + R_{L}^{q}.
\end{aligned}
\end{equation}
Here, the superscripts \(q\) and \(r\) denote quiet and regular breathing, respectively. Equations~\eqref{eq:lung_governing_regular} and~\eqref{eq:lung_governing_quiet} contain five unknowns, \(C_{1L}\), \(C_{2L}\), \(R_{L}^{r}\), \(R_{L}^{q}\), and \(h_{\mathrm{eq}}\), while only two equations are available. The system is therefore under-determined. However, we can determine the equilibrium position \(h_{\mathrm{eq}}\) analytically. The spring contribution in equations~\eqref{eq:lung_governing_regular}-\eqref{eq:lung_governing_quiet} has the form
\(K_{L}(t)\left(\frac{V_{L}(t)}{\alpha_{L}} + h_{\mathrm{eq}} - h_{\mathrm{max}}\right)\), the parameter \(h_{\mathrm{eq}}\) represents the equilibrium spring position, and the elastic element should not impose a net bias over a breathing cycle. Hence,
\begin{equation}
\label{eq:heq_condition}
\left\langle
K_{L}(t)\left(\frac{V_{L}(t)}{\alpha_{L}} + h_{\mathrm{eq}} - h_{\mathrm{max}}\right)
\right\rangle = 0
\end{equation}
where \(\langle \cdot \rangle\) denotes a time average over one representative cycle. Solving equation~\eqref{eq:heq_condition} gives
\begin{equation}
\label{eq:heq_solution}
h_{\mathrm{eq}} = h_{\mathrm{max}} - \frac{1}{\alpha_{L}}\,
\frac{\left\langle K_{L}(t)\,V_{L}(t)\right\rangle}{\left\langle K_{L}(t)\right\rangle}.
\end{equation}
Defining the FRC as
\begin{equation}
\label{eq:neutral_volume_def}
\mathrm{FRC} = \alpha_{L}\left(h_{\mathrm{max}} - h_{\mathrm{eq}}\right)
\end{equation}
and substituting equation~\eqref{eq:heq_solution} into equation~\eqref{eq:neutral_volume_def} provides the FRC as
\begin{equation}
\label{eq:neutral_volume_solution}
\mathrm{FRC} = \frac{\left\langle K_{L}(t)\,V_{L}(t)\right\rangle}{\left\langle K_{L}(t)\right\rangle}
\end{equation}

Applying equation~\eqref{eq:neutral_volume_solution} to the quiet- and regular-breathing waveforms gives
\(\mathrm{FRC}^{q}=2.79~\mathrm{L}\) and \(\mathrm{FRC}^{r}=2.93~\mathrm{L}\). Their mean value is taken as
\begin{equation}
\label{eq:frc_avg}
\mathrm{FRC} = \frac{\mathrm{FRC}^{q}+\mathrm{FRC}^{r}}{2}
= 2.86~\mathrm{L}
\end{equation}
corresponding to \(55.13\%\) of the total lung capacity \((5.19~\mathrm{L})\). The relative difference between the two breathing-specific values is
\begin{equation}
\label{eq:frc_percent_diff}
\varepsilon_{\mathrm{FRC}} =
\frac{\left|\mathrm{FRC}^{r}-\mathrm{FRC}^{q}\right|}{\frac{1}{2}\left(\mathrm{FRC}^{r}+\mathrm{FRC}^{q}\right)}\times 100
= 4.68\%
\end{equation}

Thus, we will use the average of the both functional residual capacities as a single resting operating point for subsequent simulations. Using \(\alpha_{L}=3.387300\times 10^{-2}~\mathrm{m^2}\), \(h_{\mathrm{max}}=\frac{V_{\mathrm{TLC}}}{\alpha_{L}}=0.1532~\mathrm{m}\), and  \(\mathrm{FRC}=2.86~\mathrm{L}\), equation~\eqref{eq:neutral_volume_def} gives \(h_{\mathrm{eq}} = h_{\mathrm{max}} - \frac{\mathrm{FRC}}{\alpha_{L}}= 0.06875~\mathrm{m}\). After calculating \(h_{\mathrm{eq}}\), there still remains four unknowns. The damping coefficients \(C_{1L}\) and \(C_{2L}\) are first estimated using particle swarm optimization, and the optimized values are then substituted into equations~\eqref{eq:lung_governing_regular} and~\eqref{eq:lung_governing_quiet} to compute the supportive forces \(R_{L}^{r}\) and \(R_{L}^{q}\) for regular and quiet breathing. Details of the optimization procedure are provided in the supplementary materials under the section particle swamp optimization algorithm.

\subsection*{Particle Swamp Optimization Algorithm}
The particle swarm optimization algorithm provides a population-based stochastic search method. Each particle represents a candidate solution that evolves through the search space by updating its position according to its own historical performance and the performance of neighboring particles. In the present work, particle swarm optimization is employed to identify the lung damping coefficients \(C_{1L}\) and \(C_{2L}\), which govern the linear and nonlinear dissipative mechanisms of the respiratory subsystem. The optimization framework consists of two components. The first component imposes physiological and mechanical constraints that restrict the feasible region of the parameter space. These constraints include bounds on dissipative energy expenditure per breath and limits on the supportive forces to ensure a realistic behavior. The second component describes the particle swarm algorithm itself, which explores the constrained parameter space and identifies combinations of \(C_{1L}\) and \(C_{2L}\) that minimize the objective function while satisfying all imposed physiological requirements.
\par
In the present framework, the lung damping parameters are calculated based on the mechanical energy dissipated over a single breathing cycle. The total dissipative energy associated with the lung dashpot is written as
\begin{equation}
E_d = C_{1L} \int \dot{h}(t)^2 \, dt + C_{2L} \int \lvert \dot{h}(t) \rvert^3 \, dt,
\label{eq:Ed_model}
\end{equation}
where $\dot{h}(t)$ denotes the instantaneous velocity of the lung piston, $C_{1L}$ is the linear viscous damping coefficient, and $C_{2L}$ is the nonlinear damping coefficient. The integrals are evaluated over one complete breathing cycle. In the lumped-element representation of the lung, the piston \( V_L(t) = \alpha_L \left( h_{\max} - h(t) \right)\). Differentiating this relation with respect to time gives
\begin{equation}
\dot{h}(t) = -\frac{\dot{V}_L(t)}{\alpha_L}
\end{equation}
Substituting this expression into equation~\eqref{eq:Ed_model} allows the dissipative energy to be written in terms of the lung volume rate. Thus, the dissipation integrals becomes
\begin{equation}
I_1 = \int_{0}^{T_{breath}} \frac{\dot{V}_L(t)^2}{\alpha_L^2} \, dt
\end{equation}
and
\begin{equation}
I_2 = \int_{0}^{ T_{breath}} \frac{\lvert \dot{V}_L(t) \rvert^3}{\alpha_L^3} \, dt,
\end{equation}
where $ T_{breath}$ denotes the breathing period. The dissipative energy per breath can be expressed in a compact linear form as
\begin{equation}
E_d = C_{1L} I_1 + C_{2L} I_2
\end{equation}
For a single breathing condition, the above relation may be written in matrix form as
\begin{equation}
\begin{bmatrix}
I_1 & I_2
\end{bmatrix}
\begin{bmatrix}
C_{1L} \\
C_{2L}
\end{bmatrix}
=
\begin{bmatrix}
E_d
\end{bmatrix}.
\end{equation}
In the present modeling work, time-resolved lung-volume waveforms $V_L(t)$ are available at a breathing rate of $12$ breaths per minute for quiet breathing and regular breathing. From these waveforms, the corresponding dissipation integrals $I_1^{(q)}$, $I_2^{(q)}$, $I_1^{(r)}$, and $I_2^{(r)}$ can be computed directly. If the dissipative energy loss associated with quiet and regular breathing at $12$ breaths per minute were known separately, the damping parameters could be obtained by solving
\begin{equation}
\begin{bmatrix}
I_1^{(q)} & I_2^{(q)} \\
I_1^{(r)} & I_2^{(r)}
\end{bmatrix}
\begin{bmatrix}
C_{1L} \\
C_{2L}
\end{bmatrix}
=
\begin{bmatrix}
E_{\mathrm{diss}}^{(q)} \\
E_{\mathrm{diss}}^{(r)}
\end{bmatrix}.
\end{equation}

However, the experimental study of Athanasiades \textit{et al.}~\cite{athanasiades2000energy} reports the dissipative energy loss per breathing cycle without distinguishing between quiet and regular breathing. At $12$ breaths per minute, the reported dissipative energy per breath is
\begin{equation}
E_{\mathrm{diss}}^{12} = 7.5833 \times 10^{-2}~\mathrm{J/cycle}.
\end{equation}

Because the breathing mode associated with this value is not specified, the above system cannot be applied directly. To resolve this ambiguity, additional experimental information is used. The same study reports the dissipative energy loss at a lower breathing rate of $8$ breaths per minute, given by
\begin{equation}
E_{\mathrm{diss}}^{8} = 9.25 \times 10^{-2}~\mathrm{J/cycle}.
\end{equation}

Although time-resolved lung-volume data are not available at $8$ breaths per minute, the availability of waveform data at $12$ breaths per minute enables an interpolation-based strategy to estimate the corresponding dynamics at $8$ breaths per minute. For this, we will denote $V_L^{12}(t)$ as the lung-volume waveform at $12$ breaths per minute, defined over one breathing period $T_{12} = 60/12$. For a target breathing rate $f$, the breathing period is $T(f)=60/f$. We define a time-scaling factor
\begin{equation}
a = \frac{T(f)}{T_{12}} = \frac{12}{f},
\end{equation}
and a volume-scaling factor
\begin{equation}
s = \frac{V_T(f)}{V_T(12)},
\end{equation}
where $V_T(f)$ denotes the tidal volume at breathing rate $f$. The introduction of the time-scaling factor $a$ adjusts the duration of the breathing cycle so that the waveform spans the correct period corresponding to the new breathing rate and the volume-scaling factor $s$ ensures that the interpolated lung-volume waveform at the target breathing rate preserves the correct tidal volume.  In general, the tidal volume is not constant across breathing rates. Experimental measurements show that slower breathing is accompanied by a larger tidal volume, whereas faster breathing is associated with a smaller tidal volume. Thus, to construct an approximate lung-volume waveform at a new breathing rate, we enforce the correct total volume change per breath in addition to the correct period. Using these scaling factors, the lung-volume waveform at rate $f$ is approximated as
\begin{equation}
V_L^{f}(t') = s \, V_L^{12}\left(\frac{t'}{a}\right)
\end{equation}
with $t' = a t$. The differentiation of the above equation gives the corresponding lung-volume rate
\begin{equation}
\dot{V}_L^{f}(t') = \frac{s}{a} \, \dot{V}_L^{12}\left(\frac{t'}{a}\right).
\end{equation}

Now, using this scaled waveform, the dissipation integrals at breathing rate $f$ satisfy the scaling relations
\begin{equation}
I_1^{(f)} = \frac{s^2}{a} \, I_1^{(12)}
\end{equation}
and
\begin{equation}
I_2^{(f)} = \frac{s^3}{a^2} \, I_2^{(12)}.
\end{equation}
Applying these relations for $f=8$ breaths per minute allows the dissipative energy expressions at $12$ and $8$ breaths per minute to be written as
\begin{equation}
E_{\mathrm{diss}}^{12} = C_{1L} I_1^{12} + C_{2L} I_2^{12}
\end{equation}
\begin{equation}
E_{\mathrm{diss}}^{8} = C_{1L} I_1^{8} + C_{2L} I_2^{8}
\end{equation}
or equivalently in a matrix form as
\begin{equation}
\begin{bmatrix}
I_1^{12} & I_2^{12} \\
I_1^{8} & I_2^{8}
\end{bmatrix}
\begin{bmatrix}
C_{1L} \\
C_{2L}
\end{bmatrix}
=
\begin{bmatrix}
E_{\mathrm{diss}}^{12} \\
E_{\mathrm{diss}}^{8}
\end{bmatrix}.
\end{equation}

Even with this interpolation strategy, it remains unclear which specific breathing mode (quiet or regular) the experimentally reported dissipative energy values correspond to. Thus, rather than forcing a single mode assumption, we identify a physically plausible range for $(C_{1L},C_{2L})$. Now, for each mode $m\in\{q,r\}$ and rate $f\in\{8,12\}$, the dissipated energy per breath can be written as
\begin{equation}
E_{\mathrm{dis}}^{(m,f)} = C_{1L}\,I_1^{(m,f)} + C_{2L}\,I_2^{(m,f)},
\label{eq:Wdis_mf_linear}
\end{equation}
where the dissipation integrals are computed from the mode- and rate-specific volume waveform as follows
\begin{equation}
I_1^{(m,f)} = \int_{0}^{T(f)} \frac{\dot{V}_L^{(m,f)}(t)^2}{\alpha_L^2}\,dt
\end{equation}

\begin{equation}
I_2^{(m,f)} = \int_{0}^{T(f)} \frac{\lvert\dot{V}_L^{(m,f)}(t)\rvert^3}{\alpha_L^3}\,dt
\label{eq:I1I2_mf_def}
\end{equation}
and $T(f)=60/f$.

Now, we enforce that the waveforms for quiet breathing at $12$ breaths per minute and the interpolated quiet waveform at $8$ breaths per minute reproduce the two rate-specific dissipative energy losses reported by Athanasiades \textit{et al.}~\cite{athanasiades2000energy}. Hence,
\begin{equation}
\begin{bmatrix}
I_1^{(q,12)} & I_2^{(q,12)} \\
I_1^{(q,8)}  & I_2^{(q,8)}
\end{bmatrix}
\begin{bmatrix}
C_{1L} \\
C_{2L}
\end{bmatrix}
=
\begin{bmatrix}
E_{\mathrm{dis}}^{(12)} \\
E_{\mathrm{dis}}^{(8)}
\end{bmatrix}.
\label{eq:quiet_2x2_system}
\end{equation}

Solving equation~\eqref{eq:quiet_2x2_system} gives a quiet-conditioned estimate of
\begin{equation}
\begin{bmatrix}
\widehat{C}_{1L}^{(q)} \\
\widehat{C}_{2L}^{(q)}
\end{bmatrix}
=
\begin{bmatrix}
I_1^{(q,12)} & I_2^{(q,12)} \\
I_1^{(q,8)}  & I_2^{(q,8)}
\end{bmatrix}^{-1}
\begin{bmatrix}
E_{\mathrm{dis}}^{(12)} \\
E_{\mathrm{dis}}^{(8)}
\end{bmatrix}
\label{eq:C_hat_quiet}
\end{equation}

Similarly, for the regular breathing we apply the same strategy which gives
\begin{equation}
\begin{bmatrix}
I_1^{(r,12)} & I_2^{(r,12)} \\
I_1^{(r,8)}  & I_2^{(r,8)}
\end{bmatrix}
\begin{bmatrix}
C_{1L} \\
C_{2L}
\end{bmatrix}
=
\begin{bmatrix}
E_{\mathrm{dis}}^{(12)} \\
E_{\mathrm{dis}}^{(8)}
\end{bmatrix},
\label{eq:regular_2x2_system}
\end{equation}
which gives the regular-conditioned estimate
\begin{equation}
\begin{bmatrix}
\widehat{C}_{1L}^{(r)} \\
\widehat{C}_{2L}^{(r)}
\end{bmatrix}
=
\begin{bmatrix}
I_1^{(r,12)} & I_2^{(r,12)} \\
I_1^{(r,8)}  & I_2^{(r,8)}
\end{bmatrix}^{-1}
\begin{bmatrix}
E_{\mathrm{dis}}^{(12)} \\
E_{\mathrm{dis}}^{(8)}
\end{bmatrix}.
\label{eq:C_hat_regular}
\end{equation}

These two solutions in equations~\eqref{eq:C_hat_quiet}--\eqref{eq:C_hat_regular} are both compatible with the same rate-specific dissipation targets but arise from different waveform kinematics. Therefore, rather than selecting one solution and discarding the other, we interpret them as bracketing a plausible parameter region for $C_{1L}$ and $C_{2L}$ as follows

\begin{equation}
C_{1,\min}=\min\!\bigl(\widehat{C}_{1L}^{(q)},\,\widehat{C}_{1L}^{(r)}\bigr)
\qquad
C_{1,\max}=\max\!\bigl(\widehat{C}_{1L}^{(q)},\,\widehat{C}_{1L}^{(r)}\bigr)
\label{eq:C1_bracket}
\end{equation}
\begin{equation}
C_{2,\min}=\min\!\bigl(\widehat{C}_{2L}^{(q)},\,\widehat{C}_{2L}^{(r)}\bigr)
\qquad
C_{2,\max}=\max\!\bigl(\widehat{C}_{2L}^{(q)},\,\widehat{C}_{2L}^{(r)}\bigr)
\label{eq:C2_bracket}
\end{equation}

Although equations~\eqref{eq:C1_bracket} and \eqref{eq:C2_bracket} are directly derived from experimental measurements, several sources of uncertainty are there that suggests a need for a controlled widening of the current parameter range. These include uncertainty in the reported dissipation values, the approximate nature of the waveform interpolation used to construct the $8$ breaths per minute dynamics, and the possibility that different parameter combinations may result in nearly identical dissipative energies. Thus, to tackle these uncertainties, we expand the conservative intervals by $\pm 25\%$ and define the final search bounds
\begin{equation}
C_{1L,\min}=0.75\,C_{1,\min}
\qquad
C_{1L,\max}=1.25\,C_{1,\max}
\label{eq:C1_expanded}
\end{equation}
\begin{equation}
C_{2L,\min}=0.75\,C_{2,\min}
\qquad
C_{2L,\max}=1.25\,C_{2,\max}
\label{eq:C2_expanded}
\end{equation}

With the search region defined, the next step is to construct an objective function that enforces consistency across breathing rates and breathing modes. For any candidate parameter pair $(C_{1L},C_{2L})$, the predicted dissipated energy per breath in mode $m\in\{q,r\}$ at rate $f\in\{8,12\}$ is computed from equation~\eqref{eq:Wdis_mf_linear} as
\begin{equation}
E_{f}^{(m)}(C_{1L},C_{2L}) = C_{1L}\,I_1^{(m,f)} + C_{2L}\,I_2^{(m,f)}.
\label{eq:Efm_definition}
\end{equation}

Then we define the experimental dissipation ratio
\begin{equation}
R_{\mathrm{data}}=\frac{E_{\mathrm{dis}}^{(12)}}{E_{\mathrm{dis}}^{(8)}}
\label{eq:R_data}
\end{equation}
and the corresponding predicted ratios for each breathing mode
\begin{equation}
R_{\mathrm{pred}}^{(m)}(C_{1L},C_{2L})=
\frac{E_{12}^{(m)}(C_{1L},C_{2L})}{E_{8}^{(m)}(C_{1L},C_{2L})}
\label{eq:R_pred_m}
\end{equation}
The assumption we used in this work is that the net energy loss of the respiratory subsystem should be comparable between the quiet and regular breathing at a given rate. Since the experimental data do not label mode, we enforce that both breathing modes reproduce the observed cross-rate scaling. We therefore define the objective function of 
\begin{equation}
J(C_{1L},C_{2L})=
\bigl(R_{\mathrm{pred}}^{(q)}-R_{\mathrm{data}}\bigr)^2
+
\bigl(R_{\mathrm{pred}}^{(r)}-R_{\mathrm{data}}\bigr)^2
\label{eq:cost_function_ratio}
\end{equation}

The ratio-based cost function in equation~\eqref{eq:cost_function_ratio} is adopted because it captures the frequency dependence of the dissipative energy. Moreover, by operating on ratios rather than absolute energy values, the formulation also reduces the sensitivity to uncertainty in the reported magnitudes of $E_{\mathrm{dis}}^{(12)}$ and $E_{\mathrm{dis}}^{(8)}$, since any uniform bias affecting both measurements partially cancels. In addition, this approach enforces a shared parameterization across quiet and regular breathing conditions, requiring a single pair of damping coefficients to reproduce the same cross-rate scaling for both quiet and regular breathing and thereby preventing overfitting to a specific breathing condition. Thus, in the optimization procedure, candidate pairs $(C_{1L},C_{2L})$ are sampled within the expanded bounds given by equations~\eqref{eq:C1_expanded} and \eqref{eq:C2_expanded}, to minimize the objective function in equation~\eqref{eq:cost_function_ratio}. 
\par
The next part of the optimization framework involves the application of particle swarm optimization algorithm. This approach will provide a population based stochastic search strategy for identifying the parameters that best reproduce the supportive forces during quiet and regular breathing. In this framework, each particle represents a candidate parameter vector, denoted by \(\mathbf{x}_{i}^{(t)} \in \mathbb{R}^{2}\), where the dimensionality of the parameter space corresponds to the two optimization variables \(C_{1L}\) and \(C_{2L}\), and \(j\) is the iteration index. The velocity of the particle \(\mathbf{v}_{i}^{(j)}\) determines the direction and magnitude of its movement through the search domain. The evolution of each particle is influenced by its own historically best position \((\mathbf{pbest}_{i})\) and by the best position discovered by the swarm \((\mathbf{gbest})\). The velocity update rule combines inertia with stochastic cognitive and social components and is written as
\begin{equation}
v_{i,d}^{(j+1)} = \chi \left[ w\, v_{i,d}^{(j)} + c_{1} r_{1}\big(pbest_{i,d} - x_{i,d}^{(j)}\big) + c_{2} r_{2}\big(gbest_{d} - x_{i,d}^{(j)}\big), \right]
\end{equation}
where \(\chi = 0.729\) is the constriction factor that ensures a stable convergence, \(c_{1} = c_{2} = 2.05\) are the cognitive and social coefficients, and \(r_{1}\) and \(r_{2}\) are independent stochastic variables drawn from a uniform distribution between zero and one. The particle position is updated according to
\begin{equation}
x_{i,d}^{(j+1)} = x_{i,d}^{(j)} + v_{i,d}^{(j+1)}.
\end{equation}

To ensure an effective balance between the global exploration during the early stages of the search and the local refinement as the algorithm approaches convergence, the inertia weight \(w\) decreases linearly with time according to
\begin{equation}
w = 0.9 - j\left(\frac{0.5}{T_{\max}}\right),
\end{equation}
where \(T_{\max}\) is the maximum number of iterations. This schedule is known to enhance convergence properties by gradually reducing the influence of previous velocities. In the present study, the particle swarm consists of 4000 particles and is executed for 5000 iterations to minimize the objective function. At each iteration, every particle compares its current objective value to its historical best value and updates \(\mathbf{pbest}_{i}\) if improvement is observed. The global best \(\mathbf{gbest}\) is also updated whenever a particle discovers a superior solution. After the convergence, the optimal parameter vector is obtained as
\begin{equation}
\mathbf{x}^{*} = \left[ C_{1L}^{*},\, C_{2L}^{*} \right],
\end{equation}
which represents the set of values that most accurately reproduces the supportive forces associated with quiet and regular breathing. To assess the robustness and reduce the influence of stochastic variability inherent to swarm-based optimizations, the particle swarm procedure is repeated across multiple independent runs using different random seeds. The final calibrated values of $(C_{1L}, C_{2L})$ are obtained by averaging the optimal solutions across all trials, and the associated variability is quantified through the corresponding standard deviations. The resulting coefficients are \( C_{\mathrm{1L}} = 195.79 \pm 16.05\%~\mathrm{kg\,s^{-1}} \) and
\( C_{\mathrm{2L}} = 1291.01 \pm 16.05\%~\mathrm{kg\,m^{-2}\,s^{-1}} \).
\par
With the identification of the lung damping coefficients $C_{1L}$ and $C_{2L}$, the mechanical energy dissipated by the lung dashpot during a breathing cycle can now be evaluated directly from the model. For a given breathing mode $m\in\{q,r\}$ at $12$ breaths per minute, the lung-related dissipative energy is computed as
\begin{equation}
E_{L}^{(m,12)} = C_{1L}\,I_1^{(m,12)} + C_{2L}\,I_2^{(m,12)},
\end{equation}
where $I_1^{(m,12)}$ and $I_2^{(m,12)}$ are the dissipation integrals obtained from the corresponding lung-volume waveform. This modeled lung dissipation represents only the contribution associated with tissue-level damping captured by $C_{1L}$ and $C_{2L}$ and is therefore not expected to exactly match the experimentally reported total dissipative energy $E_{\mathrm{dis}}^{(12)}$, which reflects dissipative losses from both the lungs and the compressible airways. Thus, the remaining discrepancy between the experimentally reported dissipation and the lung dissipation is attributed to additional viscous losses occurring within the compressible airways. Accordingly, the damping energy dissipated by the compressible airways over one breathing cycle is defined as
\begin{equation}
E_{cd}^{(m,12)} = E_{\mathrm{dis}}^{(12)} - \left( C_{1L}\,I_1^{(m,12)} + C_{2L}\,I_2^{(m,12)} \right).
\end{equation}

Applying this expression to the identified damping parameters gives an average mechanical energy dissipation by the compressible airways of approximately $0.015~\mathrm{J/cycle}$ for regular breathing and $0.02~\mathrm{J/cycle}$ for quiet breathing at $12$ breaths per minute.

\subsection*{Modeling of Lung Supportive Forces}

\begin{figure}[htbp]
\centering
\includegraphics[scale=0.4]{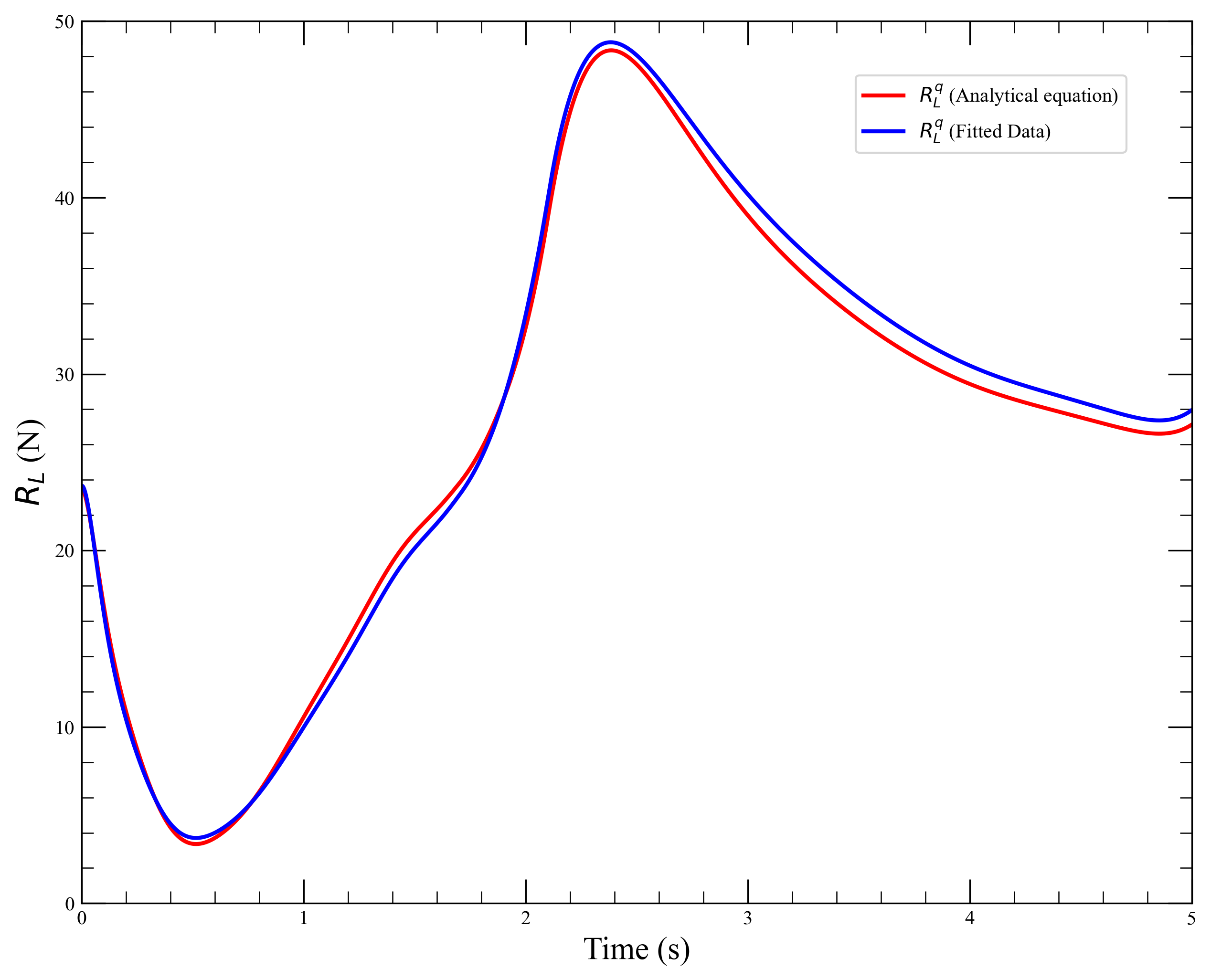}
\caption{Comparison between the lung supportive force obtained directly from the governing force balance equations and the supportive force reconstructed using the unified state-based formulation. The red curve represents the supportive force \(R_{L}^{q}(t)\) computed analytically from the full lung piston–cylinder model under quiet breathing conditions, while the blue curve shows the corresponding supportive force reconstructed from the fitted linear relation \(R_{L}(t) = \bar{A}_{R}\,\widetilde{V}_{L}(t) + \bar{B}_{R}\), where \(\widetilde{V}_{L}(t) = V_{L}(t) + \beta_{L}\dot{V}_{L}(t)\). The close agreement between the two trajectories over the entire breathing cycle demonstrates that the averaged coefficients accurately capture the underlying mechanical behavior of the lung supportive force despite the simplification.}
\label{fig:R_L_q}
\end{figure}

\begin{figure}[htbp]
\centering
\includegraphics[scale=0.4]{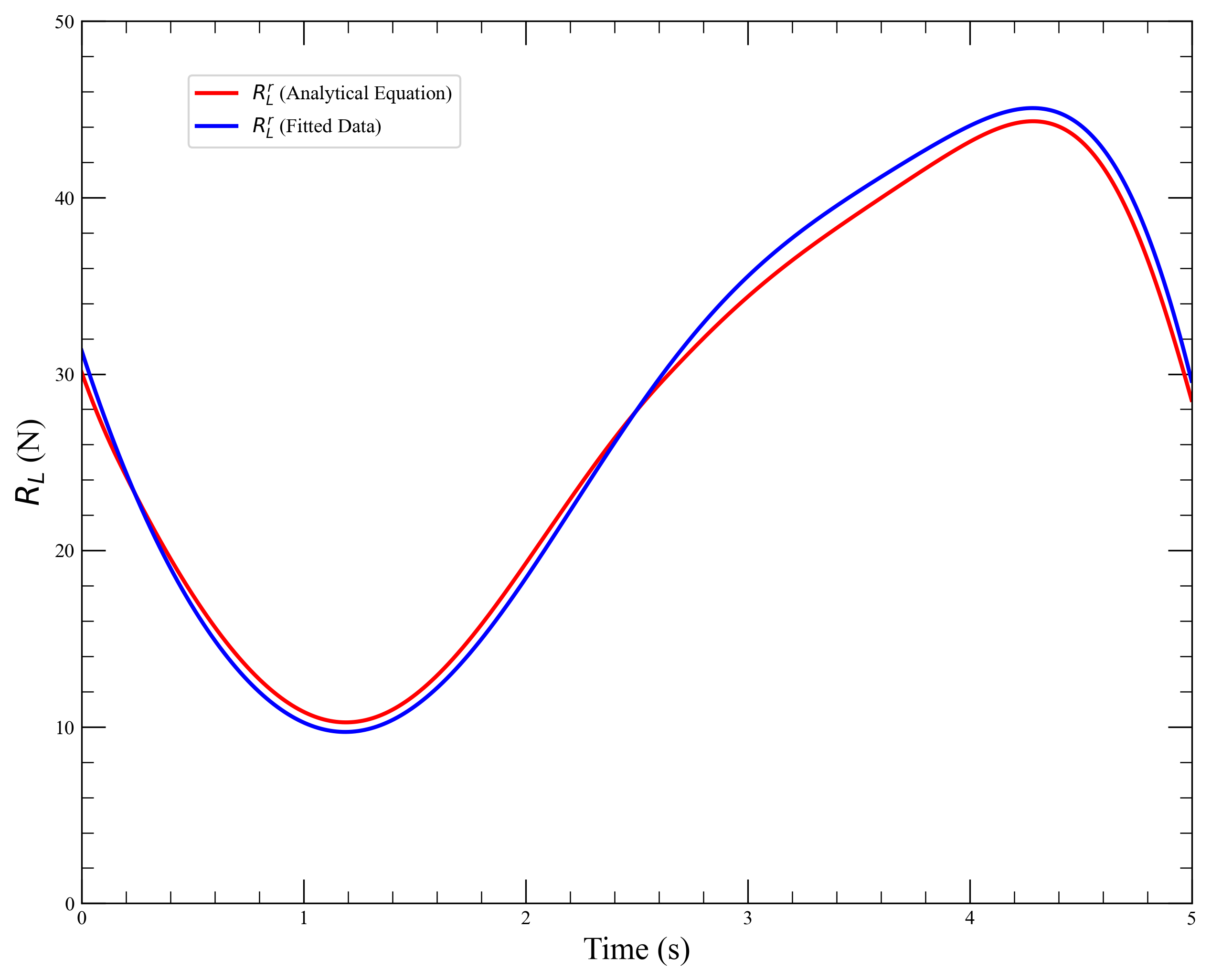}
\caption{Comparison between the lung supportive force computed from the full governing equations and the supportive force reconstructed using the unified state-based formulation under regular breathing conditions. The red curve denotes the analytically obtained supportive force \(R_{L}^{r}(t)\) evaluated directly from the lung piston–cylinder force balance, while the blue curve represents the supportive force reconstructed from the fitted linear relation \(R_{L}(t) = \bar{A}_{R}\,\widetilde{V}_{L}(t) + \bar{B}_{R}\), where the effective lung volume is defined as \(\widetilde{V}_{L}(t) = V_{L}(t) + \beta_{L}\dot{V}_{L}(t)\). The close overlap of the two curves throughout the breathing cycle demonstrates that the averaged state-based formulation accurately captures the dominant mechanical features of the lung supportive force during regular breathing, including both the inspiratory loading phase and the expiratory unloading phase.}
\label{fig:R_L_r}
\end{figure}

The particle swarm optimization algorithm provides optimal estimates of the lung damping coefficients \(C_{1L}\) and \(C_{2L}\). Substituting these optimized parameters into equations~\ref{eq:lung_governing_regular} and~\ref{eq:lung_governing_quiet} allows the lung supportive forces associated with quiet breathing, denoted by \(R_{L}^{q}\), and regular breathing, denoted by \(R_{L}^{r}\), to be computed directly from the force balance equations. The resulting supportive forces, shown in Figs.~\ref{fig:R_L_q} and~\ref{fig:R_L_r}, reflect the mechanical response of the thoracic support system under the corresponding breathing conditions. A direct use of these expressions, however, provides supportive forces that are explicitly tied to the breathing pattern from which they are derived. Such dependence limits their applicability to phonatory conditions, where pleural pressure dynamics, airflow rates, and temporal characteristics differ substantially from those observed during quiet or regular breathing. To establish a formulation that remains valid across a broader range of respiratory tasks, including sustained phonation and nonperiodic breathing patterns, the supportive force must be expressed in a form that depends only on the instantaneous mechanical state of the lung rather than on the global breathing mode.

In the current framework, the mechanical state of the lung is characterized by \(\big(V_{L}(t),\,\dot{V}_{L}(t),\,\ddot{V}_{L}(t)\big)\), which captures the instantaneous lung volume, volumetric flow rate, and inertial loading, respectively. Thus, the lung supportive force is expressed in the general functional form
\begin{equation}
R_{L}(t) = f\big(V_{L}(t),\,\dot{V}_{L}(t),\,\ddot{V}_{L}(t)\big),
\label{eq:RL_state}
\end{equation}
which implies that two respiratory conditions producing identical instantaneous lung volume, flow, and acceleration must generate identical supportive forces. To construct a unified representation of the supportive force, an effective lung volume \(\widetilde{V}_{L}(t)\) is introduced and defined as
\begin{equation}
\widetilde{V}_{L}(t) = V_{L}(t) + \beta_{L}\,\dot{V}_{L}(t) + \gamma_{L}\,\ddot{V}_{L}(t),
\label{eq:effective_lung_volume}
\end{equation}
where \(\beta_{L}\) and \(\gamma_{L}\) are scalar coefficients that quantify the relative contributions of flow-dependent and inertial effects to the effective mechanical state of the lung. In principle, a more general formulation could include higher-order temporal derivatives of lung volume to explicitly represent additional dynamic effects. However, identifying such contributions would require reliable measurements across multiple distinct breathing conditions. Since only quiet and regular breathing data are available in the present study, the representation is truncated at the second temporal derivative. The relative importance of the included terms is further assessed through the ratio \(\left|\gamma_{L}/\beta_{L}\right|\), which provides insight into the hierarchical influence of inertial versus flow-related effects in shaping the lung supportive force.
\par
Following this formulation, a consistent description of the lung supportive force across different breathing conditions is obtained when the trajectories of \(R_{L}^{q}\) and \(R_{L}^{r}\) intersect at identical values of the effective lung volume \(\widetilde{V}_{L}\). Each intersection corresponds to an instant at which quiet and regular breathing produce the same supportive force, despite arising from distinct respiratory dynamics. Intersection points between the supportive force curves were identified using cubic spline interpolation. At each intersection time \(t_{i}\), the corresponding values of lung volume \(V_{L}(t_{i})\), volumetric flow rate \(\dot{V}_{L}(t_{i})\), and volumetric acceleration \(\ddot{V}_{L}(t_{i})\) were extracted for both breathing conditions. Using two such intersection instances, denoted by \((t_{\mathrm{in}1}, t_{\mathrm{in}2})\), a system of equations was formulated to determine the coefficients \(\beta_{L}\) and \(\gamma_{L}\):
\begin{equation}
\begin{aligned}
\beta_{L} \Big(\dot{V}_L^q(t_{\mathrm{in}1}) - \dot{V}_L^r(t_{\mathrm{in}1})\Big)
+ \gamma_{L} \Big(\ddot{V}_L^q(t_{\mathrm{in}1}) - \ddot{V}_L^r(t_{\mathrm{in}1})\Big)
&= V_L^r(t_{\mathrm{in}1}) - V_L^q(t_{\mathrm{in}1}) \\
\beta_{L} \Big(\dot{V}_L^q(t_{\mathrm{in}2}) - \dot{V}_L^r(t_{\mathrm{in}2})\Big)
+ \gamma_{L} \Big(\ddot{V}_L^q(t_{\mathrm{in}2}) - \ddot{V}_L^r(t_{\mathrm{in}2})\Big)
&= V_L^r(t_{\mathrm{in}2}) - V_L^q(t_{\mathrm{in}2})
\end{aligned}
\label{eq:beta_gamma_identification}
\end{equation}

Solving this system gives \(\beta_{L} = -5.919~\mathrm{s}\) and \(\gamma_{L} = 0.0164~\mathrm{s^{2}}\). These values indicate that the volumetric flow rate contributes strongly and negatively to the effective state variable, while the contribution from the volumetric acceleration is positive but comparatively small. The ratio \(\left|\gamma_{L}/\beta_{L}\right| \approx 2.77 \times 10^{-3}\) demonstrates that the influence of the second temporal derivative is several orders of magnitude weaker than that of the first derivative. This pronounced disparity confirms that higher-order temporal effects contribute negligibly to the supportive force within the physiological range of breathing dynamics. Consequently, the acceleration term is omitted from our reduced order reconstruction without meaningful loss of accuracy. This allowed the supportive force to be described accurately using only the lung volume and its first temporal derivative. Under this simplification, the effective lung volume reduces to \(\widetilde{V}_{L}(t) = V_{L}(t) + \beta_{L}\,\dot{V}_{L}(t)\). Now, a unified representation of the lung supportive force is obtained by expressing \(R_{L}(t)\) as a linear function of \(\widetilde{V}_{L}(t)\):
\begin{equation}
R_{L}(t) = A_{R}\,\widetilde{V}_{L}(t) + B_{R}
\end{equation}

The coefficients \(A_{R}\) and \(B_{R}\) are computed from the supportive forces associated with quiet and regular breathing as
\begin{equation}
A_{R} = \frac{R_{L}^{r} - R_{L}^{q}}{\widetilde{V}_{L}^{r} - \widetilde{V}_{L}^{q}}
\end{equation}
\begin{equation}
B_{R} = R_{L}^{q} - A_{L}\,\widetilde{V}_{L}^{q}
\end{equation}

The coefficients \(A_{R}\) and \(B_{R}\) vary in time because the relationship between the supportive force and effective lung volume evolves throughout the breathing cycle. To further simplify the model, average values of these coefficients, denoted by \(\bar{A}_{R}\) and \(\bar{B}_{R}\), were computed. The validity of this approximation was assessed by reconstructing the supportive force using
\(R_{L}(t) = \bar{A}_{R}\,\widetilde{V}_{L}(t) + \bar{B}_{R}\) and comparing the results with the supportive forces obtained directly from quiet and regular breathing. As shown in Figs.~\ref{fig:R_L_q} and~\ref{fig:R_L_r}, this averaged representation achieves a fit accuracy of approximately \(94.2\%\) for quiet breathing and \(95.2\%\) for regular breathing. These results demonstrate that despite the simplification, the averaged coefficients reproduce the supportive force trajectories with high fidelity.

\subsection*{Compressible Airways Geometry and Effective Mass}

\begin{figure}[htbp]
\centering
\includegraphics[width=.6\textwidth]{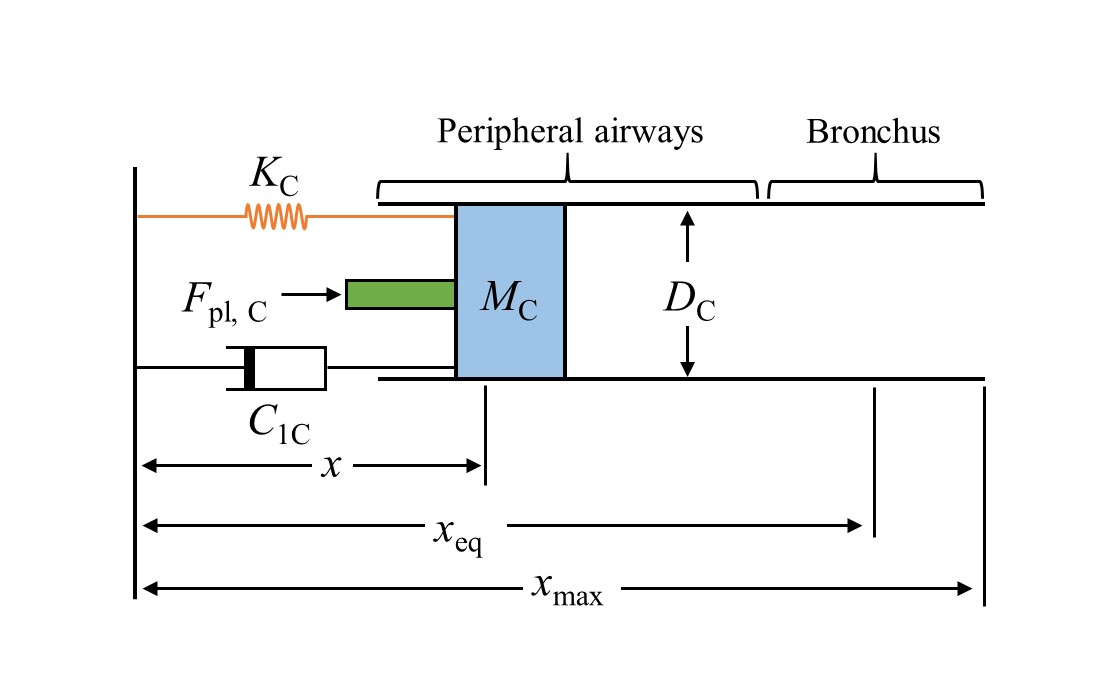}
\caption{Lumped-element representation of the compressible airways. The piston of effective mass \(M_{C}\) represents the deformable airway walls. The spring element with stiffness \(K_{C}\) models the elastic compliance of the airway tissues, while the dashpot with damping coefficient \(C_{1C}\) accounts for viscous energy dissipation associated with airway wall motion. The piston is driven by the intrapleural force \(F_{\mathrm{pl,C}}\), which induces volume changes. The instantaneous piston displacement is denoted by \(x\), with \(x_{\mathrm{eq}}\) indicating the equilibrium position and \(x_{\max}\) the maximum allowable excursion corresponding to the fully expanded compressible airways volume. The effective airway diameter is denoted by \(D_{C}\). The upper region corresponds to the bronchus, while the lower region represents the peripheral airways lacking cartilaginous support.}
\label{fig:compressible_airway_piston_model}
\end{figure}

The compressible airways represent a pressure-sensitive, variable-volume segment whose geometry changes in response to the transmural pressure~\cite{golden2007mathematical}. The complete geometric configuration is illustrated in Fig.~\ref{fig:unified_phonation_model}, where each cylindrical assembly consists of an upper segment representing a bronchus and a lower segment representing one-half of the peripheral airways region. Each piston-cylinder unit is idealized as having a constant diameter \(D_{C}\), as shown in Fig.~\ref{fig:compressible_airway_piston_model}. To maintain geometric and mechanical continuity with the proximal airways, this diameter is taken to be equal to the exit diameter of the lung piston-cylinder system, \((D_{C} = 17~\mathrm{mm})\). This choice is consistent with established anatomical averages for the right and left bronchi and avoids introducing nonphysiological flow features associated with abrupt expansions or contractions at the airway junctions. The corresponding cross-sectional area is therefore \(\alpha_{C} = \pi D_{C}^{2}/4 \approx 2.27\times10^{-4}~\mathrm{m^{2}}\).

The total compressible airways volume consists of contributions from both the two bronchi and the peripheral airways. A representative bronchial length of \(L_{B} = 3~\mathrm{cm}\) is adopted for each bronchus~\cite{lee2014comparison}, giving a combined bronchial volume of \(V_{B} = 2\,\alpha_{C}L_{B} \approx 13.65~\mathrm{mL}\). The remaining contribution arises from the peripheral airways, whose volume has been reported to vary between \(0.015~\mathrm{L}\) and \(0.185~\mathrm{L}\)~\cite{marconi2020silico}. As a result, the total compressible airways volume spans from \(28.65~\mathrm{mL}\) to \(198.65~\mathrm{mL}\). Using the upper bound of this range, the corresponding maximum piston excursion is
\(x_{\max} = \frac{V_{C,\max}}{2\alpha_{C}} == 43.76~\mathrm{cm}\) This effective maximum travel length represents the volume-equivalent axial displacement of the combined bronchi and peripheral airways structures within the lumped compressible airways model. During dynamic simulations, the instantaneous compressible airways volume is expressed as
\begin{equation}
V_{C}(t) = 2\,\alpha_{C}\bigl(x_{\max} - x(t)\bigr),
\label{eq:compressible_airway_volume}
\end{equation}
where \(x(t)\) denotes the time-dependent piston displacement. To ensure consistency with lung mechanics at functional residual capacity, the equilibrium position of the compressible airways piston is defined such that the airway volume occupies the same fraction of its maximum volume as the lung volume occupies relative to the total lung capacity. This condition gives
\(x_{\mathrm{eq}} = x_{\max}\left(\frac{h_{\mathrm{eq}}}{h_{\max}}\right)\). The inertial response of the compressible airways is represented through an effective mass \(M_{C}\), corresponding to the mass of the airway wall tissues associated with the two bronchi and the peripheral airways. Direct measurements of the mass of the peripheral airways tissue are not available. Therefore, a representative tissue density \(\rho_{C}\) is estimated by averaging reported densities of the primary structural constituents of the airways walls. Hyaline cartilage exhibits densities between \(1.1\) and \(1.2~\mathrm{g\,mL^{-1}}\)~\cite{eschweiler2021biomechanics}, connective tissue typically ranges from \(1.0\) to \(1.04~\mathrm{g\,mL^{-1}}\)~\cite{dakin2011changes}, and the smooth muscle has a reported density of approximately \(1.06~\mathrm{g\,mL^{-1}}\)~\cite{ward2005density}. Assuming comparable volumetric contributions from these tissue types, their mean value is adopted, giving \(\rho_{C} \approx 1.10~\mathrm{g\,mL^{-1}} = 1100~\mathrm{kg\,m^{-3}}\). The total airway wall tissue volume is approximated as \(V_{w,\,C} = 2\pi x_{\max}\left[(r_{C}+t_{C})^{2} -r_{C}^{2}\right]\), where the inner radius \(r_{C} = 0.0085~\mathrm{m}\) and wall thickness \(t_{C} = 0.000799~\mathrm{m}\) are taken from anatomical measurements reported by Bankier \textit{et al.}~\cite{bankier1996bronchial}. Using these values gives an effective compressible airways tissue mass of \(M_{C} = \rho_{C} V_{t} \approx 2.95\times10^{-3}~\mathrm{kg}\).

\subsection*{Compressible Airways Spring Model and Compliance}

The elastic behavior of the compressible airways is represented in the lumped-element framework by an effective spring with stiffness \(K_{C}\). Consistent with the mechanical analogy adopted for the lung piston–cylinder system, the spring stiffness is related to the airway compliance through
\begin{equation}
K_{C} = \frac{2\,\alpha_{C}^{2}}{C_{C}}
\label{eq:compressible_airway_stiffness_compliance},
\end{equation}
where \(\alpha_{C}\) is the cross-sectional area of a single airway piston and \(C_{C}\) denotes the compliance of the compressible airways compartment. The factor of 2 accounts for the presence of two identical piston–cylinder units arranged in parallel to represent the combined response of the bilateral compressible airways system. The deformation of the compressible airways is governed by the transmural pressure \(P_{tm}\), defined as the pressure difference between the airway lumen and the surrounding
intrapleural space \cite{marconi2020silico}. The static pressure–volume relationship of the compressible airways is described by the sigmoidal function
\begin{equation}
V_{C}(P_{tm}) = \frac{V_{C,\max}}{1 + \exp\!\left(-\frac{P_{tm}-A}{B}\right)},
\label{eq:compressible_airway_pressure_volume}
\end{equation}
where \(V_{C,\max} = 198.65~\mathrm{mL}\) denotes the maximum attainable compressible airways volume. The parameter \(A = 431.493~\mathrm{Pa}\) specifies the transmural pressure at which the airway compliance reaches its maximum, while \(B = 431.493~\mathrm{Pa}\) controls the steepness of the pressure–volume transition. Differentiation of equation~\eqref{eq:compressible_airway_pressure_volume} with respect to \(P_{tm}\) gives the nonlinear, pressure-dependent compliance
\begin{equation}
C_{C}(V_{C}) = \frac{dV_{C}}{dP_{tm}} = \frac{V_{C,\max}}{B} \frac{\exp\!\left(-\frac{P_{tm}-A}{B}\right)}
{\left[1 + \exp\!\left(-\frac{P_{tm}-A}{B}\right)\right]^{2}} = \frac{V_{C}}{B}
\left(1 - \frac{V_{C}}{V_{C,\max}}\right),
\label{eq:compressible_airway_compliance}
\end{equation}
which reaches its maximum near mid-range airway volumes and decreases as the airway approaches either collapse or full distension. Substitution of equation~\eqref{eq:compressible_airway_compliance} into equation~\eqref{eq:compressible_airway_stiffness_compliance} gives the corresponding pressure-dependent stiffness of the compressible airways
\begin{equation}
K_{C} = \frac{2\,\alpha_{C}^{2}\,B\,V_{C,\max}}
{V_{C}\,\bigl(V_{C,\max} - V_{C}\bigr)},
\label{eq:compressible_airway_stiffness}
\end{equation}
which increases sharply as the airways volume approaches the limits of collapse or full distension.

\subsection*{Damping Behavior of the Compressible Airways}

The dissipative behavior of the compressible airways is represented in the lumped-element framework by a single Newtonian dashpot, as illustrated in Fig.~\ref{fig:compressible_airway_piston_model}. This is different from the lung damping formulation, which incorporates both linear and nonlinear components to account for the tissue-level dissipation. This is because the compressible airways experience substantially smaller deformations than the lung parenchyma during respiration and phonation. As a result, the dominant mechanism of energy loss in this region arises from viscous friction rather than nonlinear tissue deformation or strain-dependent stiffening. Accordingly, a linear Newtonian damping representation is sufficient to capture the relevant physical behavior of the compressible airways. With this assumption, the damping force acting on each compressible airways piston is
modeled as
\begin{equation}
F_{\mathrm{damp,\,C}}(\dot{x}) = -C_{1C}\,\dot{x},
\label{eq:compressible_airway_damping_force}
\end{equation}
where \(C_{1C}\) denotes the viscous damping coefficient and \(\dot{x}\) is the axial velocity of the compressible airways piston. The mechanical energy dissipated by this damping element over a single breathing cycle is obtained by integrating the instantaneous power loss \(C_{1C}\,\dot{x}^{2}\) over the cycle duration. Because the compressible airways are represented by two identical piston–cylinder assemblies operating in parallel, energy is dissipated in both damping elements. The total dissipated energy per breathing cycle is therefore given by
\begin{equation}
E_{\mathrm{cd}} = 2\,C_{1C}\,\int_{0}^{T_{\mathrm{breath}}} \dot{x}^{2}\,dt,
\label{eq:compressible_airway_energy_dissipation}
\end{equation}
where \(T_{\mathrm{breath}}\) denotes the duration of one complete breathing cycle and \(E_{\mathrm{cd}}\) represents the total damping energy loss per cycle. Using the kinematic relation
\(\dot{x} = -\dot{V}_{C}/(2\alpha_{C})\), the dissipated energy can be written as
\begin{equation}
E_{\mathrm{cd}} = \frac{C_{1C}}{2} \int_{0}^{T_{\mathrm{breath}}} \left(\frac{\dot{V}_{C}}{\alpha_{C}}\right)^{2} dt
\label{eq:compressible_airway_energy_volume_form}
\end{equation}

The average mechanical energy dissipated by the compressible airways during one breathing cycle is provided in the supplementary materials under the section particle swamp optimization algorithm, with values of \(0.015~\mathrm{J/cycle}\) for regular breathing and \(0.02~\mathrm{J/cycle}\) for quiet breathing. Equating these values with equation~\eqref{eq:compressible_airway_energy_volume_form} allows the damping coefficient to be computed for each breathing condition. For quiet and regular breathing, the corresponding damping coefficients are given by
\begin{equation}
C_{1C}^{q} = \frac{2E_{\mathrm{cd}}^{q}} {\displaystyle \int_{0}^{T_{\mathrm{breath,\,q}}} \left(\frac{\dot{V}_{C}^{q}}{\alpha_{C}}\right)^{2} dt} 
\end{equation}
\begin{equation}
C_{1C}^{r} = \frac{2E_{\mathrm{cd}}^{r}} {\displaystyle \int_{0}^{T_{\mathrm{breath,\,r}}} \left(\frac{\dot{V}_{C}^{r}}{\alpha_{C}}\right)^{2} dt}
\label{eq:compressible_airway_damping_coefficients}
\end{equation}

A single effective damping coefficient is adopted to represent the compressible airways across all respiratory conditions. This coefficient is defined as the arithmetic mean of the values obtained for quiet and regular breathing, \(C_{1C} = (C_{1C}^{q} + C_{1C}^{r})/2 = 7.37~\mathrm{N\,s\,m^{-1}}\).

\subsection*{Governing Equation of the Compressible Airways System}

\begin{figure}[htbp]
\centering
\includegraphics[width=0.3\textwidth]{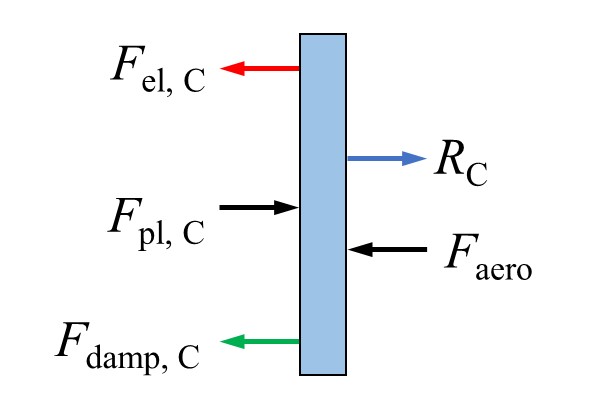}
\caption{Free-body diagram of the piston representing the compressible airways. The piston is subjected to an elastic restoring force $F_{\mathrm{el,C}}$, a viscous damping force $F_{\mathrm{damp,C}}$, an intrapleural pressure force $F_{\mathrm{pl,C}}$ acting from the surrounding pleural space, and an aerodynamic pressure force $F_{\mathrm{aero}}$ arising from the airway lumen. The supportive force $R_{C}$ represents structural support provided by the surrounding airway tree and connective tissues.}
\label{fig:compressible_airway_fbd}
\end{figure}

As illustrated in Fig.~\ref{fig:compressible_airway_fbd}, several forces act on the piston of the compressible airways piston-cylinder system. The elastic restoring force $F_{\mathrm{el,C}}$ arises from the deformation of the airways wall and is modeled as
\begin{equation}
F_{\mathrm{el,C}}(t) = -K_{C}(t)\Big(x(t) - x_{\mathrm{eq}}\Big).
\label{eq:compressible_airway_elastic_force}
\end{equation}

The viscous energy dissipation within the airway wall is represented by the damping force $F_{\mathrm{damp,C}}$, defined previously in equation~\eqref{eq:compressible_airway_damping_force}. The piston is further subjected to an intrapleural pressure force $F_{\mathrm{pl,C}} = \alpha_{C} P_{\mathrm{pl}}(t)$ acting from the surrounding pleural cavity and an opposing aerodynamic pressure force $F_{\mathrm{aero}} = \alpha_{C} P_{\mathrm{aero}}(t)$ generated by the air contained within the airway lumen. In addition, the piston experiences a supportive force $R_{C}(t)$, which accounts for the mechanical support from the surrounding airway tree and connective tissues. Summing all forces acting on the piston and applying Newton’s second law gives the governing equation of motion
\begin{equation}
\begin{aligned}
M_{C}\,\ddot{x}(t) &= -K_{C}(t)\Big(x(t) - x_{\mathrm{eq}}\Big) + \alpha_{C} P_{\mathrm{pl}}(t) - C_{1C}\,\dot{x}(t) - \alpha_{C} P_{\mathrm{aero}}(t) + R_{C}(t)
\end{aligned}
\label{eq:compressible_airway_motion}
\end{equation}

The governing equation can be expressed in terms of compressible airways volume using the kinematic relation $V_{C}(t) = 2\alpha_{C}\big(x_{\max} - x(t)\big)$,
which leads to
\begin{equation}
\begin{aligned}
-\frac{M_{C}}{2\alpha_{C}}\,\ddot{V}_{C}(t) &= K_{C}(t)\Bigg(\frac{V_{C}(t)}{2\alpha_{C}} + x_{\mathrm{eq}} - x_{\max}\Bigg) + \alpha_{C}\Big(P_{\mathrm{pl}}(t) - P_{\mathrm{aero}}(t)\Big) \\ &\quad + \frac{C_{1C}}{2\alpha_{C}}\,\dot{V}_{C}(t) + R_{C}(t)
\end{aligned}
\label{eq:compressible_airway_volume_eq}
\end{equation}

Substituting the time-resolved pressure and volume waveforms corresponding to regular and quiet breathing into equation~\eqref{eq:compressible_airway_volume_eq} produces the following system
\begin{equation}
\begin{aligned}
-\frac{M_{C}}{2\alpha_{C}}\,\ddot{V}_{C}^{r}(t) &= K_{C}^{r}(t)\Bigg(\frac{V_{C}^{r}(t)}{2\alpha_{C}} + x_{\mathrm{eq}} - x_{\max}\Bigg) + \alpha_{C}\Big(P_{\mathrm{pl}}^{r}(t) - P_{\mathrm{aero}}^{r}(t)\Big) \\ &\quad + \frac{C_{1C}}{2\alpha_{C}}\,\dot{V}_{C}^{r}(t) + R_{C}^{r}(t)
\end{aligned}
\label{eq:compressible_airway_regular}
\end{equation}

\begin{equation}
\begin{aligned}
-\frac{M_{C}}{2\alpha_{C}}\,\ddot{V}_{C}^{q}(t) &= K_{C}^{q}(t)\Bigg(\frac{V_{C}^{q}(t)}{2\alpha_{C}} + x_{\mathrm{eq}} - x_{\max}\Bigg) + \alpha_{C}\Big(P_{\mathrm{pl}}^{q}(t) - P_{\mathrm{aero}}^{q}(t)\Big) \\ &\quad + \frac{C_{1C}}{2\alpha_{C}}\,\dot{V}_{C}^{q}(t) + R_{C}^{q}(t)
\end{aligned}
\label{eq:compressible_airway_quiet}
\end{equation}

All parameters appearing in equations~\eqref{eq:compressible_airway_regular} and~\eqref{eq:compressible_airway_quiet} have been defined in the preceding sections. These governing equations therefore provide the supportive forces $R_{C}^{r}(t)$ and $R_{C}^{q}(t)$ associated with regular and quiet breathing directly from the mechanical balance of the compressible airways system.

\subsection*{Modeling of Compressible Airways Supportive Forces}

Following the same methodology developed for the lung supportive forces, the compressible airways supportive force is generalized in terms of the instantaneous mechanical state of the airways rather than the specific breathing pattern from which it is derived. For this, an effective compressible airways volume is introduced as a linear combination of the instantaneous volume and its temporal derivatives as follows
\begin{equation}
\widetilde{V}_{C}(t) = V_{C}(t) + \beta_{C}\dot{V}_{C}(t) + \gamma_{C}\ddot{V}_{C}(t),
\label{eq:effective_compressible_volume}
\end{equation}
where $\beta_{C}$ and $\gamma_{C}$ are scalar coefficients that quantify the influence of flow rate and inertial acceleration on the effective mechanical state of the compressible airways. 

The coefficients $\beta_{C}$ and $\gamma_{C}$ are identified by enforcing the condition that quiet and regular breathing must produce identical supportive forces whenever they share the same effective compressible airways volume. This condition is satisfied at the intersection points of the supportive-force trajectories $R_{C}^{q}(t)$ and $R_{C}^{r}(t)$. Intersection times are identified using a cubic spline interpolation of the two force signals. At each intersection time $t_{i}$, the corresponding values of $V_{C}(t_{i})$, $\dot{V}_{C}(t_{i})$, and $\ddot{V}_{C}(t_{i})$ are extracted for both breathing conditions. Using two such intersection points $t_{i,1}$ and $t_{i,2}$, the following linear system is constructed
\begin{equation}
\begin{aligned}
\beta_{C}\big(\dot{V}_{C}^{q}(t_{i,1})-\dot{V}_{C}^{r}(t_{i,1})\big) + \gamma_{C}\big(\ddot{V}_{C}^{q}(t_{i,1})-\ddot{V}_{C}^{r}(t_{i,1})\big) &= V_{C}^{r}(t_{i,1}) - V_{C}^{q}(t_{i,1}) \\ 
\beta_{C}\big(\dot{V}_{C}^{q}(t_{i,2})-\dot{V}_{C}^{r}(t_{i,2})\big) + \gamma_{C}\big(\ddot{V}_{C}^{q}(t_{i,2})-\ddot{V}_{C}^{r}(t_{i,2})\big) &= V_{C}^{r}(t_{i,2}) - V_{C}^{q}(t_{i,2})
\end{aligned}
\label{eq:compressible_state_matching}
\end{equation}

Solving this system gives $\beta_{C} = 14.762\,\mathrm{s}$ and $\gamma_{C} = 0.014\,\mathrm{s^{2}}$. These values indicate that the compressible airways supportive force is strongly influenced by the airflow rate, while inertial effects play a minor role. The ratio $\lvert\gamma_{C}/\beta_{C}\rvert \approx 9.5\times10^{-4}$ confirms that the contribution of the second temporal derivative is nearly three orders of magnitude smaller than that of the first derivative. As a result, the acceleration term can be neglected without noticeable loss of accuracy, leading to the reduced effective volume definition
\begin{equation}
\widetilde{V}_{C}(t) = V_{C}(t) + \beta_{C}\dot{V}_{C}(t)
\label{eq:reduced_effective_volume}
\end{equation}

With $\beta_{C}$ determined, the unified supportive force law is written as
\begin{equation}
R_{C}(t) = A_{C}\widetilde{V}_{C}(t) + B_{C}
\label{eq:compressible_supportive_final}
\end{equation}

The coefficients $A_{C}$ and $B_{C}$ are obtained from
\begin{equation}
A_{C} = \frac{R_{C}^{r} - R_{C}^{q}}{\widetilde{V}_{C}^{r} - \widetilde{V}_{C}^{q}}
\end{equation}

\begin{equation}
B_{C} = R_{C}^{q} - A_{C}\widetilde{V}_{C}^{q}
\label{eq:AC_BC_def}
\end{equation}

\begin{figure}[htbp]
\centering
\includegraphics[scale=0.45]{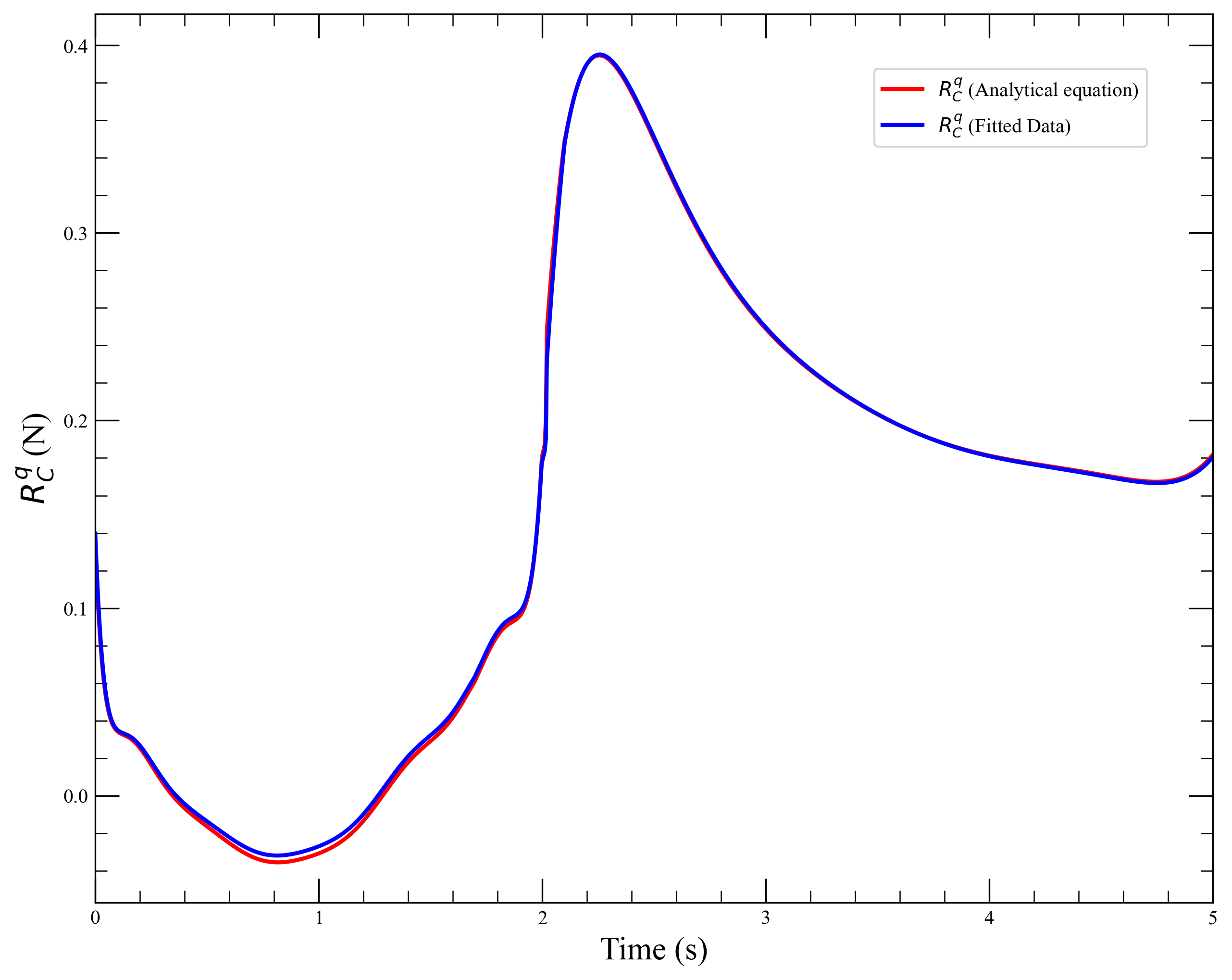}
\caption{Time-resolved compressible airways supportive force during quiet breathing. The red curve corresponds to the supportive force computed directly from the governing equation of the compressible airways piston–cylinder system, while the blue curve shows the force reconstructed using the unified state-based formulation. The close agreement demonstrates that the averaged linear representation accurately captures the compressible airways supportive force.}
\label{fig:compressible_airway_supportive_quiet}
\end{figure}

\begin{figure}[htbp]
\centering
\includegraphics[scale=0.45]{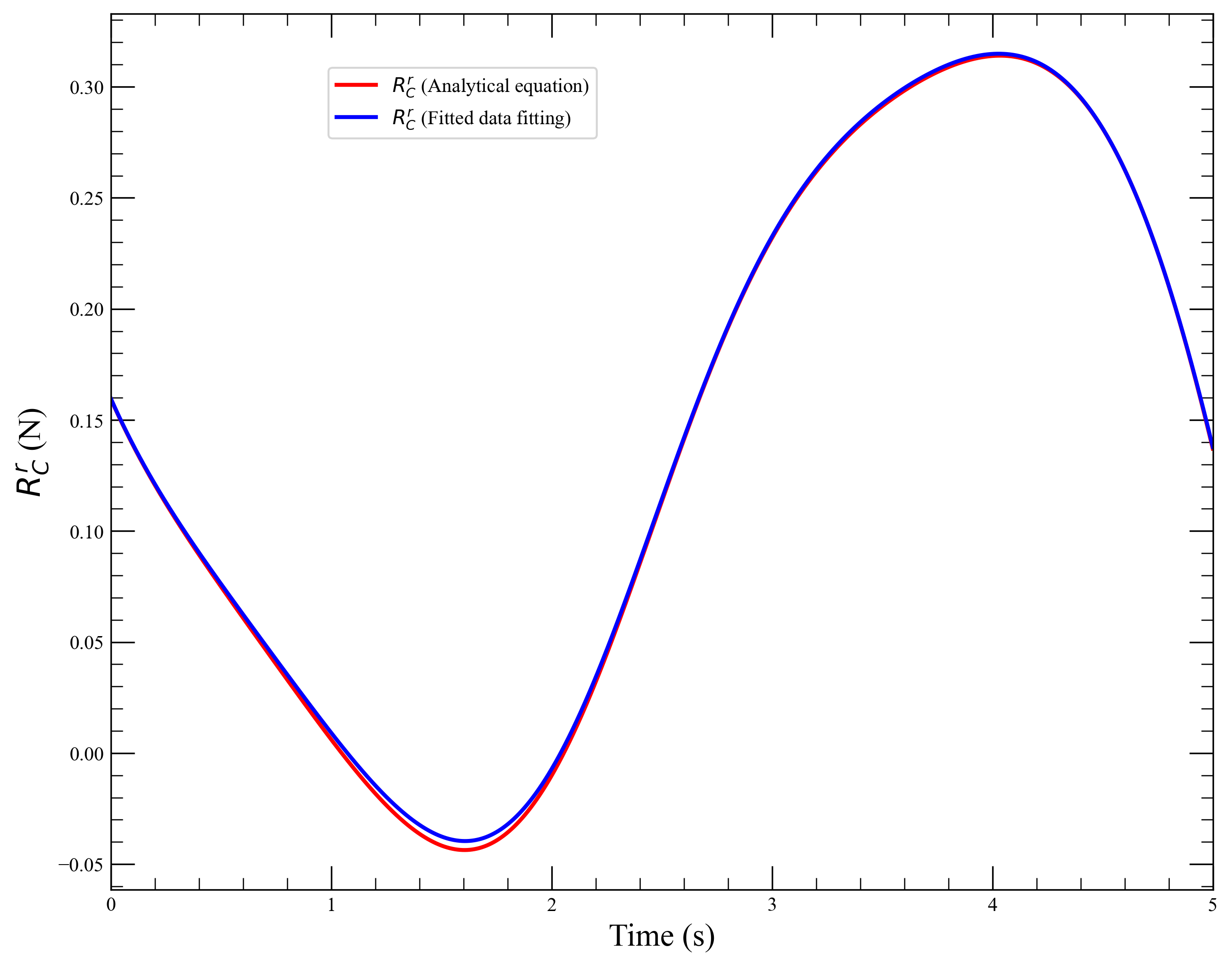}
\caption{Time-resolved compressible airways supportive force during regular breathing. The analytical solution obtained from the governing equation is shown in red, and the force reconstructed from the unified state-based model is shown in blue. The strong overlap between the two curves confirms that the reduced-order formulation accurately reproduces the supportive force dynamics.}
\label{fig:compressible_airway_supportive_regular}
\end{figure}

Although the coefficients \(A_{C}\) and \(B_{C}\) are time-dependent quantities that vary over the breathing cycle, we replace these time-varying coefficients by their cycle-averaged values, denoted by \(\bar{A}_{C} = -1076.888\,\mathrm{N\,m^{-3}}\) and \(\bar{B}_{C} = 0.219\,\mathrm{N}\). This averaging strategy mirrors the approach adopted for the lung supportive forces. Using these averaged coefficients, the compressible airways supportive force is approximated as \(R_{C}(t) \approx \bar{A}_{C}\,\widetilde{V}_{C}(t) + \bar{B}_{C}\), which provides a linear state-based description of the supportive force in terms of the effective compressible airways volume. The predictive accuracy of this approximation is evaluated by comparing the reconstructed supportive force with the forces obtained directly from the governing equations under quiet and regular breathing conditions. As shown in Figs.~\ref{fig:compressible_airway_supportive_quiet} and~\ref{fig:compressible_airway_supportive_regular}, the reconstructed force achieves a fit accuracy of \(98.7\%\) for quiet breathing and \(97.9\%\) for regular breathing.

\subsection*{Flow Resistances}\label{Note14}

The flow resistance quantifies the pressure drop required to drive a unit volumetric airflow through the airway network. In the present lumped-element formulation, seven distinct resistive components are incorporated to represent pressure losses occurring along different anatomical regions of the respiratory tract, as was illustrated in Fig.~\ref{fig:unified_phonation_model}. These components are the lung tissue resistance \(R_{LT}\), the small airways resistance \(FR_{S}\), the compressible airways resistance \(FR_{C}\), the tracheal resistance \(FR_{T}\), the vocal folds resistance \(FR_{VF}\), the upper airways resistance \(FR_{U}\), and the expiratory resistance \(FR_{E}\). Together, these elements account for viscous dissipation, geometric constrictions, and deformation-induced losses encountered by the airflow between the lungs and the atmosphere.
\par
The lung tissue resistance represents energy losses associated with the deformation of the lung parenchyma during volume change. This resistance is assumed to be constant and is taken as \(1.96 \times 10^{4}~\mathrm{Pa\,s\,m^{-3}}\) as reported by \cite{marconi2020silico}. This resistance captures the energy loss due to the tissue deformation and is independent of the airway geometry. The resistive contribution of the small airways reflects the flow limitation imposed by bronchioles. This volume-dependent behavior is modeled using an exponential relationship of the form
\begin{equation}
FR_{S} = R_{s,d}\, \exp\!\left[ - K_{S}\, \frac{V_{L} - RV}{TLC - RV} \right] + R_{s,m},
\label{eq:small_airway_resistance}
\end{equation}
where \(R_{s,d} = 215.75~\mathrm{kPa\,s\,m^{-3}}\) characterizes the dynamic component of small airways resistance that decreases with the lung inflation, \(K_{S} = 10.9\) governs the sensitivity of this resistance to the lung volume, and \(R_{s,m} = 1.96~\mathrm{kPa\,s\,m^{-3}}\) represents the minimum resistance attained at high lung volumes. The residual volume is taken as \(RV = 23\times10^{-6}~\mathrm{m^{3}}\). This formulation captures the physiological opening of small airways as the lung volume increases.
\par
In addition to this empirical description, viscous losses within the bronchi are evaluated using Poiseuille's law. The resistance of a single bronchus of length \(L_{B}\) and diameter \(D_{C}\) is given by
\begin{equation}
FR_{B}
= \frac{128\mu L_{B}}{\pi (D_{C})^{4}}
\label{eq:bronchial_poiseuille}
\end{equation}
Since the compressible airways representation consists of two identical bronchus, arranged in parallel, the effective resistance becomes one half of the single-path resistance, giving
\(R_{B,\mathrm{eff}}=\frac{64\mu L_{B}}{\pi (D_{C})^{4}}\).

The compressible airways resistance incorporates both geometric constriction effects of the peripheral airways and viscous losses within the two bronchi. Thus, it is modeled as
\begin{equation}
FR_{C} = \bar{k}_{C} \left( \frac{V_{C,\max}}{V_{C}} \right)^{2} + \frac{64\mu L_{B}}{\pi (D_{C})^{4}},
\label{eq:compressible_airway_resistance}
\end{equation}
where \(\bar{k}_{C} = 20.59~\mathrm{kPa\,s\,m^{-3}}\) denotes the minimum resistance corresponding to the maximal airway expansion of the peripheral airways. Next, is tracheal flow resistance, which is also evaluated using Poiseuille’s law by modeling the trachea as a rigid cylindrical conduit supporting the laminar airflow. In the present study, the trachea is represented as a straight tube with an average length of \(L = 11.8~\mathrm{cm}\) and a mean internal diameter of \(D = 17~\mathrm{mm}\). The dynamic viscosity of air at body temperature is taken as \(\mu = 1.89\times10^{-5}~\mathrm{Pa\,s}\). Substituting these physiological parameters into Poiseuille’s formulation gives the tracheal flow resistance
\begin{equation}
FR_{T} = \frac{8\mu L}{\pi r^{4}} = 1.15~\mathrm{kPa\,s\,m^{-3}}
\label{eq:tracheal_resistance}
\end{equation}
This resistance accounts for viscous dissipation along the tracheal segment and remains constant under the assumption of laminar flow and negligible tracheal deformation. At the downstream of the trachea, an additional and highly dynamic resistive element arises during sustained phonation due to the periodic opening and closing of the glottis. Such a behavior imposes a time-varying aerodynamic impedance on the airflow, which is represented by the glottal flow resistance \(FR_{\mathrm{VF}}\). The instantaneous geometry of the glottis is quantified by the glottal area waveform \(A_{g}(t)\), extracted from high-speed videoendoscopy data using the segmentation framework described earlier. Since the vocal folds form a rapidly varying constriction, the associated resistance is defined as
\begin{equation}
FR_{\mathrm{VF}}(t) = \frac{P_{\mathrm{sub}}(t) - P_{\mathrm{sup}}(t)}{Q_{g}(t)},
\label{eq:glottal_resistance}
\end{equation}
where \(P_{\mathrm{sub}}\) and \(P_{\mathrm{sup}}\) denote the spatially averaged subglottal and supraglottal pressures and \(Q_{g}(t)\) is the instantaneous glottal flow rate. Unlike the resistive elements associated with respiration, which vary slowly with the lung volume or transmural pressure, the glottal resistance oscillates at the fundamental frequency of vocal folds vibration and can change substantially within a single vibratory cycle. When the glottis is widely open, the resistance approaches a minimum value, whereas during near closure the resistance increases sharply and becomes infinity when the vocal folds are fully closed, reflecting the temporary interruption of airflow. The detailed computational procedure for evaluating \(FR_{\mathrm{VF}}(t)\) is described in the supplementary materials under the section phonation dynamics within the lumped-element architecture.
\par
The upper airways flow resistance represents the cumulative pressure losses contributed by the anatomical structures extending from the supraglottal region to the oral cavity. These losses arise from a combination of geometric curvature, gradual and abrupt area changes, flow separation, and local turbulence within the pharyngeal and oral passages. In previous work, Marconi and De~Lazzari~\cite{marconi2020silico} modeled the combined resistance of the two bronchi, trachea, and upper airways using a nonlinear flow-dependent formulation of the form
\begin{equation}
FR_{U,\mathrm{combined}} = A_{U} + K_{U}\,|F|,
\label{eq:upper_airway_combined}
\end{equation}
where \(A_{U}\) represents the minimal resistance at low flow rates and \(K_{U}\) characterizes the increase in resistance with airflow magnitude \(F\). In the present model, resistances offered by the trachea and the two bronchi are treated separately. Thus, the upper airways resistance is reformulated to represent only the supraglottal-to-oral segment. The resulting modified upper airways resistance is expressed as
\begin{equation}
FR_{U} = A_{U,\mathrm{mod}} + K_{U}\,|F|,
\label{eq:upper_airway_resistance}
\end{equation}
where \(A_{U,\mathrm{mod}} = 31.04~\mathrm{kPa\,s\,m^{-3}}\) denotes the minimal upper airways resistance after removal of the tracheal component and \(K_{U} = 45.11~\mathrm{MPa\,s^{2}\,m^{-6}}\) characterizes the nonlinear increase in upper airways resistance with the airflow magnitude. The final resistive component introduced during phonation occurs at the exit of the oral cavity and is referred to as the expiratory flow resistance \(FR_{E}\). This resistance quantifies the additional pressure required to discharge airflow from the lips into the surrounding environment. Unlike the upstream airway segments, where pressure losses are primarily associated with viscous friction along compliant or rigid walls, the expiratory resistance arises from a rapid geometric transition at the mouth opening. As air exits the oral cavity, it undergoes a sudden expansion from the finite lip aperture into the effectively unbounded external domain. This abrupt expansion prevents the airflow from remaining attached to the boundaries, causing flow separation at the lip edges and the formation of a free jet. The kinetic energy carried by this jet is partially dissipated through turbulent mixing with the surrounding quiescent air, leading to irreversible mechanical energy losses. These losses manifest as a pressure drop across the lip opening. A direct physical manifestation of this process is the perceptible air puff felt when placing a hand in front of the mouth during phonation or forced exhalation. This sensation reflects the momentum flux of the emerging jet, which indicates a positive gauge pressure at the lip exit. These processes are described by the classical head-loss relation for a sudden expansion, in which a flow passes from a conduit of cross-sectional area \(A_{1}\) and mean velocity \(V_{1}\) into a downstream region of area \(A_{2}\) and mean velocity \(V_{2}\). The associated pressure drop can be written as
\begin{equation}
\Delta P = \frac{1}{2}\,\rho_{a}\,\left(V_{1}^{2}-V_{2}^{2}\right)
\left[1-\left(\frac{A_{1}}{A_{2}}\right)^{2}\right],
\label{eq:sudden_expansion_pressure_drop}
\end{equation}
where \(\rho_{a}\) is the air density. In the phonatory setting, the upstream section corresponds to the lip opening and the downstream region is the external environment, which during sustained phonation can be treated as effectively unbounded, corresponding to the limiting case \(A_{2}\to\infty\). In this limit, \(A_{1}/A_{2}\to 0\) and the downstream mean velocity approaches \(V_{2}\to 0\), so equation~\eqref{eq:sudden_expansion_pressure_drop} reduces to
\begin{equation}
\Delta P = \frac{1}{2}\,\rho_{a}\,V_{1}^{2}
\label{eq:lip_exit_pressure_drop}
\end{equation}
showing that the required pressure surplus at the lips is set by the kinetic energy per unit volume of the emerging jet. Using continuity at the lip opening, the mean exit velocity is
\begin{equation}
V_{1}(t)=\frac{Q_{g}(t)}{A_{l}},
\label{eq:lip_exit_velocity}
\end{equation}
where \(Q_{g}(t)\) is the glottal flow rate and \(A_{l}\) is the lip opening area. Substituting equation~\eqref{eq:lip_exit_velocity} into equation~\eqref{eq:lip_exit_pressure_drop} gives
\begin{equation}
\Delta P(t) = \frac{1}{2}\,\rho_{a}\left(\frac{Q_{g}(t)}{A_{l}}\right)^{2}
\label{eq:lip_exit_pressure_drop_flow}
\end{equation}

Based on this formulation, the expiratory flow resistance can be defined as
\begin{equation}
FR_{E}(t) = \frac{\Delta P(t)}{Q_{g}(t)} = \frac{1}{2}\,\rho_{a}\,\frac{|Q_{g}(t)|}{A_{l}^{2}},
\label{eq:expiratory_flow_resistance}
\end{equation}
where the absolute value ensures that \(FR_{E}(t)\) remains nonnegative and that the associated pressure loss always opposes the direction of flow.

\subsection*{Phonation Dynamics within the Lumped-Element Architecture}

\begin{figure}[htbp]
    \centering
    \includegraphics[width=0.3\textwidth]{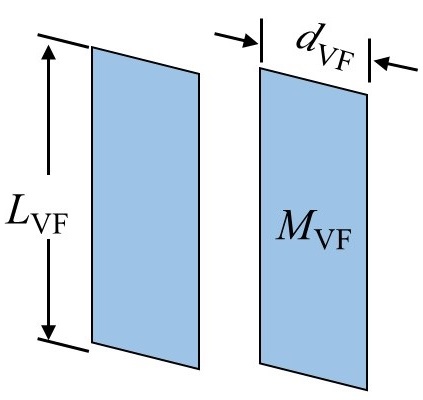}
    \caption{Schematic representation of the single-mass vocal fold model in the present modeling framework. Each vocal fold is represented as a rigid rectangular body with effective mass \(M_{\mathrm{VF}}\), superior-inferior length \(L_{\mathrm{VF}}\), and medial-lateral thickness \(d_{\mathrm{VF}}\). The two folds are positioned symmetrically about the glottal midline.}
    \label{fig:vocal_fold_single_mass}
\end{figure}

The unified lumped-element framework incorporates a mechanical representation of the vocal folds that is driven by subglottal airflow. Since the glottis forms the narrowest constriction in the phonatory airway, it governs the instantaneous impedance to the airflow and plays a dominant role in shaping the aerodynamic loading. In this study, phonatory dynamics are represented using a single-mass vocal fold model, illustrated schematically in Fig.~\ref{fig:vocal_fold_single_mass}. In this formulation, each vocal fold is idealized as a three-dimensional rectangular body whose distributed tissue mechanics are reduced to a single mass, which is governed by a linear spring and a nonlinear damping element. Despite its reduced order, such a single mass model is capable of capturing the essential features of vocal folds vibration, including the self-oscillation, asymmetric loading during opening and closing phases, and the sharp increase in mechanical resistance near collision. The geometric, inertial, and constitutive parameters employed here are adopted directly from the study of Bin Ali \textit{et al.}\cite{ali2025highspeed}. Accordingly, the superior–inferior length of each vocal fold is set to \(L_{\mathrm{VF}} = 1.60~\mathrm{cm}\), the medial–lateral thickness is taken as \(d_{\mathrm{VF}} = 0.32~\mathrm{cm}\), and the effective mass of each fold is specified as \(M_{\mathrm{VF}} = 0.120~\mathrm{g}\). The elastic response of the tissue is modeled using a linear stiffness constant \(K_{\mathrm{VF}} = 396.6550~\mathrm{N\,m^{-1}}\) and the dissipative characteristics of the vocal folds are modeled by the following nonlinear damping formulation
\begin{equation}
C_{VF}(x) = C_{\min} + C_{f}\,\tanh\!\left(x_{c,\mathrm{VF}} - x_{\mathrm{VF}}\right)
\label{eq:vocal_fold_damping}
\end{equation}

Here, \(C_{\min}=0\) denotes the minimum viscous resistance when vocal folds have the maximum distance from each other, and \(C_{f}=7.5362\times10^{3}\) controls the magnitude of damping amplification as the folds approach collision. The variable \(x_{\mathrm{VF}}\) represents the instantaneous medial displacement of each vocal fold measured from the midline, while \(x_{c,\mathrm{VF}}\) denotes the critical displacement at which the initial contact between the two folds occurs. This definition follows the modeling convention that \(x_{c,\mathrm{VF}}\) marks the transition between the open and closed phases of the glottal cycle. When \(x_{\mathrm{VF}} > x_{c,\mathrm{VF}}\), the vocal folds remain separated and the glottis is open. When \(x_{\mathrm{VF}} = x_{c,\mathrm{VF}}\), the folds are just in contact, and when \(x_{\mathrm{VF}} < x_{c,\mathrm{VF}}\), a compressive interaction between the folds takes place. The change in damping is modeled using the positive portion of the hyperbolic tangent function, which provides a smooth yet rapidly increasing damping profile that reflects the sharp rise in tissue resistance near the contact without introducing numerical discontinuities.
\par
During sustained phonation, the periodic opening and closing of the vocal folds produce a strongly time-varying pressure difference between the subglottal and supraglottal regions. This transglottal pressure difference arises from the combined effects of the airflow acceleration through the glottal constriction and viscous losses associated with the narrowing of the glottal channel as the folds approach contact. Within the present lumped-element formulation, the instantaneous pressure drop across the glottis is expressed as
\begin{equation}
P_{\mathrm{sub}}(t) - P_{\mathrm{sup}}(t) = \frac{k_t \rho_a}{2}\,\frac{|Q_g(t)|\,Q_g(t)}{A_g(t)^2} + \gamma\,\frac{Q_g(t)}{A_g(t)^3},
\label{eq:transglottal_pressure}
\end{equation}
where \(P_{\mathrm{sub}}\) and \(P_{\mathrm{sup}}\) denote the spatially averaged subglottal and supraglottal pressures, respectively, \(Q_g(t)\) is the glottal volume flow rate, and \(A_g(t)\) is the instantaneous glottal area extracted from high-speed videoendoscopy data. The coefficient \(k_t = 1\) represents the transglottal pressure coefficient, and \(\rho_a\) is the air density. The first term in equation~\eqref{eq:transglottal_pressure} represents the inertial component of the pressure drop and dominates when the vocal folds are widely separated and the airflow accelerates rapidly through the glottis. This quadratic dependence on the flow reflects the conversion of pressure energy into the kinetic energy of the jet formed within the glottal constriction. The second term captures the viscous dissipation within the glottal channel and becomes dominant as the glottal area decreases during the closing phase of vibration. This term rises sharply near the closure and accounts for the rapid increase in the flow resistance that accompanies narrowing of the glottis. To complete the aerodynamic coupling between the vocal folds and the surrounding airway system, the glottal flow rate \(Q_g(t)\) is computed using the nonlinear formulation proposed by Lucero and Schoentgen~\cite{lucero2015smoothness} as follows
\begin{equation}
Q_g(t) = \frac{2\,A_g(t)^3\,\delta_p(t)}{\dfrac{\rho_a c\,A_g(t)^3}{A^{*}}+\gamma+\sqrt{\left(\dfrac{\rho_a c\,A_g(t)^3}{A^{*}} + \gamma\right)^2+2\,k_t\,\rho_a\,A_g(t)^4\,|\delta_p(t)|}}
\label{eq:glottal_flow}
\end{equation}
Here, \(\delta_p(t)\) denotes the effective driving pressure acting across the glottis and is defined as
\(\delta_p(t) = (1 + r_s)\,P_{\mathrm{sub}}^{+}(t) - (1 + r_e)\,P_{\mathrm{sup}}^{-}(t)\), where \(P_{\mathrm{sub}}^{+}\) and \(P_{\mathrm{sup}}^{-}\) represent the forward-traveling pressure waves in the subglottal and supraglottal tracts, respectively, and \(r_s\) and \(r_e\) are the corresponding acoustic reflection coefficients. The parameter \(c \approx 35000~\mathrm{cm/s}\) denotes the speed of sound in air. The effective acoustic coupling area \(A^{*}\) is defined through \(\frac{1}{A^{*}} = \frac{1}{A_s} + \frac{1}{A_e}\), where \(A_s\) and \(A_e\) denote the cross-sectional areas of the subglottal and supraglottal airways, respectively.
\par
In the present tudy, the effective driving pressure \(\delta_p\) is selected based on both the experimental validation and physiological considerations. For benchmarking against the excised-larynx measurements reported by Lehoux \textit{et al.}~\cite{lehoux2021subglottal}, \(\delta_p\) was set to \(2.72~\mathrm{kPa}\), which corresponds to the upper range of steady subglottal pressure levels reported for sustained phonation in their anechoic configuration. For general phonation simulations, however, the mean subglottal pressure typically lies within the range of approximately \(5\)–\(10~\mathrm{cmH_2O}\). Accordingly, a target mean subglottal pressure of \(7.5~\mathrm{cmH_2O}\) is adopted following Hoffman \textit{et al}.~\cite{hoffman2009reliable}. Under these conditions, \(\delta_p(t)\) is determined iteratively. The procedure begins with an initial estimate corresponding to \(7.5~\mathrm{cmH_2O}\), after which the resulting subglottal pressure waveform \(P_{\mathrm{sub}}(t)\) is computed and its cycle-averaged value is evaluated. If the computed mean subglottal pressure is lower than the target value, \(\delta_p\) is increased; if it exceeds the target, \(\delta_p\) is reduced. This iterative adjustment is repeated until the relative error between the simulated and target mean subglottal pressures falls below \(0.01\%\). Based on this iterative procedure, the effective driving pressure converges to \(\delta_p = 8.82~\mathrm{cmH_2O}\).

\subsection*{Numerical Method}

From the governing equations of the lung and the compressible airways given by equations~\eqref{eq:lung_volume_eom} and~\eqref{eq:compressible_airway_volume}, they can be combined into a single coupled dynamic constraint that governs the simultaneous evolution of lung volume and compressible airways volume during phonation. Eliminating the pleural pressure and rearranging the inertial elastic dissipative and supportive force contributions gives the following composite governing equation
\begin{equation}
\begin{gathered}
\frac{M_L}{\alpha_L^{2}}\,\ddot{V}_L(t)
- \frac{M_C}{2\alpha_C^{2}}\,\ddot{V}_C(t)
+ F R_s\,\dot{V}_L(t)
+ \frac{C_{1L}}{\alpha_L^{2}}\,\dot{V}_L(t)
- \frac{C_{1C}}{2\alpha_C^{2}}\,\dot{V}_C(t)
+ \frac{C_{2L}}{\alpha_L^{3}} 
\frac{\dot{V}_L(t)^{3}}{\lvert \dot{V}_L(t)\rvert}
\\
+ \frac{K_L(t)}{\alpha_L}
\left(
\frac{V_L(t)}{\alpha_L} + h_{\mathrm{eq}} - h_{max}
\right)
- \frac{K_C(t)}{\alpha_C}
\left(
\frac{V_C(t)}{2\alpha_C}+x_{\mathrm{eq}}-x_{max}
\right) - \frac{M_L g}{\alpha_L}
+ \frac{R_L}{\alpha_L}
- \frac{R_C}{\alpha_C}
= 0
\end{gathered}
\label{eq:combined_lung_airway_constraint}
\end{equation}

Equation~\eqref{eq:combined_lung_airway_constraint} represents a strongly nonlinear second order differential algebraic system that couples lung mechanics, compressible airways mechanics, tissue viscoelasticity, gravity, and flow dependent resistive losses. The unknowns \(V_L(t)\) and \(V_C(t)\) are not independent but are additionally constrained by volumetric continuity across the glottis. During phonation, the total volumetric balance is governed by \(\dot{V}_L(t) + \dot{V}_C(t) + \dot{V}_{\mathrm{VF}}(t) = - Q_g(t)\) or equivalently \(\dot{V}_L(t) + \dot{V}_C(t) = - \widetilde{Q}_g(t)\), where \(\widetilde{Q}_g(t) = Q_g(t) + \dot{V}_{\mathrm{VF}}(t)\) denotes the adjusted glottal flow rate that accounts for the instantaneous volumetric change of the vocal folds. This kinematic constraint introduces an additional coupling between the lung and airways accelerations through the relation \(\ddot{V}_C(t) = - \ddot{V}_L(t) - \dot{\widetilde{Q}}_g(t)\). Together equation~\eqref{eq:combined_lung_airway_constraint} and the continuity constraint form a closed dynamical system in which the lung acceleration \(\ddot{V}_L(t)\) becomes the primary dynamic variable. Due to the presence of nonlinear stiffness terms, exponential volume dependent resistances, cubic velocity dependent damping, and externally prescribed glottal flow, the system does not have a closed form analytical solution. Thus, a direct numerical time-integration strategy is required.
\par
In the present work, the coupled lung–airways dynamics are solved using a fourth order Runge–Kutta (RK4) time integration scheme. The RK4 method is particularly well suited for this problem because it provides high order accuracy for smooth but strongly nonlinear systems while maintaining numerical stability under the high temporal resolution. To numerically integrate the coupled lung-compressible airways dynamics defined by equation~\eqref{eq:combined_lung_airway_constraint} together with the volumetric continuity constraint, the governing equations are first transformed into an explicit first order state-space representation suitable for time marching. The primary state variables are chosen as the lung volume and its first derivative together with the compressible airways volume and its first derivative. Accordingly, the state vector is defined as 
\begin{equation}
\mathbf{y}(t) =\begin{bmatrix}V_L(t) \\\dot{V}_L(t) \\V_C(t) \\\dot{V}_C(t)\end{bmatrix}
\end{equation}

The time derivatives of the state variables follow directly from kinematics and the governing equations. The first and third components satisfy the identities
\(\frac{dV_L}{dt} = \dot{V}_L \qquad \text{and} \qquad \frac{dV_C}{dt} = \dot{V}_C\), while the second and fourth components require evaluation of the dynamic force balance. Using the continuity constraint
\(\dot{V}_L(t) + \dot{V}_C(t) = - \widetilde{Q}_g(t)\), we get \(\dot{V}_C(t) = - \widetilde{Q}_g(t) -\dot{V}_L(t)\) and \(\ddot{V}_C(t) = - \ddot{V}_L(t) - \dot{\widetilde{Q}}_g(t)\). Substitution of this expression into equation~\eqref{eq:combined_lung_airway_constraint} produces a single scalar equation that is solved explicitly for the lung acceleration \(\ddot{V}_L(t)\). All remaining terms are treated as known functions of the current state \(\mathbf{y}(t)\) and time \(t\), including nonlinear elastic forces, velocity dependent damping, gravity, supportive forces, and the glottal flow rate. Once \(\ddot{V}_L(t)\) is obtained, the airway acceleration \(\ddot{V}_C(t)\) follows directly from the constraint equation above. The resulting system can therefore be written compactly as\(\frac{d\mathbf{y}}{dt} = \mathbf{f}\bigl(t,\mathbf{y}(t)\bigr)\), where the vector valued function \(\mathbf{f}\) contains both the kinematic relations and the dynamically evaluated accelerations. This formulation is well suited for explicit Runge-Kutta integration. Time is discretized uniformly with a constant step size \(\Delta t = \frac{1}{4000}\ \mathrm{s}\), which matches the temporal resolution of the high-speed videoendoscopy data. At each discrete time level \(t_n = n \Delta t\), the fourth order Runge-Kutta method advances the solution according to
\(\mathbf{y}_{n+1}=\mathbf{y}_n+\frac{\Delta t}{6}\left(\mathbf{k}_1+2\mathbf{k}_2+2\mathbf{k}_3 + \mathbf{k}_4 \right)\)
where the intermediate slope evaluations are given by
\(\mathbf{k}_1 = \mathbf{f}(t_n,\mathbf{y}_n)\)
\(\mathbf{k}_2 = \mathbf{f}\!\left(t_n+\tfrac{\Delta t}{2},\mathbf{y}_n+\tfrac{\Delta t}{2}\mathbf{k}_1\right)\)
\(\mathbf{k}_3 = \mathbf{f}\!\left(t_n+\tfrac{\Delta t}{2},\mathbf{y}_n+\tfrac{\Delta t}{2}\mathbf{k}_2\right)\)
\(\mathbf{k}_4 = \mathbf{f}\!\left(t_n+\Delta t,\mathbf{y}_n+\Delta t\,\mathbf{k}_3\right)\).

At each intermediate stage the adjusted glottal flow rate \(\widetilde{Q}_g(t)\) and its time derivative are evaluated. After each full time step, the continuity constraint is explicitly enforced at the velocity level to prevent the accumulation of numerical drift between the lung and compressible airways volumes. All nonlinear terms, resistances, stiffnesses, damping contributions, and supportive forces are recomputed at each Runge-Kutta substep using the intermediate states \(\mathbf{y}_n+\tfrac{\Delta t}{2}\mathbf{k}_i\). 

\subsection*{Validation}
\begin{figure}[ht]
\centering
\includegraphics[width=0.5\linewidth]{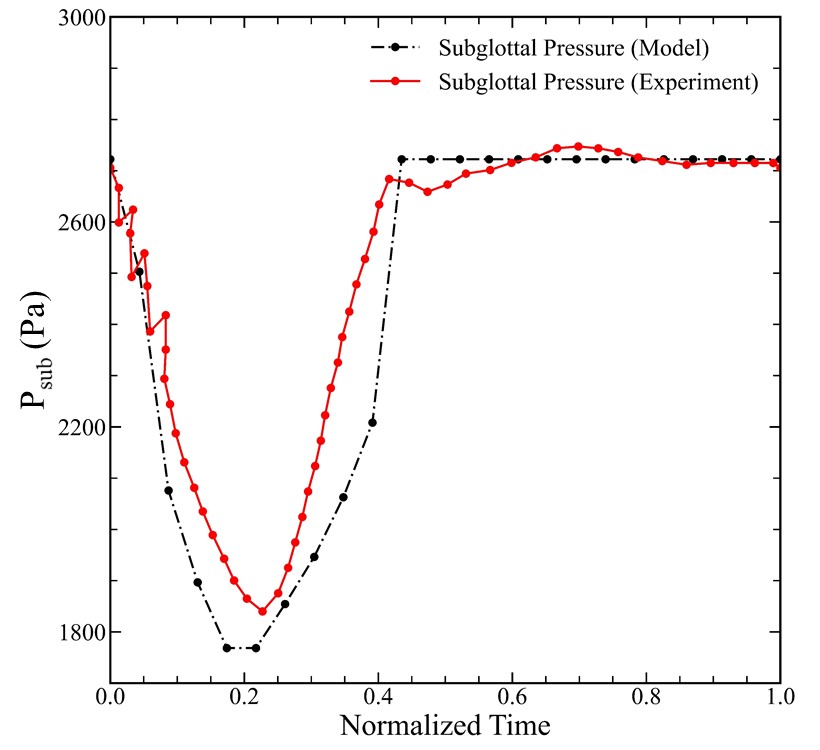}
\caption{Comparison of the subglottal pressure waveform predicted by the unified lumped-element model with experimentally measured subglottal pressure from excised-larynx experiments reported by Lehoux \textit{et al.}~\cite{lehoux2021subglottal}. Both waveforms are shown over a normalized phonatory cycle.}
\label{fig:Psub_validation}
\end{figure}

The predictive performance of the unified lumped–element model was assessed by comparing its subglottal pressure predictions with excised-larynx measurements obtained by Lehoux \textit{et al}.~\cite{lehoux2021subglottal}. Their experiment used a red deer larynx mounted on a straight Plexiglas tube configured as an anechoic, purely resistive subglottal tract. The red deer larynx is widely regarded as an excellent biomechanical surrogate for the human larynx because its geometry, tissue composition, and vibratory behavior closely mimic those of adult male human vocal folds, making it a standard model for phonatory mechanics.
\par
During steady phonation in the reference experiment, the subglottal pressure was approximately \(2500\,\mathrm{Pa}\), with oscillation frequencies between \(100\)–\(120\,\mathrm{Hz}\). To replicate these conditions, the present model was driven with an effective pressure input \(\delta_{P}=2700\,\mathrm{Pa}\), producing a simulated subglottal pressure waveform immediately below the glottis. The vocal-fold kinematics were prescribed from the HSV-derived glottal area waveform, yielding an oscillation frequency of \(127.2\,\mathrm{Hz}\). For comparison, both model-predicted and experimental waveforms were nondimensionalized over a single phonatory cycle.
\par
The agreement was evaluated using the point-wise relative error  
\(e(t)=|P_{\mathrm{num}}(t)-P_{\mathrm{exp}}(t)|/P_{\mathrm{exp}}(t)\times 100\). The maximum discrepancy (\(e_{\max}=13.33\%\)) occurs near peak glottal opening, where the experimental flow exhibits a three-dimensional jet separation and vortical structures that cannot be captured by a one-dimensional lumped-element representation. Despite this localized deviation, the model reproduces the canonical pressure waveform pattern, including the closed-phase plateau, the monotonic pressure drop during the opening, and the subsequent rise during the closing. The timing and amplitude of the oscillations align closely with measurements, and the mean cycle-averaged error is only \(e_{\mathrm{mean}}=2.96\%\), demonstrating that the model reliably captures the dominant subglottal pressure dynamics.

\section*{Results}

In this work, we develop a unified lumped-element model of human phonation that integrates the respiratory, phonatory, and articulatory subsystems within a single mechanical framework, as illustrated in Fig.~\ref{fig:unified_phonation_model}. The model represents the airflow pathway from the lungs to the atmosphere using spring-mass-damper elements along with distributed flow resistances. The flow resistances quantify the pressure loss per unit volumetric flow rate and account for energy dissipation across different anatomical segments. The respiratory subsystem comprises both the lungs and the compressible airways, with the two lungs combined into a single equivalent cylindrical chamber. Volumetric changes of this chamber are modelled by piston motion governed by a spring-mass-damper system. The airflow generated by the lungs first encounters the resistance due to lung tissue deformation \((R_{\mathrm{LT}})\), followed by viscous losses within the bronchioles, represented as the small airways flow resistance \((FR_{\mathrm{s}})\). Upstream of this region, the airflow passes through the compressible airways, which physiologically correspond to the peripheral airways and the two primary bronchi. Pressure losses within these segments are represented by the compressible airways flow resistance \((FR_{\mathrm{c}})\). In the lumped-element representation, this subsystem is modeled using two horizontal cylindrical chambers arranged in parallel. Downstream of the respiratory subsystem, the airflow passes through the trachea and experiences pressure losses quantified by the tracheal flow resistance \((FR_{\mathrm{T}})\). The flow then enters the phonatory subsystem, which dynamically modulate the incoming respiratory airflow. This modulation produces a transglottal pressure drop, represented by the vocal fold flow resistance \((FR_{\mathrm{VF}})\). The modulated airflow subsequently enters the articulatory subsystem, represented by the upper airways, which extends from the supraglottal region to the oral cavity. Aerodynamic losses that occur within this segment prior to the sound radiation at the lips are represented by the upper airways flow resistance $(FR_{\mathrm{U}})$. Finally, the airflow is then expelled through the oral outlet when labial pressure exceeds atmospheric pressure. This effect is modeled by an effective expiratory flow resistance $(FR_{\mathrm{E}})$, which completes the flow resistance network (see Supplementary Materials, Sec.~``Lumped-Element System Overview''). Through the serial coupling of these mechanical and aerodynamic elements, the present model provides a complete representation of the energy pathway for phonation. 
\begin{figure}[htbp]
    \centering
    \begin{subfigure}[t]{0.48\linewidth}
        \captionsetup{labelformat=empty}
        \centering
        \includegraphics[width=\linewidth]{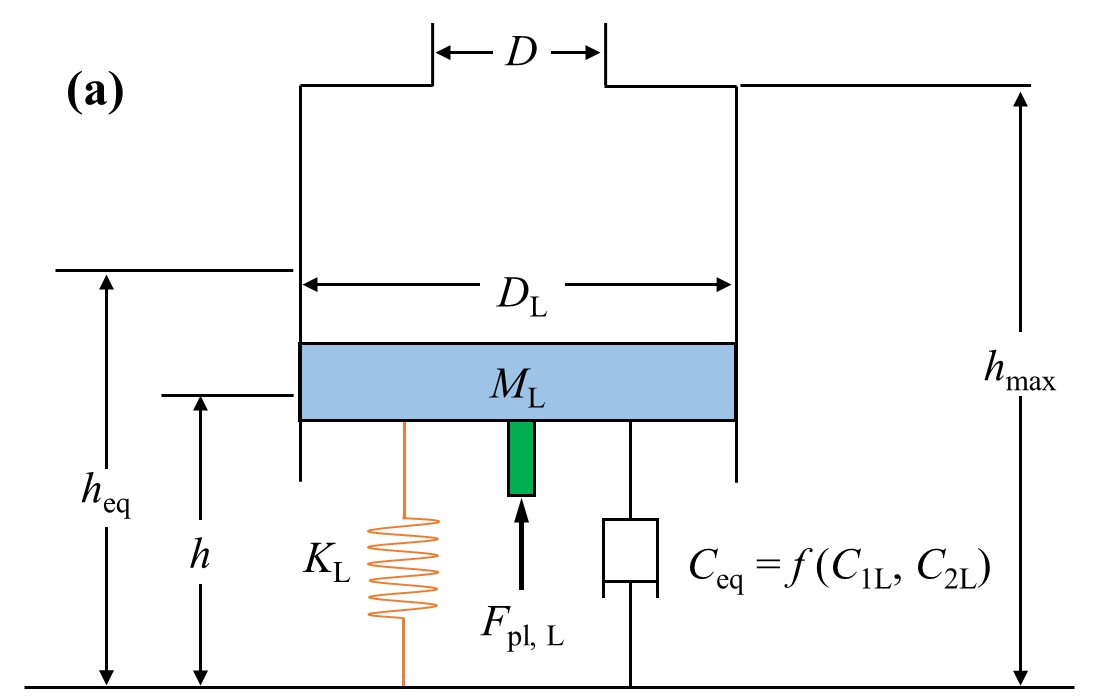}
        \phantomcaption
        \label{fig:lung_model_res}
    \end{subfigure}
    \hfill
    \begin{subfigure}[t]{0.48\linewidth}
        \captionsetup{labelformat=empty}
        \centering
        \includegraphics[width=\linewidth]{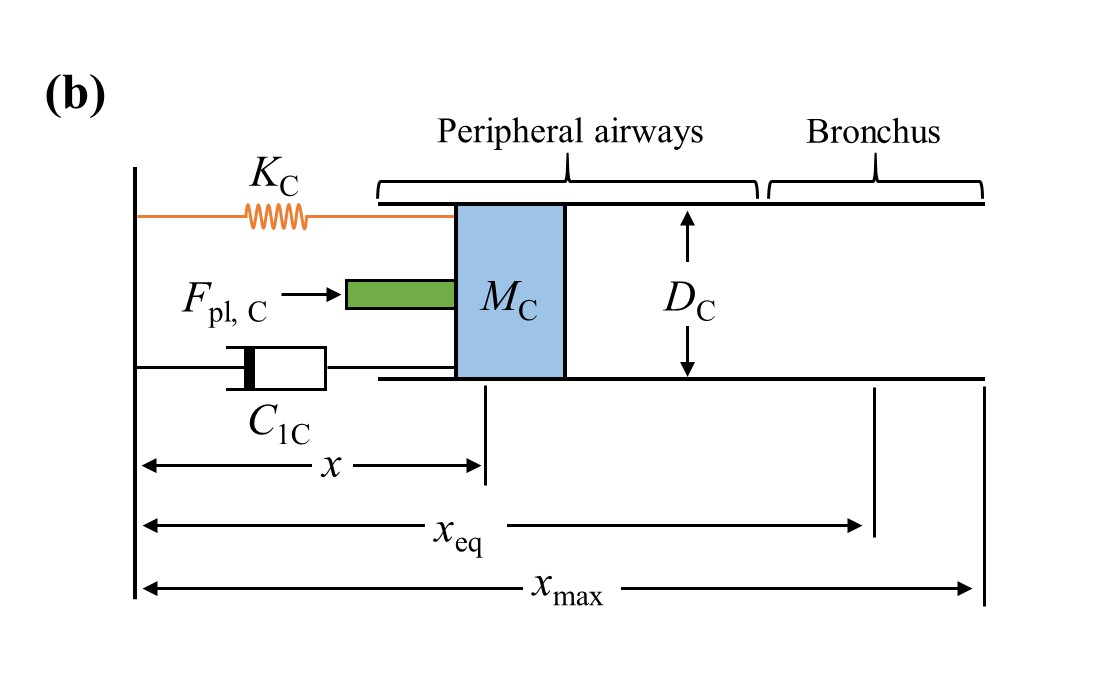}
        \phantomcaption
        \label{fig:compressible_airway_model}
    \end{subfigure}

    \caption{Lumped-element representations of the respiratory subsystem. \textbf{(a)} Unified lung model in which the left and right lungs are represented by a single equivalent piston–cylinder system with lumped mass \(M_{\mathrm{L}}\). The axial piston position \(h\) governs the instantaneous lung volume. Spring stiffness \(K_{\mathrm{L}}\) represents elastic recoil of the lung parenchyma, while damping elements \((C_{1\mathrm{L}},\,C_{2\mathrm{L}})\) capture linear and nonlinear dissipation. The piston is driven by the intrapleural force \(F_{\mathrm{pl,L}}\). Geometric parameters \(D_{\mathrm{L}}\), \(h_{\mathrm{eq}}\), and \(h_{\max}\) denote the effective lung diameter, equilibrium position, and maximum piston excursion, respectively. \textbf{(b)} Lumped-element model of the compressible airways subsystem representing the peripheral airways and primary bronchi (one assembly shown for clarity). The piston displacement \(x\) with mass \(M_{\mathrm{C}}\) governs the instantaneous airways volume. Spring stiffness \(K_{\mathrm{C}}\) characterizes wall elasticity, while damping \(C_{1\mathrm{C}}\) accounts for viscous dissipation. The piston is driven by intrapleural force \(F_{\mathrm{pl,C}}\). Geometric parameters \(D_{\mathrm{C}}\), \(x_{\mathrm{eq}}\), and \(x_{\max}\) denote the effective diameter, equilibrium position, and maximum excursion, respectively.}
    \label{fig:respiratory_subsystem_models}
\end{figure}
\subsection*{Lung Subsystem Parameter Estimation}
Figure~\ref{fig:lung_model_res} illustrates the lung subsystem modeled with a lumped mass \(M_{\mathrm{L}} = 1.2~\mathrm{kg}\)~\cite{mubbunu2018correlation}. The instantaneous air volume within the lungs (\(V_{\mathrm{L}}\)) is governed by the piston displacement as follows.
\begin{equation}
V_{\mathrm{L}} = \alpha_{\mathrm{L}} \left( h_{\max} - h \right),
\label{eq:lung_volume}
\end{equation}
where \(h\) is the piston position, measured from the reference, and \(h_{\max}\) denotes the maximum allowable piston's traveling distance corresponding to the full lung inflation. The effective cross-sectional area of the piston, \(\alpha_{\mathrm{L}}\), was identified as \(0.0356~\mathrm{m}^2\). The maximum piston displacement was found to be \(h_{\max} = 19.1~\mathrm{cm}\) for a total lung capacity of \(\mathrm{TLC} = \alpha_{\mathrm{L}} h_{\max} = 5.19~\mathrm{L}\) (see Supplementary materials, Sec.~ "Lung Geometry and Kinematics"). The elastic behavior of the lung tissue is incorporated into this model through a non-linear spring element and the corresponding elastic restoring force acting on the lung piston is written as
\begin{equation}
F_{\mathrm{el,L}} = -K_{\mathrm{L}} \left( h - h_{\mathrm{eq}} \right),
\label{eq:lung_restoring_force}
\end{equation}
where \(K_{\mathrm{L}}\) denotes the effective lung stiffness, which is inversely related to the lung compliance. \(h_{\mathrm{eq}}\) represents the equilibrium piston position, which corresponds to the functional residual capacity (FRC) of the lung, defined as the lung volume at which the inward elastic recoil of the lung parenchyma is balanced by the outward recoil of the thoracic cage. The present analysis gives a functional residual capacity of \(\mathrm{FRC} = 2.86~\mathrm{L}\) (see Supplementary materials, Sec.~"Lung Spring Model and Compliance "). Along with the elastic force, dissipative forces, representing tissue viscosity, pleural surface friction, and internal structural rearrangements during deformation, also act on the piston and are modeled as
\begin{equation}
F_{\mathrm{damp,L}} = - C_{\mathrm{1L}}\,\dot{h} - C_{\mathrm{2L}}\,\frac{\dot{h}^{3}}{|\dot{h}|},
\label{eq:lung_damping}
\end{equation}
where \(C_{\mathrm{1L}}\) and \(C_{\mathrm{2L}}\) denote the linear and nonlinear viscous damping coefficients, respectively. In the present study, particle swarm optimization was used to estimate the lung damping coefficients, resulting in \( C_{\mathrm{1L}} = 195.79 \pm 16.05\%~\mathrm{kg\,s^{-1}} \) and
\( C_{\mathrm{2L}} = 1291.01 \pm 16.05\%~\mathrm{kg\,m^{-2}\,s^{-1}} \) (see Supplementary materials, Sec.~"Lung Damping Model and Tissue Dissipation" and "Particle Swamp Optimization Algorithm").

\subsection*{Compressible Airways Subsystem Parameter Estimation}

Figure~\ref{fig:compressible_airway_model} shows the compressible airways subsystem. Unlike the single piston-cylinder representation of the lungs model, the compressible airways is modeled using two parallel identical piston-cylinder assemblies, with each piston-cylinder unit having a mass of \(M_{\mathrm{C}}\) 2.95 g (see Supplementary materials, Sec.~"Compressible Airways Geometry and Effective Mass"). The compressible airways volume \(V_{\mathrm{C}}\) is related to the piston displacement by
\begin{equation}
V_{\mathrm{C}}(t) = 2\,\alpha_{\mathrm{C}}\left(x_{\max} - x(t)\right),
\label{eq:VC_volume}
\end{equation}
where $x(t)$ denotes the instantaneous position of each compressible-airways piston and $x_{\max}$ is the maximum allowable piston excursion corresponding to the fully distended airways configuration. The piston diameter is \(D_{\mathrm{C}} = 17~\mathrm{mm}\) and the maximum compressible airways volume \(V_{\mathrm{C,max}}\) is  198.65 mL. The elastic behavior of the compressible airways is modeled by a non-linear spring and the corresponding elastic restoring force is expressed as
\begin{equation}
F_{\mathrm{el,C}} = -K_{\mathrm{C}}\left(x - x_{\mathrm{eq}}\right),
\label{eq:CA_restoring_force}
\end{equation}
where \(K_{\mathrm{C}}\) denotes the effective compressible airways stiffness and \(x_{\mathrm{eq}} = x_{\max}\,(h_{\mathrm{eq}}/h_{\max})\) is the equilibrium piston position (see Supplementary materials, Sec.~" Compressible Airways Spring Model and Compliance"). The dissipative forces, corresponding to the viscous friction within the airways wall and surrounding tissues, act on the compressible airways pistons and is modeled by a Newtonian damper as
\begin{equation}
F_{\mathrm{damp,C}} = -C_{\mathrm{1C}}\,\dot{x}
\label{eq:CA_damping}
\end{equation}
where \(C_{\mathrm{1C}} = 7.37~\mathrm{N\,s\,m^{-1}}\) is the compressible-airways damping coefficient (see Supplementary materials, Sec.~"Damping Behavior of the Compressible Airways").

\subsection*{Vocal Folds Mechanics}

\begin{figure}[htbp]
    \centering
    \includegraphics[width=.3\linewidth]{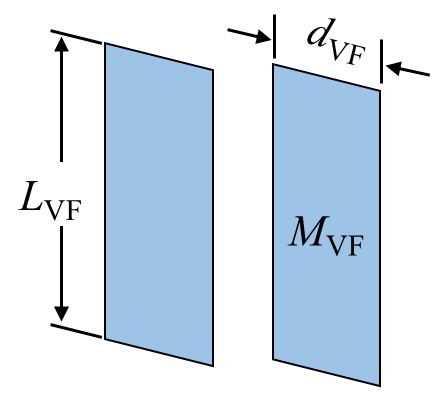}
    \caption{Schematic representation of the single-mass vocal fold model in the present modeling framework. Each vocal fold is represented as a rigid rectangular body with effective mass \(M_{\mathrm{VF}}\), superior-inferior length \(L_{\mathrm{VF}}\), and medial-lateral thickness \(d_{\mathrm{VF}}\). The two folds are positioned symmetrically about the glottal midline.}
    \label{fig:vocal_fold_model}
\end{figure}

In the current lumped-element framework, the vocal folds are represented using a single-mass model, as shown in Fig.~\ref{fig:vocal_fold_model}, following the formulation proposed by Ali \textit{et al.}~\cite{ali2025highspeed}. Accordingly, the superior-inferior length of each vocal fold is taken as \(L_{\mathrm{VF}} = 1.60~\mathrm{cm}\), the medial-lateral width is \(d_{\mathrm{VF}} = 0.32~\mathrm{cm}\), and the effective mass of each vocal fold is \(M_{\mathrm{VF}} = 0.120~\mathrm{g}\).
The instantaneous vocal fold volume is defined as \(V_{\mathrm{VF}}(t) = A_g(t)\,L_{\mathrm{VF}}\), where \(A_g(t)\) denotes the time-varying glottal area. For the present modeling work, the glottal area waveform was extracted from laryngeal high-speed videoendoscopy (HSV) recordings of a 49-year-old male subject during sustained phonation of \textbackslash i\textbackslash using a machine learning–based segmentation framework (see Supplementary materials, Sec.~"Data Collection and Processing"). The elastic response of the vocal folds is modeled using a linear stiffness constant \(K_{\mathrm{VF}} = 396.6550~\mathrm{N\,m^{-1}}\), while viscous dissipation is represented through a nonlinear damping coefficient \(C_{\mathrm{VF}}\), expressed as \(C_{\mathrm{VF}} = C_{\mathrm{f}} \tanh\!\left(x_{\mathrm{c,\,VF}} - x_{\mathrm{VF}}\right)\). Here, \(C_{\mathrm{f}} = 7.5362 \times 10^{3}\), and $x_{\mathrm{VF}}$ denotes the instantaneous medial displacement of each vocal fold measured from the midline position, while $x_{\mathrm{c,\,VF}}$ represents the critical displacement at which the two vocal folds first come into contact. This contact threshold defines the onset of the closed phase. When $x_{\mathrm{VF}} > x_{\mathrm{c,\,VF}}$, the glottis remains open, when $x_{\mathrm{VF}} = x_{\mathrm{c,\,VF}}$, the initial contact occurs, and when $x_{\mathrm{VF}} < x_{\mathrm{c,\,VF}}$, the vocal folds are in collision. Upstream of the glottis, the airflow is driven by the subglottal pressure \((P_{\mathrm{sub}})\), defined as the spatially averaged pressure immediately below the vocal folds. Downstream, the pressure is the supraglottal pressure \((P_{\mathrm{sup}})\), which denotes the pressure immediately above the glottis within the vocal tract. The difference between these pressures constitutes the transglottal driving pressure, which governs the resulting glottal airflow \(Q_{\mathrm{g}}(t)\).
\subsection*{Time-Resolved Phonatory Dynamics Evaluation}
\begin{figure}[htbp]
    \centering
    \begin{subfigure}[t]{0.48\linewidth}
        \captionsetup{labelformat=empty}
        \centering
        \includegraphics[width=\linewidth]{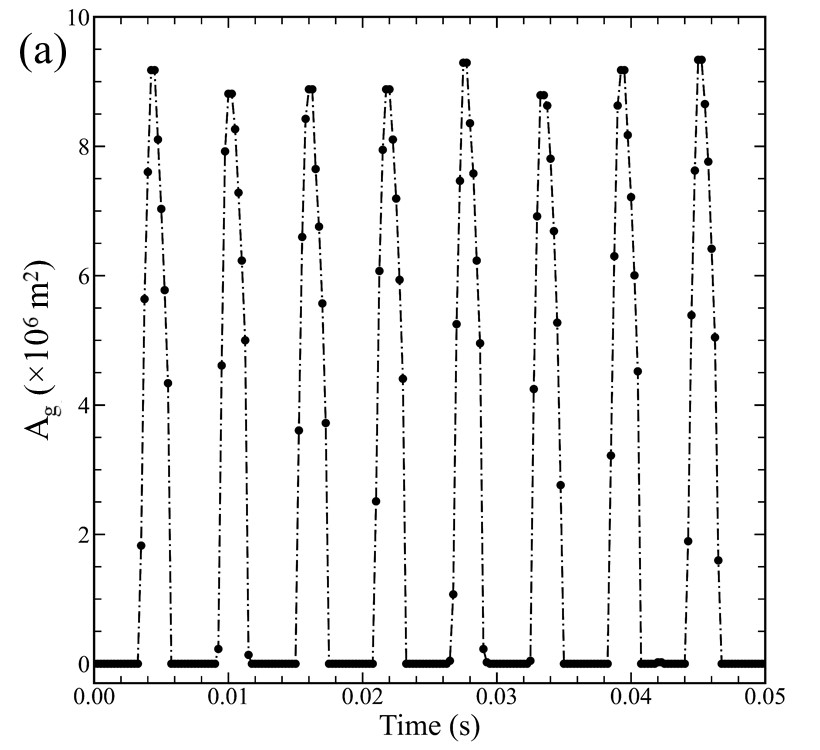}
        \phantomcaption
        \label{fig:glottal_area_time}
    \end{subfigure}
    \hfill
    \begin{subfigure}[t]{0.48\linewidth}
        \captionsetup{labelformat=empty}
        \centering
        \includegraphics[width=\linewidth]{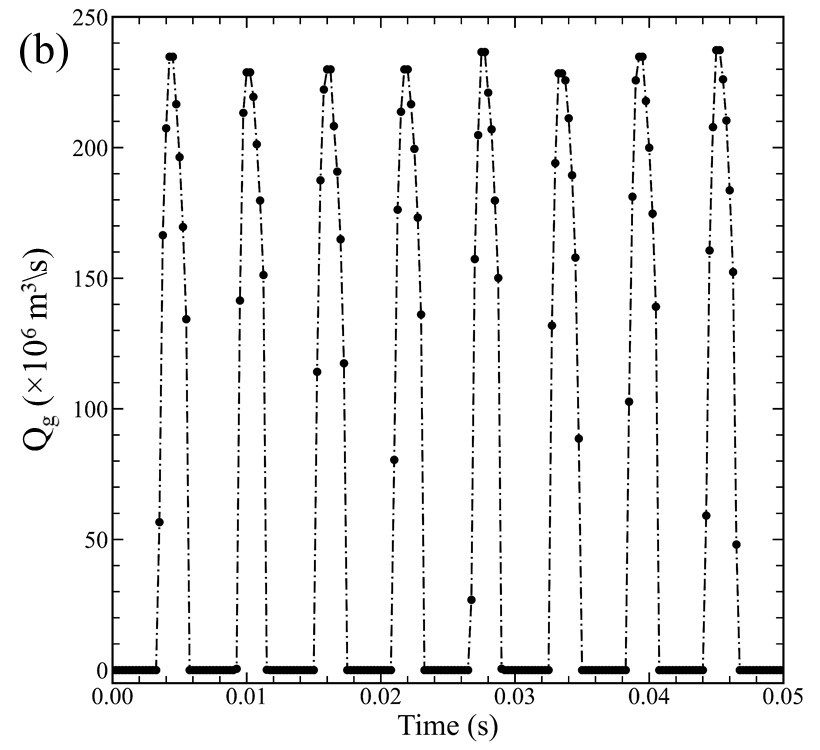}
        \phantomcaption
        \label{fig:glottal_flow_time}
    \end{subfigure}

    \vspace{0.8em}

    \begin{subfigure}[t]{0.6\linewidth}
        \captionsetup{labelformat=empty}
        \centering
        \includegraphics[width=\linewidth]{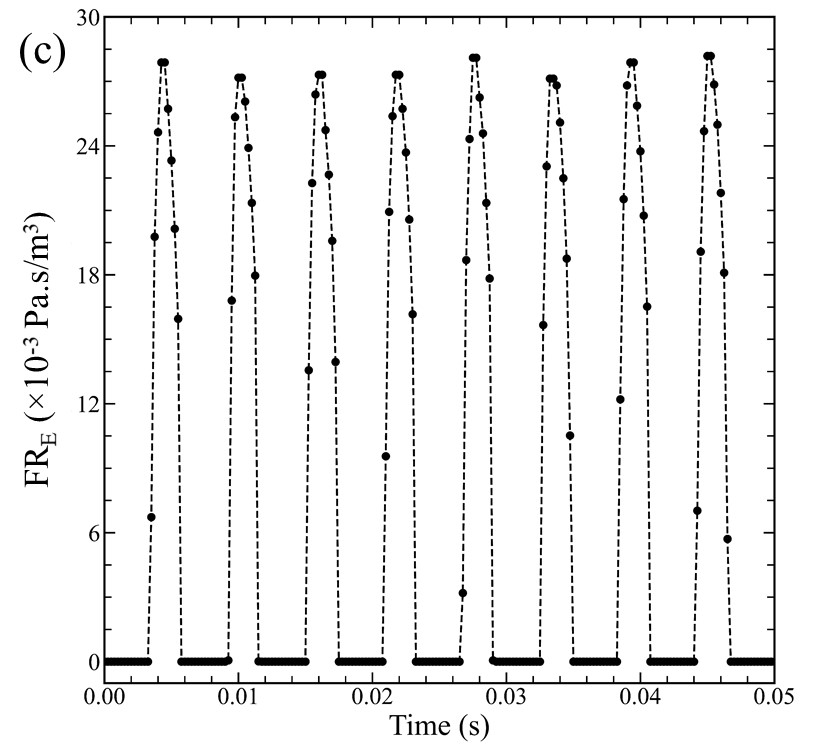}
        \phantomcaption
        \label{fig:expiratory_resistance_time}
    \end{subfigure}

    \caption{Time-resolved phonatory variables obtained from the unified lumped-element model during the sustained phonation: 
    \textbf{(a)} Temporal evolution of the glottal area $(A_g(t))$ over multiple phonatory cycles, illustrating the periodic opening and closure of the vocal folds; 
    \textbf{(b)} Time history of the glottal volumetric flow rate $Q_g(t)$ associated with the dynamic modulation of the glottal aperture, reflecting the pulsatile nature of phonatory airflow; and 
    \textbf{(c)} Instantaneous expiratory flow resistance $(F_{RE}(t))$ generated by the coupled respiratory-phonatory-articulatory subsystems, capturing the time-varying labial loading.}
    \label{fig:phonation_time_series}
\end{figure}

Having established the unified lumped-element framework, we now examine the time-resolved phonatory variables predicted by the model. to evaluate the model’s ability to reproduce key temporal characteristics of sustained phonation. Figure~\ref{fig:phonation_time_series} presents three principal flow-related features: the glottal area waveform, the glottal flow rate, and the time-varying expiratory flow resistance. The temporal evolution of the GAW, shown in Fig.~\ref{fig:glottal_area_time}, describes the instantaneous, time-varying area between the left and right vocal folds during phonation over a \(50~\mathrm{ms}\) interval. The waveform exhibits a quasi-periodic pattern of repeated glottal opening and closure. The peak glottal area reaches \(8\text{-}9~\mathrm{mm}^2\), while the minimum glottal area reaches zero, indicating a complete glottal closure during each vibratory cycle. The amplitude and periodicity remains stable across the analyzed interval with minor cycle-to-cycle variations in peak amplitude and opening slopes. This indicates small fluctuations in vocal fold biomechanics and the associated aerodynamic loading. The nearly uniform temporal spacing between successive peaks further indicates a stable fundamental frequency.

The glottal flow rate in Fig.~\ref{fig:glottal_flow_time} reflects the kinematic features of the glottal area waveform. It represents the volumetric airflow passing through the glottis and varies cyclically with vocal fold motion. Each cycle produces a distinct flow pulse that rises rapidly and then decays more gradually. Periods of zero flow occur during complete glottal closure, indicating the absence of leakage with peak flow rates ranging from \(200\text{ to }240~\mathrm{ml/s}\). The waveform exhibits mild asymmetry across different cycles, with small variations in peak magnitude and waveform shape. This signifies subtle fluctuations in the underlying biomechanical and aerodynamic interactions during the sustained phonation.
 \par
This glottal flow then passes through the supraglottal vocal tract and is ultimately expelled into the atmosphere. Prior to expulsion, the airflow has to overcome the expiratory flow resistance \((FR_{\mathrm{E}})\). Figure~\ref{fig:expiratory_resistance_time} shows the temporal evolution of the expiratory flow resistance. Within each phonatory cycle, \(FR_{\mathrm{E}}\) rises rapidly from zero to peak values of approximately \(25\text{-}29 \times 10^{3}~\mathrm{Pa}/(\mathrm{m}^3\!\cdot\!\mathrm{s})\), followed by a more gradual decay back toward zero. This produces a mildly asymmetric pulse shape, characterized by steep rising edges and broader falling segments that coincide with those observed in the glottal flow waveform. Higher glottal flow rates correspond to increased expiratory flow resistance, reflecting the greater pressure required to drive the airflow through the oral outlet. During intervals of zero glottal flow, the expiratory flow resistance collapses to zero, indicating that no pressure gradient is required between the lips and the surrounding atmosphere when the airflow is absent.
\begin{figure}[htbp]
    \centering

    \begin{subfigure}[t]{0.48\linewidth}
        \captionsetup{labelformat=empty}
        \centering
        \includegraphics[width=\linewidth]{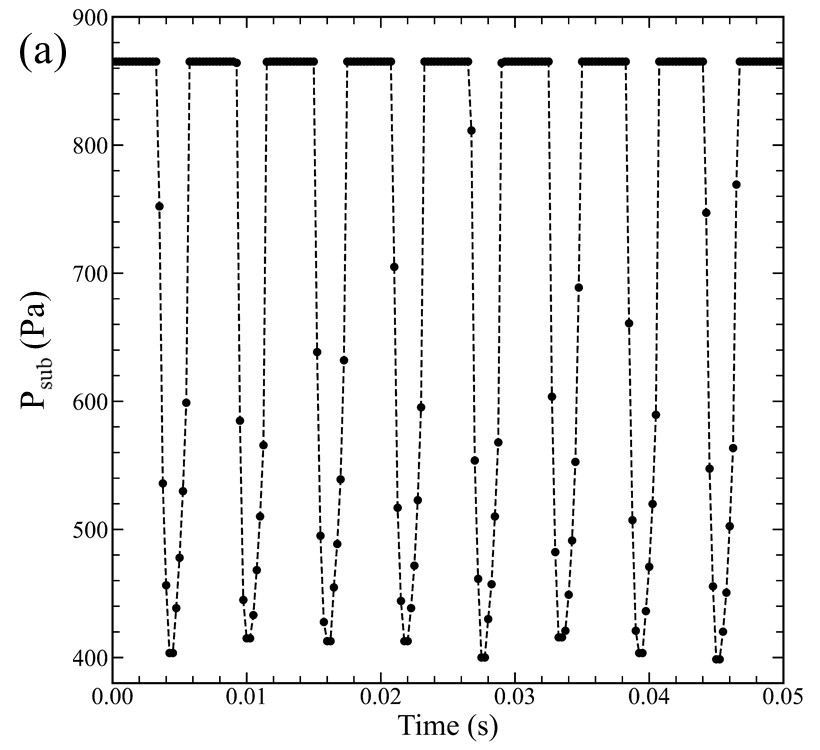}
        \phantomcaption
        \label{fig:subglottal_pressure_time}
    \end{subfigure}
    \hfill
    \begin{subfigure}[t]{0.48\linewidth}
        \captionsetup{labelformat=empty}
        \centering
        \includegraphics[width=\linewidth]{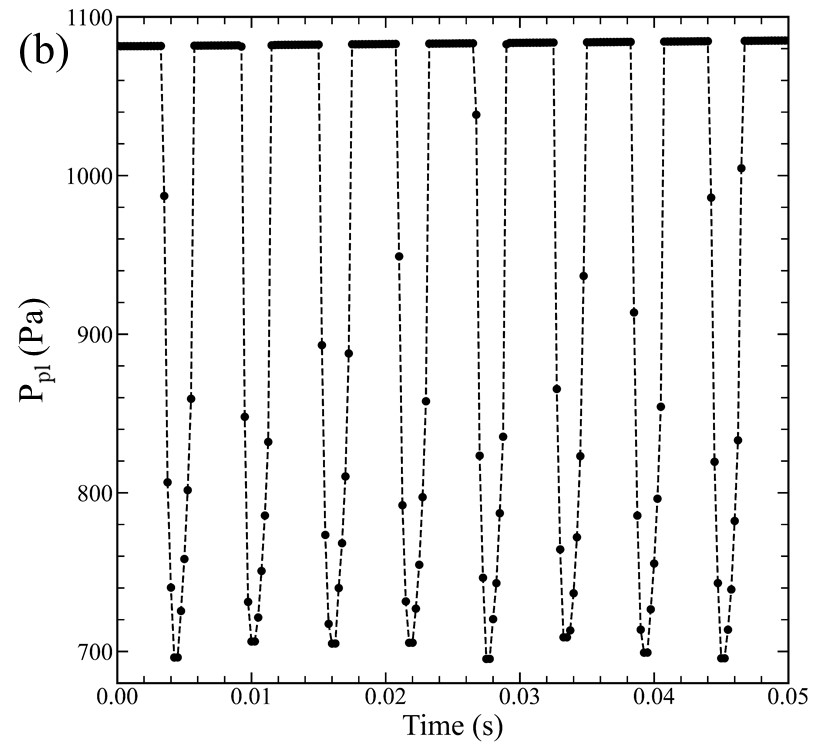}
        \phantomcaption
        \label{fig:pleural_pressure_time}
    \end{subfigure}

    \caption{Time-resolved pressure dynamics obtained from the unified lumped-element model during sustained phonation. The figure shows subglottal pressure $P_{\mathrm{sub}}(t)$ within the compressible airways compartment and pleural pressure $P_{\mathrm{pl}}(t)$ applied to the lung piston over multiple oscillation cycles.}
    \label{fig:pressure_time_series}
\end{figure}

The glottal area waveform, glottal flow rate, and expiratory flow resistance waveforms share a common qualitative structure, characterized by a rapid transition from zero to a peak value followed by a decay back to zero within each phonatory cycle. This pattern reflects the opening and closing behavior of the glottal valve and the resulting modulation of airflow and expiratory loading. However, when moving upstream of the vocal folds into the subglottal and respiratory domains, this waveform pattern becomes inverted. Instead of exhibiting isolated pulses that rise from zero, the pressure signals remain elevated for the corresponding time interval of each cycle and undergo sharp decreases during glottal openings. This inverted behavior is clearly observed in the subglottal pressure waveform shown in Fig.~\ref{fig:subglottal_pressure_time}. Subglottal pressure serves as the primary aerodynamic driving force for phonation. It is generated by expiratory effort in the lungs and provides the energy required to initiate and sustain the vocal folds vibration. The temporal variation of the subglottal pressure during sustained phonation exhibits a quasi-periodic structure that is synchronized with the glottal flow rate and GAW dynamics. Across the analyzed time window, the subglottal pressure oscillates between \(400~\mathrm{Pa}\) and \(850{-}880~\mathrm{Pa}\). A defining feature of the waveform is the presence of extended pressure plateaus at the upper pressure level, followed by rapid pressure drops and subsequent recovery phases within each phonatory cycle. These high-pressure plateaus occur at uniform intervals. Minor cycle-to-cycle variations are observed in the minimum pressure values and in the slope of the recovery phase. The temporal spacing between successive pressure minima corresponds to the temporal spacing between the maxima of the glottal area waveform. No systematic drift in the mean subglottal pressure level is observed across cycles, suggesting a steady respiratory drive during the recorded phonation.
\par
A similar but more pronounced trend is observed in the intrapleural pressure waveform, shown in Fig.~\ref{fig:pleural_pressure_time}. In comparison to the subglottal pressure, the intrapleural pressure exhibits a higher overall magnitude, reflecting its role as the primary mechanical load responsible for the lung compression and expiratory drive. Over the analyzed time interval, the intrapleural pressure remains within the range of approximately \(700~\mathrm{Pa}\) to \(1080\text{-}1090~\mathrm{Pa}\). The overall temporal structure of the intrapleural pressure waveform closely parallels that of the subglottal pressure, characterized by extended pressure plateaus followed by rapid decreases and more gradual recovery phases. The elevated plateau levels indicate stronger compressive forces acting on the lungs, which in turn sustain the subglottal pressure necessary for a stable phonation.

\section*{Discussion}
The following discussion explains the physical mechanisms underlying the observed temporal patterns of the predicted variables and examines the energy transfer across the coupled respiratory–phonatory–articulatory subsystems during sustained phonation. Each predicted quantity is interpreted in relation to its physiological role and modeling assumptions. Deviations from the experimentally observed fine-scale behavior are also discussed to clarify the implications of these assumptions and to indicate directions for future refinements.
\par
The discussion starts by examining the glottal area waveform, shown in Fig.~\ref{fig:glottal_area_time}, as it provides the most direct kinematic description of vocal fold motion. The observed quasi-periodic oscillations arise from nonlinear fluid–structure interaction between the airflow driven by the subglottal pressure and the elastic tissues of the vocal folds, which is consistent with the general vibratory characteristics of the vocal folds~\cite{patel2022glottal, krausert2011mucosal}. Increasing subglottal pressure initiates vocal fold separation, where aerodynamic driving forces dominate over tissue restoring forces. As airflow accelerates through the glottis, intraglottal pressure decreases due to inertial and convective effects~\cite{alipour1997pressure}. This pressure reduction, together with the elastic recoil of the vocal fold tissues, initiates the closing phase, leading to a complete glottal closure. Apart from minor cycle-to-cycle variations in the peak glottal area and waveform shape, the oscillations remain regular, indicating that the energy supplied by the airflow is balanced by the energy dissipated through tissue damping and collision losses. The absence of irregular oscillations and incomplete closures suggests an adequate vocal fold adduction during sustained phonation.
\par
The glottal flow rate waveform shown in Fig.~\ref{fig:glottal_flow_time} follows the oscillatory motion of the vocal folds. At the beginning of each phonatory cycle, the vocal folds are fully closed, resulting in maximal flow resistance and zero glottal airflow. During this phase, the expiratory effort leads to the accumulation of pressure beneath the vocal folds. Once the aerodynamic forces overcome the elastic and inertial resistance of the tissues, the glottis begins to open. As the vocal folds separate, the glottal area increases and the resistance to airflow decreases, allowing air to accelerate through the glottal constriction. This produces the rapid rise in glottal flow rate observed at the onset of each cycle. Continued expansion of the glottal opening increases the aerodynamic conductance of the glottis, leading to a further increase in the airflow. Consequently, the peak glottal flow rate occurs at the instant of maximum glottal area, demonstrating the strong influence of glottal geometry on the magnitude and shape of the flow waveform. The relatively consistent peak flow values across cycles indicate stable aerodynamic conditions and effective neuromuscular control of vocal folds' posture during sustained phonation. Following the peak, the glottal flow rate gradually decreases as the vocal folds enter the closing phase. During this phase, the elastic recoil within the vocal fold tissues, together with reductions in the intraglottal pressure associated with the flow inertia and pressure recovery, initiate and sustain the glottal closure. As the glottal area decreases, the airflow is progressively restricted until the vocal folds come into contact and the flow rate drops to zero. The presence of distinct zero-flow intervals indicates an effective glottal closure, indicating efficient phonatory function.
\par
An important discrepancy is observed between the glottal flow rate predicted by the present model and experimentally reported measurements. Previous studies show that the peak glottal flow does not necessarily coincide exactly with the instant of maximum glottal area \cite{krane2007unsteady}. Rather, previous work on glottal inverse filtering and scaled vocal fold models have demonstrated that the peak glottal flow occurrs slightly after the maximum glottal opening \cite{palaparthi2020analysis, dejonckere2017dynamics, rothenberg1977nonlinear}. This phase lag is attributed to vocal-tract acoustic inertance, which introduces unsteady aerodynamic effects such as a delayed flow buildup and asymmetric flow shutoff. As a result, the airflow does not respond instantaneously to the geometric opening and closing of the vocal folds, leading to a temporal offset between the extrema of glottal area and glottal flow. However, in the present study, glottal flow dynamics are modeled within a lumped-element framework that emphasizes the dominant mechanical and aerodynamic interactions. This formulation therefore neglects distributed acoustic effects and frequency-dependent inertance of the supraglottal vocal tract. Consequently, subtle phase shifts between the glottal area and glottal flow that arise from the acoustic wave propagation, pressure reflections, and inertive loading are not explicitly captured. Despite this simplification, the model successfully reproduces the essential features of glottal flow, including its quasi-periodic structure, effective closure, and stable cycle-to-cycle behaviors. The absence of explicit acoustic inertance therefore highlights future extension of the present work. Incorporating distributed vocal-tract acoustics and inertive effects into the current framework will further refine phase relationships between the glottal area, pressure, and glottal flow.
\par
The time-varying expiratory flow resistance shown in Fig.~\ref{fig:expiratory_resistance_time} reflects how the unsteady glottal airflow interacts with the oral outlet during sustained phonation. Unlike resistive elements associated with fixed anatomical constrictions, the expiratory loading arises because the airflow reaching the lips is inherently pulsatile. Even for a nearly constant mouth opening, the instantaneous pressure required to expel air into the surrounding atmosphere varies within each phonatory cycle, as it depends directly on the magnitude and acceleration of the glottal flow arriving at the oral outlet. As seen in Fig.~\ref{fig:expiratory_resistance_time}, the expiratory resistance increases during phases of elevated glottal flow, indicating that a greater aerodynamic effort is required to sustain the outflow when the glottal flow rate is higher. As the phonatory cycle progresses and the vocal folds begin to close, the reduction in the glottal flow leads to a corresponding decrease in the expiratory loading. When the glottis is fully closed and the airflow ceases, the expiratory resistance effectively vanishes, indicating that no pressure difference is required to maintain the outflow in the absence of flow. In the present framework, the expiratory resistance does not act as an independent control element but instead emerges as a downstream response to the pulsatile flow generated at the glottis. Small variations in the magnitude and timing of the resistance waveform primarily reflect variations in the upstream glottal flow, reinforcing the interpretation that expiratory loading is governed by instantaneous flow conditions. However, it should be noted that, in real life phonation, the pressure response at the lips is also influenced by distributed acoustic propagation and inertive effects within the vocal tract, which can introduce phase delays between changes in the glottal flow and supraglottal pressure. These effects are not explicitly resolved in the present lumped-element model. Nevertheless, the behavior shown in Fig.~\ref{fig:expiratory_resistance_time} captures the dominant, the cycle-resolved relationship between the glottal flow and the expiratory flow resistance.
\par
The subglottal pressure waveform shown in Fig.~\ref{fig:subglottal_pressure_time} reflects the dynamic balance between respiratory driving forces and the time-varying aerodynamic load imposed by the vocal folds vibration. The subglottal pressure acts as the primary energy source for phonation, and during the sustained phonation, it is modulated by the cyclic opening and closing of the glottis. The extended plateaus observed at elevated pressure levels correspond to phases of complete glottal closure. During these intervals, the airflow is strongly impeded, and the continued expiratory effort leads to the accumulation of pressure within the subglottal system. These plateau regions therefore represent periods of energy storage, during which the elastic recoil of the lungs sustains a nearly constant driving pressure. In physiological phonation, however, such plateaus are not perfectly flat, as subglottal pressure typically exhibits small oscillations even during the apparent closure. The absence of these fine-scale pressure fluctuations in the present results arises from the modeling assumptions adopted in this work. In the present work, the vocal folds are represented using a single-mass lumped-element model in which a complete vocal fold contact is assumed whenever the glottal area measured at the superior edges reaches zero. This assumption provides a clear and computationally efficient representation of glottal closure. However, this approach fails to capture the three-dimensional nature of vocal folds' motion. In realistic phonation, glottal opening and closing do not occur simultaneously along the entire inferior–superior extent of the vocal folds. Instead, separation and contact progress gradually, initiating at the inferior edges and propagating toward the superior edges. Our HSV recordings provide measurements of the glottal area only from a superior viewing perspective. As a result, a zero glottal area in the extracted glottal area waveform corresponds specifically to contact at the superior edges of the vocal folds. When this occurs, the present model assumes that the entire medial surfaces of the vocal folds are in contact. In reality, during the opening phase, the vocal folds start to separate from each other at the inferior edges while the superior edges remains closed. Therefore, even when the glottal area measured at the superior edge is zero, volumetric changes is already happening in the inferior portion of the glottis. These subglottal-to-infraglottal geometric changes allow the airflow acceleration and pressure redistribution below the vocal folds, leading to subtle oscillations in the subglottal pressure. Since the current modeling framework relies on two-dimensional glottal area measurements extracted from the superior edges only, it lacks direct information about the dynamic behavior of the glottal geometry at the inferior edge. As a result, the model is inherently blind to volumetric changes and pressure fluctuations occurring behind the closed superior edges. As a consequence of this limitation, the model predicts an exactly constant subglottal pressure during the closed phase, while in real life phonation, the subglottal pressure displays cycle-synchronous oscillations. Although the current framework does not capture every temporal nuance associated with the three-dimensional vocal fold motion, the time-dependent behavior obtained in the model represents a physiologically meaningful approximation of the underlying respiratory mechanics. From an applicational point of view, the resulting subglottal pressure waveform can be used as an inlet condition for aerodynamic and aeroacoustic modeling of phonation. In most of the existing literature, subglottal pressure is assumed to be constant for such modeling work \cite{luo2009analysis, chang2013role, granados2017numerical, yang2010biomechanical, perrine2023using}. This assumption can hide key aspects of flow–structure interaction, energy exchange between the airflow and the vocal folds, and parameters associated with voice quality. By capturing the time-varying behavior of the subglottal pressure, the present framework serves as a strong foundation for subject-specific phonation modeling and for future extensions of the work that will incorporate more detailed three-dimensional and acoustic effects.
\par
The intrapleural pressure waveform shown in Fig.~\ref{fig:pleural_pressure_time} characterizes the loading conditions imposed on the lungs during sustained phonation. Within the respiratory system, variations in intrapleural pressure directly regulate the lung volume by controlling the compressive forces acting on lung tissues. A change in the intrapleural pressure therefore corresponds to a change in the compressive load on the lungs. In our present lumped-element modeling, the intrapleural pressure is the primary driving force, which links the respiratory mechanics with the glottal airflow. Thus, the temporal variation of intrapleural pressure arises from the strong coupling between the respiratory subsystem and the oscillatory dynamics of the vocal folds. As the vocal folds periodically restrict and release airflow at the glottis, the downstream aerodynamic load experienced by the respiratory system varies, producing corresponding modulations in intrapleural pressure. During phases of increased glottal opening, the overall airways resistance decreases, allowing the airflow to pass more readily through the system. As a result, a lower driving pressure is required at the level of the lungs, leading to a reduction in intrapleural pressure. Conversely, as the glottis approaches closure and the airways resistance increases, a higher intrapleural pressure is necessary to sustain the expiratory airflow. This cycle-dependent adjustment reflects the dynamic redistribution of pressure and energy across the coupled respiratory-phonatory system during sustained phonation. Similar to the subglottal pressure, extended upper plateaus are observed in the intrapleural pressure waveform, which would not be expected in real-life phonation. As in the case of the subglottal pressure, this behavior arises from the limitations of the present one-dimensional modeling framework, which does not resolve fine-scale pressure fluctuations. Thus, these plateau regions are interpreted as intervals during which the respiratory system maintains a quasi-steady operating state. One prominent approach to further extend the proposed modeling paradigm is to take into account the power-law rheology and viscoelasticity of bio-tissues. This will be looked into in our future works, following our earlier developmental research in applied and computational mathematics \cite{suzuki2021anomalous, suzuki2021data, suzuki2022general, suzuki2023fractional, Zayernouri_Wang_Shen_Karniadakis_2024}.  

\section*{Conclusion}

This study presents a comprehensive, physics-based, and data-driven framework for modeling human voice production by integrating the respiratory, phonatory, and articulatory subsystems within a unified lumped-element modeling (LEM) paradigm. Unlike conventional approaches that rely on simplified or prescribed inlet conditions, the present work introduces a subject-specific, dynamically coupled system capable of reproducing the time-dependent aerodynamic and biomechanical interactions governing sustained phonation. The major findings of this study are summarized as follows. 

\begin{itemize}

\item The proposed model successfully captures the essential temporal dynamics of phonation using glottal area waveforms (GAW) extracted from high-speed videoendoscopy data. The simulated glottal area and glottal airflow exhibit stable, quasi-periodic oscillations with strong temporal synchronization, reflecting consistent cycle-to-cycle vocal fold vibration in a normophonic subject. The peak glottal area (approximately 8–9 mm$^2$) and peak glottal flow rate (200–240 ml/s) occur nearly simultaneously within each phonatory cycle, demonstrating the tight coupling between vocal fold kinematics and aerodynamic response.

\item The analysis further reveals that expiratory flow resistance is not a fixed parameter but a dynamically evolving quantity that closely follows the glottal airflow waveform. This indicates that downstream aerodynamic resistance is governed by the pulsatile nature of the glottal jet rather than being independently prescribed. In addition, the upper airway resistance exhibits pronounced nonlinear, flow-dependent behavior, increasing significantly during phases of high airflow due to complex aerodynamic mechanisms such as flow separation, vortex formation, and pressure recovery in the supraglottal region.

\item In the subglottal domain, the model predicts time-varying pressure dynamics characterized by plateau regions during glottal closure and rapid pressure drops during glottal opening. These oscillations reflect the cyclic storage and release of aerodynamic energy within the respiratory system. Similarly, intrapleural pressure and lung volume demonstrate smooth and continuous variations, capturing passive expiratory mechanics and the adaptive regulation of respiratory effort in response to time-varying phonatory loading. Together, these results confirm that phonation is governed by a strongly coupled interaction between lung compliance, elastic recoil, and vocal fold resistance, emphasizing that accurate modeling requires simultaneous consideration of all subsystems rather than isolated components.

\item While the model successfully reproduces the dominant features of phonatory dynamics, certain limitations arise from its lumped and one-dimensional formulation. In particular, the assumption of complete glottal closure and the absence of distributed acoustic effects limit the ability to capture fine-scale phenomena such as phase differences between glottal area and flow, as well as intracycle pressure fluctuations. These limitations highlight the need for future extensions incorporating three-dimensional vocal fold motion, inferior–superior phase differences, and vocal tract acoustic inertance.

\end{itemize}

Overall, this work establishes a unified, subject-specific lumped-element framework capable of simulating respiratory-driven phonation dynamics using experimentally derived inputs. The model provides a robust and physiologically grounded foundation for advancing computational voice modeling. It also enables systematic investigation of disordered phonation by offering a quantitative baseline for normophonic conditions. Future work will focus on extending the present framework to pathological voice conditions, incorporating fractional viscoelastic tissue behavior, and integrating distributed acoustic models to capture higher-fidelity aeroacoustic interactions. These advancements will further enhance the predictive capability and clinical relevance of the proposed modeling approach.

\bibliography{sample}

@book{Zayernouri_Wang_Shen_Karniadakis_2024,
  author    = {Zayernouri, Mohsen and Wang, Li-Lian and Shen, Jie and Karniadakis, George Em},
  title     = {Spectral and Spectral Element Methods for Fractional Ordinary and Partial Differential Equations},
  publisher = {Cambridge University Press},
  year      = {2024}
}

@article{suzuki2023fractional,
  title={Fractional modeling in action: A survey of nonlocal models for subsurface transport, turbulent flows, and anomalous materials},
  author={Suzuki, Jorge L and Gulian, Mamikon and Zayernouri, Mohsen and D’Elia, Marta},
  journal={Journal of Peridynamics and Nonlocal modeling},
  volume={5},
  number={3},
  pages={392--459},
  year={2023},
  publisher={Springer}
}

@article{suzuki2022general,
  title={A general return-mapping framework for fractional visco-elasto-plasticity},
  author={Suzuki, Jorge L and Naghibolhosseini, Maryam and Zayernouri, Mohsen},
  journal={Fractal and fractional},
  volume={6},
  number={12},
  pages={715},
  year={2022},
  publisher={MDPI}
}

@article{suzuki2021data,
  title={A data-driven memory-dependent modeling framework for anomalous rheology: Application to urinary bladder tissue},
  author={Suzuki, Jorge L and Tuttle, Tyler G and Roccabianca, Sara and Zayernouri, Mohsen},
  journal={Fractal and Fractional},
  volume={5},
  number={4},
  pages={223},
  year={2021},
  publisher={MDPI}
}

@article{suzuki2021anomalous,
  title={Anomalous nonlinear dynamics behavior of fractional viscoelastic beams},
  author={Suzuki, Jorge L and Kharazmi, Ehsan and Varghaei, Pegah and Naghibolhosseini, Maryam and Zayernouri, Mohsen},
  journal={Journal of computational and nonlinear dynamics},
  volume={16},
  number={11},
  pages={111005},
  year={2021},
  publisher={American Society of Mechanical Engineers}
}

@article{marconi2020silico,
  title={In silico study of airway/lung mechanics in normal human breathing},
  author={Marconi, Silvia and De Lazzari, Claudio},
  journal={Mathematics and computers in simulation},
  volume={177},
  pages={603--624},
  year={2020},
  publisher={Elsevier}
}

@article{bersani2017interaction,
  title={Interaction between the respiratory and cardiovascular system: a simplified 0-D mathematical model},
  author={Bersani, Alberto Maria and Bersani, Enrico and De Lazzari, Claudio and others},
  journal={Cardiovascular and Pulmonary Artificial Organs: Educational Training Simulators. CNR Edizioni},
  pages={87--103},
  year={2017}
}

@article{athanasiades2000energy,
  title={Energy analysis of a nonlinear model of the normal human lung},
  author={Athanasiades, A and Ghorbel, F and Clark Jr, JW and Niranjan, SC and Olansen, J and Zwischenberger, JB and Bidani, A},
  journal={Journal of Biological Systems},
  volume={8},
  number={02},
  pages={115--139},
  year={2000},
  publisher={World Scientific}
}

@article{sharp1969total,
  title={The Total Work of Breathing in Normal and Obese Men},
  author={Sharp, JT and Henry, JP and Sweany, SK and Meadows, WR and Pietras, RJ},
  journal={Journal of Occupational and Environmental Medicine},
  volume={11},
  number={8},
  pages={453--454},
  year={1969},
  publisher={LWW}
}

@article{mubbunu2018correlation,
  title={Correlation of internal organ weights with body weight and body height in normal adult Zambians: a case study of Ndola Teaching Hospital},
  author={Mubbunu, Lumamba and Bowa, Kasonde and Petrenko, Volodymyer and Silitongo, Moono},
  journal={Anatomy research international},
  volume={2018},
  number={1},
  pages={4687538},
  year={2018},
  publisher={Wiley Online Library}
}

@article{mogensen2011physiological,
  title={A model of ventilation of the healthy human lung},
  author={Steimle, Kristoffer Lindegaard and Mogensen, Mads Lause and Karbing, Dan S and de la Serna, J Bernardino and Andreassen, Steen},
  journal={computer methods and programs in biomedicine},
  volume={101},
  number={2},
  pages={144--155},
  year={2011},
  publisher={Elsevier}
}

@inproceedings{ronneberger2015u,
  title={U-net: Convolutional networks for biomedical image segmentation},
  author={Ronneberger, Olaf and Fischer, Philipp and Brox, Thomas},
  booktitle={Medical image computing and computer-assisted intervention--MICCAI 2015: 18th international conference, Munich, Germany, October 5-9, 2015, proceedings, part III 18},
  pages={234--241},
  year={2015},
  organization={Springer}
}

@article{vsvec2025application,
  title={Application of nonlinear dynamics theory to understanding normal and pathologic voices in humans},
  author={{\v{S}}vec, Jan G and Zhang, Zhaoyan},
  journal={Philosophical Transactions B},
  volume={380},
  number={1923},
  pages={20240018},
  year={2025},
  publisher={The Royal Society}
}

@article{inoue2024nonlinear,
  title={Nonlinear dynamics and chaos in a vocal-ventricular fold system},
  author={Inoue, Takumi and Shiozawa, Kota and Matsumoto, Takuma and Kanaya, Mayuka and Tokuda, Isao T},
  journal={Chaos: An Interdisciplinary Journal of Nonlinear Science},
  volume={34},
  number={2},
  year={2024},
  publisher={AIP Publishing}
}

@article{sadeghi2019computational,
  title={Computational models of laryngeal aerodynamics: Potentials and numerical costs},
  author={Sadeghi, Hossein and Kniesburges, Stefan and Kaltenbacher, Manfred and Sch{\"u}tzenberger, Anne and D{\"o}llinger, Michael},
  journal={Journal of Voice},
  volume={33},
  number={4},
  pages={385--400},
  year={2019},
  publisher={Elsevier}
}

@article{dollinger2023computational,
  title={Computational fluid dynamics of upper airway aerodynamics for exercise-induced laryngeal obstruction: A feasibility study},
  author={D{\"o}llinger, Michael and Jakuba{\ss}, Bernhard and Cheng, Hu and Carter, Stephen J and Kniesburges, Stefan and Aidoo, Bea and Lee, Chi Hwan and Milstein, Claudio and Patel, Rita R},
  journal={Laryngoscope Investigative Otolaryngology},
  volume={8},
  number={5},
  pages={1294--1303},
  year={2023},
  publisher={Wiley Online Library}
}

@article{cveticanin2012review,
  title={Review on mathematical and mechanical models of the vocal cord},
  author={Cveticanin, Livija},
  journal={Journal of Applied Mathematics},
  volume={2012},
  number={1},
  pages={928591},
  year={2012},
  publisher={Wiley Online Library}
}

@article{sidlof2010finite,
  title={Finite element modeling of airflow during phonation},
  author={Sidlof, Petr and Lun{\'e}ville, Eric and Chambeyron, Colin and Doar{\'e}, Olivier and Chaigne, Antoine and Hor{\'a}{\v{c}}ek, Jarom{\'\i}r},
  journal={Journal of Computational and Applied Mechanics},
  volume={4},
  pages={121--132},
  year={2010}
}

@article{de2015computational,
  title={Computational study of false vocal folds effects on unsteady airflows through static models of the human larynx},
  author={de Luzan, Charles Farbos and Chen, Jie and Mihaescu, Mihai and Khosla, Sid M and Gutmark, Ephraim},
  journal={Journal of biomechanics},
  volume={48},
  number={7},
  pages={1248--1257},
  year={2015},
  publisher={Elsevier}
}

@article{xue2010computational,
  title={A computational study of the effect of vocal-fold asymmetry on phonation},
  author={Xue, Q and Mittal, R and Zheng, X and Bielamowicz, S},
  journal={The Journal of the Acoustical Society of America},
  volume={128},
  number={2},
  pages={818--827},
  year={2010},
  publisher={AIP Publishing}
}

@article{tao2007asymmetric,
  title={Asymmetric airflow and vibration induced by the Coanda effect in a symmetric model of the vocal folds},
  author={Tao, Chao and Zhang, Yu and Hottinger, Daniel G and Jiang, Jack J},
  journal={The Journal of the Acoustical Society of America},
  volume={122},
  number={4},
  pages={2270--2278},
  year={2007},
  publisher={AIP Publishing}
}

@article{xue2014subject,
  title={Subject-specific computational modeling of human phonation},
  author={Xue, Qian and Zheng, Xudong and Mittal, Rajat and Bielamowicz, Steven},
  journal={The Journal of the Acoustical Society of America},
  volume={135},
  number={3},
  pages={1445--1456},
  year={2014},
  publisher={AIP Publishing}
}

@article{zheng2011computational,
  title={A computational study of asymmetric glottal jet deflection during phonation},
  author={Zheng, X and Mittal, R and Bielamowicz, S},
  journal={The Journal of the Acoustical Society of America},
  volume={129},
  number={4},
  pages={2133--2143},
  year={2011},
  publisher={AIP Publishing}
}

@article{scherer2001intraglottal,
  title={Intraglottal pressure profiles for a symmetric and oblique glottis with a divergence angle of 10 degrees},
  author={Scherer, Ronald C and Shinwari, Daoud and De Witt, Kenneth J and Zhang, Chao and Kucinschi, Bogdan R and Afjeh, Abdollah A},
  journal={The Journal of the Acoustical Society of America},
  volume={109},
  number={4},
  pages={1616--1630},
  year={2001},
  publisher={Acoustical Society of America}
}

@article{jiang2017computational,
  title={Computational modeling of fluid--structure--acoustics interaction during voice production},
  author={Jiang, Weili and Zheng, Xudong and Xue, Qian},
  journal={Frontiers in bioengineering and biotechnology},
  volume={5},
  pages={7},
  year={2017},
  publisher={Frontiers Media SA}
}

@article{jiang2022computational,
  title={Computational modeling of voice production using excised canine larynx},
  author={Jiang, Weili and Farbos de Luzan, Charles and Wang, Xiaojian and Oren, Liran and Khosla, Sid M and Xue, Qian and Zheng, Xudong},
  journal={Journal of Biomechanical Engineering},
  volume={144},
  number={2},
  pages={021003},
  year={2022},
  publisher={American Society of Mechanical Engineers}
}

@article{zhang2016cause,
  title={Cause-effect relationship between vocal fold physiology and voice production in a three-dimensional phonation model},
  author={Zhang, Zhaoyan},
  journal={The Journal of the Acoustical Society of America},
  volume={139},
  number={4},
  pages={1493--1507},
  year={2016},
  publisher={AIP Publishing}
}

@article{ibarra2021estimation,
  title={Estimation of subglottal pressure, vocal fold collision pressure, and intrinsic laryngeal muscle activation from neck-surface vibration using a neural network framework and a voice production model},
  author={Ibarra, Emiro J and Parra, Jes{\'u}s A and Alzamendi, Gabriel A and Cort{\'e}s, Juan P and Espinoza, V{\'\i}ctor M and Mehta, Daryush D and Hillman, Robert E and Za{\~n}artu, Mat{\'\i}as},
  journal={Frontiers in physiology},
  volume={12},
  pages={732244},
  year={2021},
  publisher={Frontiers Media SA}
}

@article{howe2013voicing,
  title={Voicing produced by a constant velocity lung source},
  author={Howe, MS and McGowan, RS},
  journal={The Journal of the Acoustical Society of America},
  volume={133},
  number={4},
  pages={2340--2349},
  year={2013},
  publisher={AIP Publishing}
}

@article{abur2022impact,
  title={Impact of vocal effort on respiratory and articulatory kinematics},
  author={Abur, Defne and Perkell, Joseph S and Stepp, Cara E},
  journal={Journal of Speech, Language, and Hearing Research},
  volume={65},
  number={1},
  pages={5--21},
  year={2022},
  publisher={American Speech-Language-Hearing Association}
}

@article{luizard2023flow,
  title={Flow-induced oscillations of vocal-fold replicas with tuned extensibility and material properties},
  author={Luizard, Paul and Bailly, Lucie and Yousefi-Mashouf, Hamid and Girault, Rapha{\"e}l and Org{\'e}as, Laurent and Henrich Bernardoni, Nathalie},
  journal={Scientific reports},
  volume={13},
  number={1},
  pages={22658},
  year={2023},
  publisher={Nature Publishing Group UK London}
}

@article{weerathunge2022auditory,
  title={Auditory and somatosensory feedback mechanisms of laryngeal and articulatory speech motor control},
  author={Weerathunge, Hasini R and Voon, Tiffany and Tardif, Monique and Cilento, Dante and Stepp, Cara E},
  journal={Experimental Brain Research},
  volume={240},
  number={7},
  pages={2155--2173},
  year={2022},
  publisher={Springer}
}

@article{naseri2023towards,
  title={Towards modeling of phonation and its recovery in unilateral vocal fold paralysis by fluid-structure interaction},
  author={Naseri, MohammadAmin and Razavi, Seyed Esmail},
  journal={BioImpacts: BI},
  volume={13},
  number={6},
  pages={488},
  year={2023}
}

@article{bodaghi2021effect,
  title={Effect of supraglottal acoustics on fluid--structure interaction during human voice production},
  author={Bodaghi, Dariush and Jiang, Weili and Xue, Qian and Zheng, Xudong},
  journal={Journal of Biomechanical Engineering},
  volume={143},
  number={4},
  pages={041010},
  year={2021},
  publisher={American Society of Mechanical Engineers}
}

@article{schickhofer2019compressible,
  title={Compressible flow simulations of voiced speech using rigid vocal tract geometries acquired by MRI},
  author={Schickhofer, Lukas and Malinen, Jarmo and Mihaescu, Mihai},
  journal={The Journal of the Acoustical Society of America},
  volume={145},
  number={4},
  pages={2049--2061},
  year={2019},
  publisher={AIP Publishing}
}

@article{zhang2016mechanics,
  title={Mechanics of human voice production and control},
  author={Zhang, Zhaoyan},
  journal={The journal of the acoustical society of america},
  volume={140},
  number={4},
  pages={2614--2635},
  year={2016},
  publisher={AIP Publishing}
}

@article{lehoux2021subglottal,
  title={Subglottal pressure oscillations in anechoic and resonant conditions and their influence on excised larynx phonations},
  author={Lehoux, Sarah and Hampala, V{\'\i}t and {\v{S}}vec, Jan G},
  journal={Scientific reports},
  volume={11},
  number={1},
  pages={28},
  year={2021},
  publisher={Nature Publishing Group UK London}
}

@article{alipour2004flow,
  title={Flow separation in a computational oscillating vocal fold model},
  author={Alipour, Fariborz and Scherer, Ronald C},
  journal={The Journal of the Acoustical Society of America},
  volume={116},
  number={3},
  pages={1710--1719},
  year={2004},
  publisher={Acoustical Society of America}
}

@article{hofmans2003unsteady,
  title={Unsteady flow through in-vitro models of the glottis},
  author={Hofmans, GCJ and Groot, G and Ranucci, M and Graziani, Giorgio and Hirschberg, A},
  journal={The Journal of the Acoustical Society of America},
  volume={113},
  number={3},
  pages={1658--1675},
  year={2003},
  publisher={Acoustical Society of America}
}

@article{alipour1995experimental,
  title={An experimental study of pulsatile flow in canine larynges},
  author={Alipour, F and Scherer, RC and Patel, VC},
  journal={Journal of fluids engineering},
  volume={117},
  number={4},
  pages={577--581},
  year={1995},
  publisher={American Society of Mechanical Engineers Digital Collection}
}

@article{kempster2009consensus,
  title={Consensus auditory-perceptual evaluation of voice: development of a standardized clinical protocol},
  author={Kempster, Gail B and Gerratt, Bruce R and Abbott, Katherine Verdolini and Barkmeier-Kraemer, Julie and Hillman, Robert E},
  journal={Clinical Focus},
  year={2009}
}

@article{fairbanks1940voice,
  title={Voice and articulation drillbook},
  author={Fairbanks, Grant},
  journal={(No Title)},
  year={1940}
}

@article{yousef2023spatial,
  title={Spatial segmentation for laryngeal high-speed videoendoscopy in connected speech},
  author={Yousef, Ahmed M and Deliyski, Dimitar D and Zacharias, Stephanie RC and de Alarcon, Alessandro and Orlikoff, Robert F and Naghibolhosseini, Maryam},
  journal={Journal of Voice},
  volume={37},
  number={1},
  pages={26--36},
  year={2023},
  publisher={Elsevier}
}

@article{yousef2022deep,
  title={A deep learning approach for quantifying vocal fold dynamics during connected speech using laryngeal high-speed videoendoscopy},
  author={Yousef, Ahmed M and Deliyski, Dimitar D and Zacharias, Stephanie RC and de Alarcon, Alessandro and Orlikoff, Robert F and Naghibolhosseini, Maryam},
  journal={Journal of Speech, Language, and Hearing Research},
  volume={65},
  number={6},
  pages={2098--2113},
  year={2022},
  publisher={American Speech-Language-Hearing Association}
}

@article{yousef2021hybrid,
  title={A hybrid machine-learning-based method for analytic representation of the vocal fold edges during connected speech},
  author={Yousef, Ahmed M and Deliyski, Dimitar D and Zacharias, Stephanie RC and de Alarcon, Alessandro and Orlikoff, Robert F and Naghibolhosseini, Maryam},
  journal={Applied Sciences},
  volume={11},
  number={3},
  pages={1179},
  year={2021},
  publisher={Mdpi}
}

@article{le2013mathematical,
  title={Mathematical modeling of respiratory system mechanics in the newborn lamb},
  author={Le Rolle, Virginie and Samson, Nathalie and Praud, Jean-Paul and Hern{\'a}ndez, Alfredo I},
  journal={Acta Biotheoretica},
  volume={61},
  number={1},
  pages={91--107},
  year={2013},
  publisher={Springer}
}

@article{golden2007mathematical,
  title={Mathematical modeling of pulmonary airway dynamics},
  author={Golden, James F and Clark, John W and Stevens, Paul M},
  journal={IEEE Transactions on Biomedical Engineering},
  number={6},
  pages={397--404},
  year={2007},
  publisher={IEEE}
}

@article{lee2014comparison,
  title={The comparison of the lengths and diameters of main bronchi measured from two-dimensional and three-dimensional images in the same patients},
  author={Lee, Jeong Woo and Son, Ji-Seon and Choi, Jin-Wook and Han, Young-Jin and Lee, Jun-Rae},
  journal={Korean journal of anesthesiology},
  volume={66},
  number={3},
  pages={189},
  year={2014}
}

@article{bankier1996bronchial,
  title={Bronchial wall thickness: appropriate window settings for thin-section CT and radiologic-anatomic correlation.},
  author={Bankier, Alexander A and Fleischmann, Dominik and Mallek, Reinhold and Windisch, Alfred and Winkelbauer, Friedrich W and Kontrus, Manfred and Havelec, Lieselotte and Herold, Christian J and H{\"u}bsch, Peter},
  journal={Radiology},
  volume={199},
  number={3},
  pages={831--836},
  year={1996}
}

@article{eschweiler2021biomechanics,
  title={The biomechanics of cartilage—An overview},
  author={Eschweiler, Joerg and Horn, Nils and Rath, Bjoern and Betsch, Marcel and Baroncini, Alice and Tingart, Markus and Migliorini, Filippo},
  journal={Life},
  volume={11},
  number={4},
  pages={302},
  year={2021},
  publisher={MDPI}
}

@article{dakin2011changes,
  title={Changes in lung composition and regional perfusion and tissue distribution in patients with ARDS},
  author={Dakin, Jonathan and Jones, Andrew T and Hansell, David M and Hoffman, Eric A and Evans, Timothy W},
  journal={Respirology},
  volume={16},
  number={8},
  pages={1265--1272},
  year={2011},
  publisher={Wiley Online Library}
}

@article{ward2005density,
  title={Density and hydration of fresh and fixed human skeletal muscle},
  author={Ward, Samuel R and Lieber, Richard L},
  journal={Journal of biomechanics},
  volume={38},
  number={11},
  pages={2317--2320},
  year={2005},
  publisher={Elsevier}
}

@article{hoffman2009reliable,
  title={Reliable time to estimate subglottal pressure},
  author={Hoffman, Matthew R and Baggott, Christopher D and Jiang, Jack},
  journal={Journal of Voice},
  volume={23},
  number={2},
  pages={169--174},
  year={2009},
  publisher={Elsevier}
}

@article{lucero2015smoothness,
  title={Smoothness of an equation for the glottal flow rate versus the glottal area},
  author={Lucero, Jorge C and Schoentgen, Jean},
  journal={The Journal of the Acoustical Society of America},
  volume={137},
  number={5},
  pages={2970--2973},
  year={2015},
  publisher={AIP Publishing}
}

@article{ali2025highspeed,
  author       = {Ali, S. N. B. and Naghibolhosseini, M. and Zayernouri, M.},
  title        = {High Speed Videoendoscopy-Driven Lumped Modeling of Vocal Fold Dynamics},
  journal      = {Biomechanics and Modeling in Mechanobiology},
  note         = {Under review},
  year         = {2026}
}

@article{palaparthi2020analysis,
  title={Analysis of glottal inverse filtering in the presence of source-filter interaction},
  author={Palaparthi, Anil and Titze, Ingo R},
  journal={Speech communication},
  volume={123},
  pages={98--108},
  year={2020},
  publisher={Elsevier}
}

@article{dejonckere2017dynamics,
  title={Dynamics of the driving force during the normal vocal fold vibration cycle},
  author={DeJonckere, Philippe Henri and Lebacq, Jean and Titze, Ingo R},
  journal={Journal of Voice},
  volume={31},
  number={6},
  pages={649--661},
  year={2017},
  publisher={Elsevier}
}

@article{rothenberg1977nonlinear,
  title={Nonlinear inverse filtering technique for estimating the glottal-area waveform},
  author={Rothenberg, Martin and Zahorian, Stephen},
  journal={the Journal of the Acoustical Society of America},
  volume={61},
  number={4},
  pages={1063--1071},
  year={1977},
  publisher={Acoustical Society of America}
}

@article{krane2007unsteady,
  title={Unsteady behavior of flow in a scaled-up vocal folds model},
  author={Krane, Michael and Barry, Michael and Wei, Timothy},
  journal={The Journal of the Acoustical Society of America},
  volume={122},
  number={6},
  pages={3659--3670},
  year={2007},
  publisher={AIP Publishing}
}

@article{luo2009analysis,
  title={Analysis of flow-structure interaction in the larynx during phonation using an immersed-boundary method},
  author={Luo, Haoxiang and Mittal, Rajat and Bielamowicz, Steven A},
  journal={The Journal of the Acoustical Society of America},
  volume={126},
  number={2},
  pages={816--824},
  year={2009},
  publisher={AIP Publishing}
}

@article{chang2013role,
  title={The role of finite displacements in vocal fold modeling},
  author={Chang, Siyuan and Tian, Fang-Bao and Luo, Haoxiang and Doyle, James F and Rousseau, Bernard},
  journal={Journal of biomechanical engineering},
  volume={135},
  number={11},
  pages={111008},
  year={2013},
  publisher={American Society of Mechanical Engineers}
}

@article{granados2017numerical,
  title={A numerical strategy for finite element modeling of frictionless asymmetric vocal fold collision},
  author={Granados, Alba and Misztal, Marek Krzysztof and Brunskog, Jonas and Visseq, Vincent and Erleben, Kenny},
  journal={International Journal for Numerical Methods in Biomedical Engineering},
  volume={33},
  number={2},
  pages={e02793},
  year={2017},
  publisher={Wiley Online Library}
}

@article{yang2010biomechanical,
  title={Biomechanical modeling of the three-dimensional aspects of human vocal fold dynamics},
  author={Yang, Anxiong and Lohscheller, J{\"o}rg and Berry, David A and Becker, Stefan and Eysholdt, Ulrich and Voigt, Daniel and D{\"o}llinger, Michael},
  journal={The Journal of the Acoustical Society of America},
  volume={127},
  number={2},
  pages={1014--1031},
  year={2010},
  publisher={AIP Publishing}
}

@article{perrine2023using,
  title={Using a vertical three-mass computational model of the vocal folds to match human phonation of three adult males},
  author={Perrine, Brittany L and Scherer, Ronald C},
  journal={The Journal of the Acoustical Society of America},
  volume={154},
  number={3},
  pages={1505--1525},
  year={2023},
  publisher={AIP Publishing}
}

@article{patel2022glottal,
  title={Glottal airflow and glottal area waveform characteristics of flow phonation in untrained vocally healthy adults},
  author={Patel, Rita R and Sundberg, Johan and Gill, Brian and L{\~a}, Filipa MB},
  journal={Journal of Voice},
  volume={36},
  number={1},
  pages={140--e1},
  year={2022},
  publisher={Elsevier}
}

@article{alipour1997pressure,
  title={Pressure-flow relationships during phonation as afunction of adduction},
  author={Alipour, Fariborz and Scherer, Ronald C and Finnegan, Eileen},
  journal={Journal of Voice},
  volume={11},
  number={2},
  pages={187--194},
  year={1997},
  publisher={Elsevier}
}

@article{krausert2011mucosal,
  title={Mucosal wave measurement and visualization techniques},
  author={Krausert, Christopher R and Olszewski, Aleksandra E and Taylor, Lindsay N and McMurray, James S and Dailey, Seth H and Jiang, Jack J},
  journal={Journal of Voice},
  volume={25},
  number={4},
  pages={395--405},
  year={2011},
  publisher={Elsevier}
}

\section*{Acknowledgments}
The authors would like to acknowledge the support from the National Institutes of Health (NIH), National Institute on Deafness and Other Communication Disorders (NIDCD) [R21DC020003 and K01DC017751], Army Research Office (ARO) [W911NF-19-1-0444], National Science Foundation [DMS-1923201], and Michigan State University Discretionary Funding Initiative. The computational resources were provided by the Institute for Cyber-Enabled Research (ICER) at Michigan State University. The authors would also like to thank Dr. Dimitar D Deliyski, Dr. Stephanie RC Zacharias, and Mayo Clinic for their help with the data collection, and Sardar Nafis Bin Ali for his help with part of the data analysis.

\end{document}